\documentclass[10pt,letterpaper]{article}
\usepackage[top=0.85in,left=0.9in,footskip=0.75in]{geometry}

\usepackage{amsmath,amssymb}

\usepackage{mathtools}           

\usepackage{changepage}

\usepackage[utf8x]{inputenc}

\usepackage{textcomp,marvosym}

\usepackage{cite}

\usepackage{nameref,hyperref}

\usepackage[right]{lineno}

\usepackage{microtype}
\DisableLigatures[f]{encoding = *, family = * }

\usepackage[table]{xcolor}

\usepackage{array}

\newcolumntype{+}{!{\vrule width 2pt}}

\newlength\savedwidth

\raggedright
\usepackage[aboveskip=1pt,labelfont=bf,labelsep=period,justification=raggedright,singlelinecheck=off]{caption}
\renewcommand{\figurename}{Fig}

\makeatletter
\renewcommand{\@biblabel}[1]{\quad#1.}
\makeatother

\date{}

\usepackage{lastpage,fancyhdr,graphicx}
\usepackage{epstopdf}

\pdfoutput=1

\newcommand{\dg}{^{\circ}}
\newcommand\ve[1]{\boldsymbol{#1}}
\newcommand\subfiglabel[1]{\textbf{\textsf{#1}}}

\begin{document}
\vspace*{0.2in}

\begin{flushleft}
{\Large
  \textbf\newline{Grid cells show field-to-field variability and this explains the aperiodic response of inhibitory interneurons}
}
\newline
\\
Benjamin Dunn\textsuperscript{1\dag*},
Daniel Wennberg\textsuperscript{1,2\dag},
Ziwei Huang\textsuperscript{1,3},
Yasser Roudi\textsuperscript{1,4}
\\
\bigskip
\textbf{1} Kavli Institute for Systems Neuroscience / Centre for Neural Computation, NTNU, 7030 Trondheim, Norway
\\
\textbf{2} Department of Applied Physics, Stanford University, Stanford, California 94305, USA
\\
\textbf{3} Werner Reichardt Centre for Integrative Neuroscience, University of Tubingen, D-72076 Tubingen, Germany
\\
\textbf{4} Simons Center for Systems Biology, Institute for Advanced Study, Princeton, New Jersey 08540, USA
\\
\bigskip
\dag These authors contributed equally to this work.

* benjamin.dunn@ntnu.no
\end{flushleft}

\section*{Abstract}
Research on network mechanisms and coding properties of grid cells assume that the firing rate of a grid cell in each of its fields is the same.  
Furthermore, proposed network models predict spatial regularities in the firing of inhibitory interneurons that are inconsistent with experimental data.
In this paper, by analyzing the response of grid cells recorded from rats
during free navigation, we first show that there are strong variations in the
mean firing rate of the fields of individual grid cells and
thus show that the data is inconsistent with the theoretical models that
predict similar peak magnitudes. 
We then build a two population excitatory-inhibitory network model in which
sparse spatially selective input to the excitatory cells, presumed to arise
from e.g. salient external stimuli, hippocampus or a combination of both, leads
to the variability in the firing field
amplitudes of grid cells. We show that, when combined with appropriate
connectivity between the excitatory and inhibitory neurons, the variability in
the firing field amplitudes of grid cells results in inhibitory neurons that do not exhibit regular spatial firing, consistent with experimental data.  
Finally, we show that even if the spatial positions of the fields are maintained, variations in the firing rates of the fields of grid cells are enough to cause remapping of hippocampal cells.


\section{Introduction}

Grid cells in the medial entorhinal cortex (MEC) of mammals are selectively
active in positions that together form a hexagonal pattern tessellating the
space in which an animal moves~\cite{hafting2005microstructure}.  These cells,
along with other cells in the MEC that code for various spatial quantities,
e.g.\ speed~\cite{kropff2015speed}, head
direction~\cite{taube1990head,sargolini2006conjunctive},
borders~\cite{solstad2008representation}, as well as hippocampal place cells
\cite{o1971hippocampus,o1976place}, are believed to form the neural basis of
spatial navigation in mammals. Over the past decade, a great deal of knowledge
has been collected about the properties of grid cells, thereby gaining better
insight into the neural mechanisms involved in spatial navigation
\cite{rowland2016ten}.

Since the discovery of grid cells, a significant amount of work has been
devoted to building models for explaining the underlying neural basis of the
spatial selectivity of grid cells~\cite{mcnaughton2006path,fuhs2006spin,burak2009accurate,couey2013recurrent,pastoll2013feedback,widloski2014model,solanka2015noise,burgess2007oscillatory,hasselmo2007grid,blair2007scale,bush2014hybrid,mhatre2012grid,kropff2008emergence}.
Among these studies, network models based on the concept of continuous attractor dynamics explain the hexagonal firing of grid cells by proposing the generation of a hexagonal pattern of activity on a two dimensional neural array, which is then moved with the animal using velocity input. These continuous attractor models as well as studies on the coding of position by grid cells \cite{solstad2006grid,rolls2006entorhinal,fiete2008grid,sreenivasan2011grid,mathis2012resolution}
assume or predict that the different firing fields of a grid cell are periodic replicas of one another, and that the mean firing
rate, or amplitude, of the firing fields are similar. 

In this paper, by analyzing $373$ experimentally recorded grid cells, we first
demonstrate that, contrary to the predictions and assumptions in previous work,
there is significant field-to-field variability in the response of
grid cells, indicating that individual fields of a cell enjoy a significant degree of
independence even though they are laid out on a periodic lattice. We then show that the variability in the fields is instrumental in answering one of the problems of 
existing implementations of continuous attractor models of grid cells: that contrary to experimental findings, they predict regular firing for inhibitory interneurons ~\cite{buetfering2014parvalbumin}. Finally, by considering a feedforward network in which summation of grid cell activity of various scales leads to spatially localized activity akin to that
of hippocampus place cells, we show that changing the relative amplitude of the
fields of grid cells, while maintaining the spatial relationship between the
firing of the cells, leads to global remapping, as has been observed in
a recent experimental study by Kanter et al.~\cite{kanter2016}.

Continuous attractor models mainly consider a single population of grid cells, and the role of
inhibitory neurons is included in the effective interaction between the
neurons in this single population~\cite{mcnaughton2006path,guanella2007model,burak2009accurate,couey2013recurrent,fuhs2006spin}.
Those cases which do consider separate excitatory and inhibitory populations
\cite{pastoll2013feedback,widloski2014model,solanka2015noise} rely on
connectivity rules which result in inhibitory interneurons with grid- or
``antigrid''-like spatial tuning. This was shown to not hold true in
experimental recordings from the interneurons in the MEC, which did show
spatial selectivity but little or no spatial periodicity in the firing patterns
of the interneurons~\cite{buetfering2014parvalbumin}. Our analysis of a network of inhibitory and excitatory neurons shows that field-to-field variability together with sparse connectivity between inhibitory and excitatory neurons can explain this experimental finding. Grid cells in the network simulations that we report here express a hexagonal firing pattern through the interacting effect of recurrent connectivity and sparse spatially selective input: some of the neurons are activated by place-like input when the animal visits a certain position, and these neurons activate other cells that should be active at the position via effective inhibitory interactions. The model is thus different from existing continuous attractor models in that the generation of grid activity does not require moving a hexagonal pattern of activity on the network by integrating velocity \cite{mcnaughton2006path}.


\section{Results}

\subsection{Recorded grid cells show field-to-field variability}

We analyzed grid cells recorded from rats with multisite (hyperdrive) implants,
reported in Stensola et al.~\cite{stensola2012entorhinal}.
The cells were recorded from the MEC while the animals explored $1$ or $1.5
\ \mathrm{m}$ square, open-field environments.  Analysis was performed per grid
cell module. As in \cite{stensola2012entorhinal}, modules were identified
according to the firing field spacing, orientation and deformation relative to
a perfect hexagonal pattern. In total, we analyzed recordings from $373$ cells
distributed among $13$ modules and five rats (see sections \ref{supp:Data}-\ref{supp:Idealized} for
details).

Firing fields were detected in grid cell rate maps via watershed segmentation
(see section \ref{sec:firing_field_detection} for details). We defined the amplitude of a firing
field as the ratio of the number of spikes recorded to the time spent within
a circular region enclosing the field. 

\begin{figure}
  \centering
  \makebox[\linewidth][c]{
   \begin{minipage}{0.49\linewidth}
    \centering
    \includegraphics[width=0.9\linewidth]{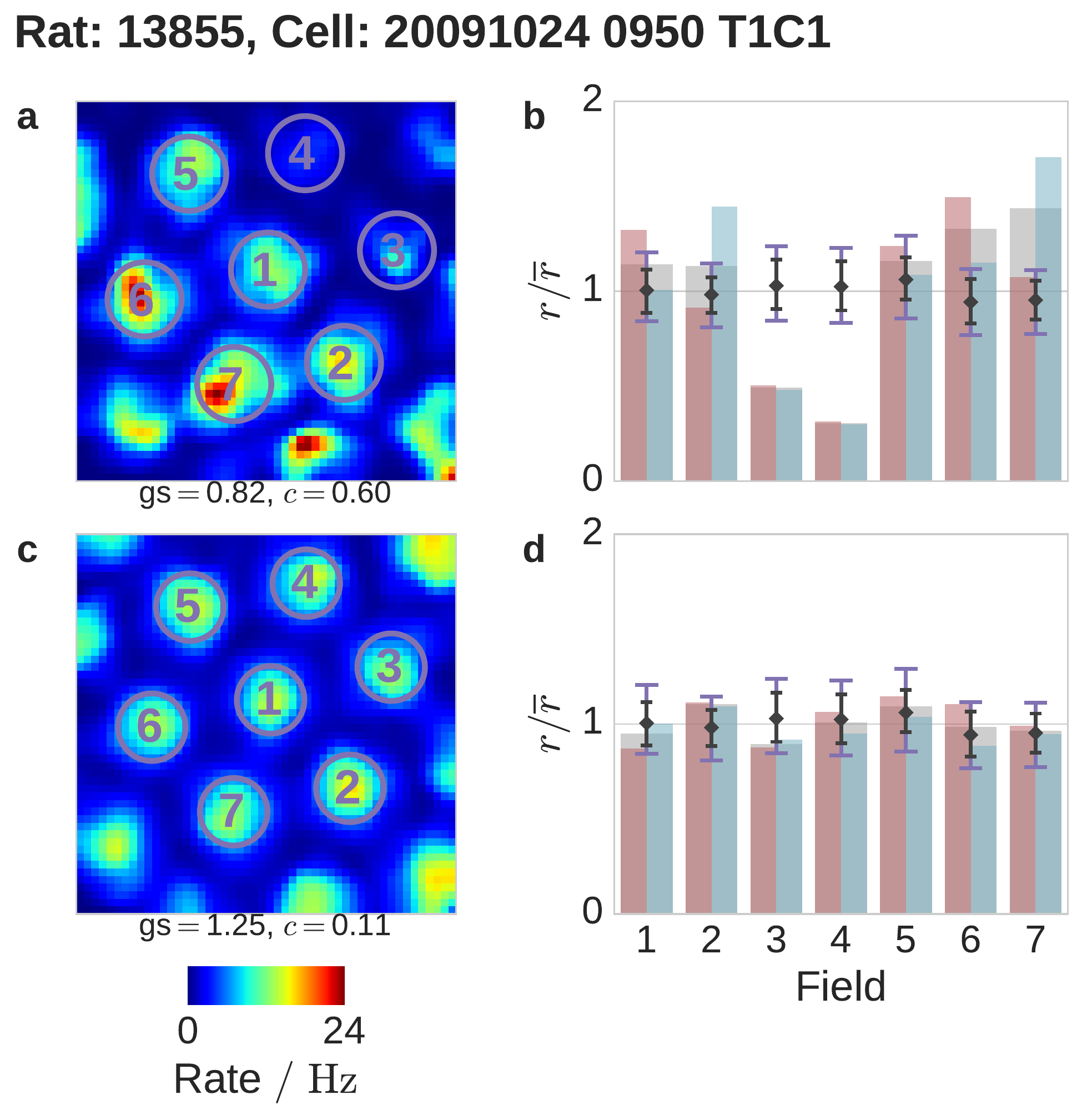}
    \caption{\label{fig:synt-real-ratemap-amplitudes}
      \textbf{Example grid cell rate map and firing field amplitudes, with
      simulated counterpart.}
      \subfiglabel{a}: A typical grid cell rate map with detected firing fields
      circled and numbered. The grid score ($\text{gs}$)
      and field amplitude coefficient of variation ($c$) are
      quoted below the rate map.
      \subfiglabel{b}: The amplitude of each field in
      \subfiglabel{a}, normalized to the mean amplitude of all
      fields from the same rate map, is shown (gray wide bars in the
      background) together with the amplitudes measured in only the first and
      second halves of the session (red and light blue bars in the foreground).
      On top of this, markers and error bars visualize the distribution of
      normalized amplitudes from $1000$ synthetic spike trains, with the black
      diamond showing the mean amplitude, the small black whiskers showing the
      central $95 \ \%$ of the full-length amplitudes (to be compared with the
      gray wide bars), and the large purple whiskers showing the central $95
      \ \%$ of the half-length amplitudes (to be compared with the red and
      light blue narrow bars). For each cell and synthetic spike train,
      amplitudes are normalized to the mean of the full-length amplitudes from
      that cell or spike train when creating this plot.
      \subfiglabel{c-d}: Equivalent to \subfiglabel{a-b}, but using synthetic
      spikes generated from the real animal trajectory and a tuning profile
      consisting of identical firing fields. Another $1000$ synthetic spike
      trains like this were used to generate the error bars in \subfiglabel{b}
      and \subfiglabel{c}, and to perform the statistical analysis of the
    variability.}
  \end{minipage}
  \begin{minipage}{0.49\linewidth}
    \centering
    \includegraphics[width=0.9\linewidth]{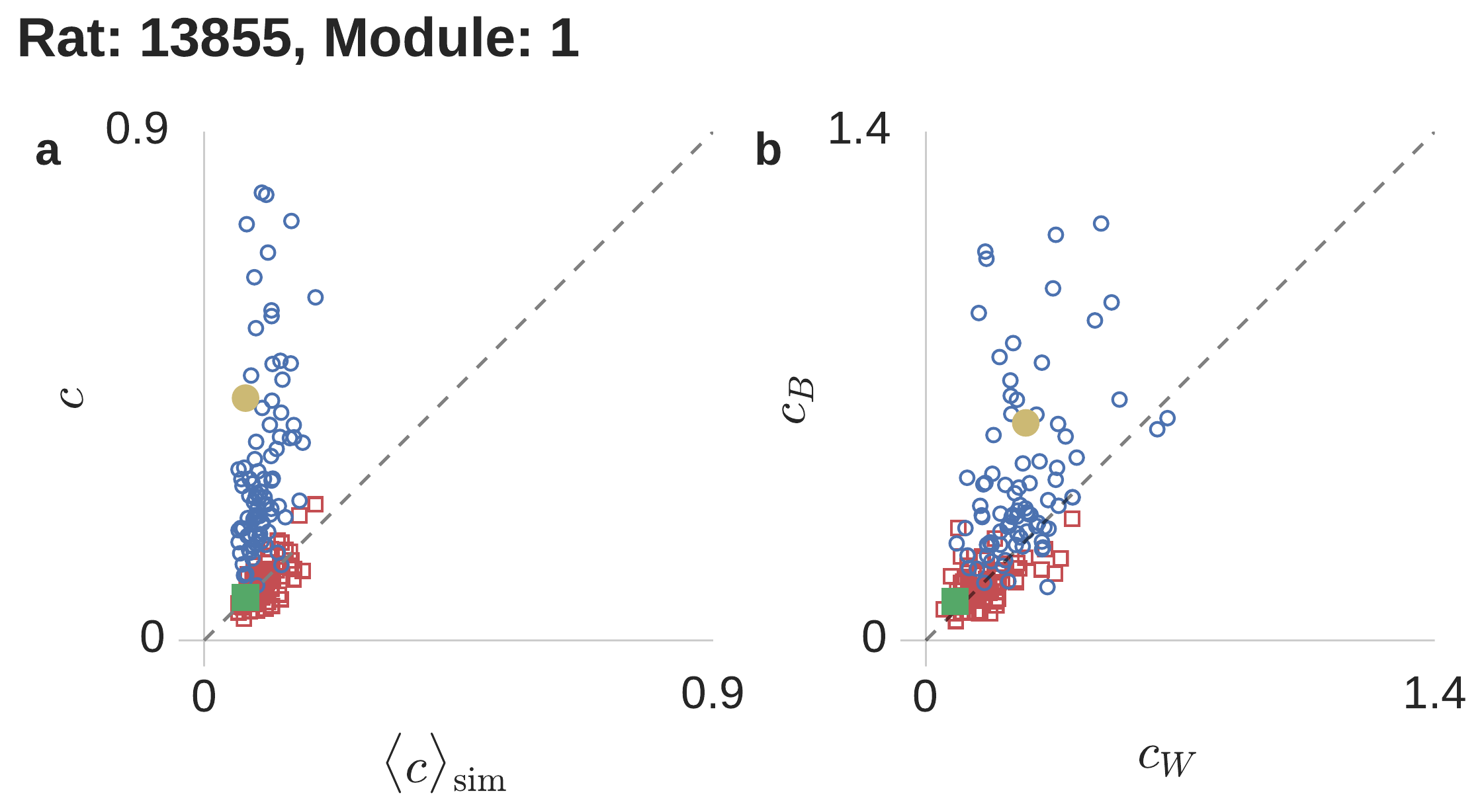}
    \caption{\label{fig:within-between-c}
      \textbf{Firing field amplitude variability.}
      \subfiglabel{a}: Blue circles show the field amplitude coefficient of
      variation of each cell in a module, scattered against the mean
      coefficient of variation from the $1000$ synthetic spike trains
      corresponding to that cell. Red squares show the same quantity for one
      particular synthetic spike train per real cell.
      \subfiglabel{b}: Between-field coefficient of variability scattered
      against within-field coefficient of variability from the same real cells
      and synthetic spikes as in \subfiglabel{a}. Variabilities are computed
      from field amplitudes measured in the first and second halves of the
      experimental sessions separately.
      For each cell, we only included the amplitudes of firing fields that were
      detected in both the real cell rate map and all corresponding synthetic
      rate maps. If less than $3$ such fields were found, the cell was
      discarded.
      The enlarged, filled markers in \subfiglabel{a} and \subfiglabel{b}
  correspond to the cell (yellow circle) and synthetic spikes (green square)
  shown in figure \ref{fig:synt-real-ratemap-amplitudes}.}
\end{minipage}
}
\end{figure}

Although grid cell rate maps often show clear field-to-field variations, as
can be seen in figure \ref{fig:synt-real-ratemap-amplitudes} \subfiglabel{a}, it is not
obvious whether this variability is mainly induced by some combination of
temporal fluctuations in the cell response, stochastic spiking dynamics, and
the non-uniform occupancy typically obtained in open-field experiments, or if
it is a stable property of the spatial tuning of the cell. We performed
a careful statistical analysis to assess the degree of inherent variability in
the firing rate of the fields, by measuring the mean firing rate within each
firing field, both over the whole experiment and in the first and second halves
of the experiment separately. These data are compared, on a cell-by-cell basis,
against a synthetic population derived from a spatial tuning profile with
identical firing fields, but having the same spacing and spatial phase as the
corresponding real cell. Synthetic spike trains were generated from this tuning
profile using the real trajectory and assuming Poisson spiking with spatially
modulated rate (full details of the procedure is given in section
\ref{supp:SampledSpikes}.) An example of how a real cell and synthetic spikes
compare is shown in figure \ref{fig:synt-real-ratemap-amplitudes}
\subfiglabel{b}.  The figure demonstrates that, when compared to the synthetic
spikes, the real cell shows considerably more variation in firing rate from one
field to another, and these variations can be substantially larger than the
variation from the first to the second half of the experiment. This indicates
that field-to-field variations are stable and cannot be explained by occupancy
and fluctuations alone.

To make this observation quantitative, we also computed the coefficient of variation $c$
of the amplitudes of each cell, and $1000$ synthetically generated spike trains
per real cell (see Eq.~\ref{eq:CVdef} in section \ref{sec:stanalysis_supp}).
Figure \ref{fig:within-between-c} \subfiglabel{a} shows the measured coefficient
of variation of all cells in one module, scattered against the corresponding
mean $\langle c \rangle_{\mathrm{sim}}$ computed from synthetic spikes, and
compared directly to a single synthetic spike train realization per real cell.
The figure clearly shows that the amplitudes of real cells vary much more than
those of the synthetic spike trains.

While this comparison gives a first indication of the degree of variation in
grid cell field amplitudes, it does not take into account the possibility that
the measured field amplitudes in a finite-time experiment may be affected by
factors not accounted for by the synthetic spike trains, such as the simultaneous
encoding of other kinds of information in the grid cell (i.e.\ head direction),
as well as non-Poisson and possibly temporally varying spiking statistics. We therefore
considered data from the first and second halves of the experiments
separately, and asked whether amplitude variation between fields exceeded what
would be expected based on the variability over time of the amplitudes of the
same cell.  Figure \ref{fig:within-between-c} \subfiglabel{b} shows the
between-field coefficient of variability $c_B$ scattered against the
within-field coefficient $c_W$ for each cell in a module, as well as for
a single synthetic spike train per real cell. The coefficients $c_B$ and $c_W$
are computed from between-field and within-field variabilities, quantities that
are standard in one-way ANOVA  (see section \ref{sec:stanalysis_supp} for
details). Hence, a $c_B$ that is significantly greater than $c_W$ indicates
that we see more variation between the field amplitudes than what would be
expected based on the temporal fluctuations we observe. From the figure we see
that this is the case for a large fraction of the real cells, but certainly not
all. On the other hand, all the simulated cells have $c_B$ and $c_W$ of
comparable magnitudes and somewhat smaller than the typical values for the real
cells.  Figure \ref{fig:synt-real-ratemap-amplitudes} \subfiglabel{b} shows what this
can look like in terms of actual field amplitudes: the difference between the
first and second half is quite large for certain firing fields when compared to
the synthetic data, but still much smaller than the variation between different
fields in the real data.

To condense this notion into a single number, we computed the $F$-statistic $F
= c_B^2 / c_W^2$. We used the $1000$ synthetic spike trains to generate
a cell-specific null distribution for the $F$-statistic, which was then used to
compute a  $p$-value by ranking the real $F$-statistic against this
distribution. Note that although we once again invoke the synthetic data here,
we only compare between-field variabilities to within-field variabilities from
the same spike train, such that the effects of factors not accounted for in the
synthetic spike generation should cancel out to a first approximation. Thus we
expect that, regardless of additional factors affecting the between-field
variation, each generated distribution for the $F$-statistic will predict with
reasonable accuracy the outcome for a real cell with this particular
trajectory, firing field locations, and mean firing rate, assuming that the
firing fields are identical in terms of the strength of the underlying spatial
tuning. This assumption could be rejected with $p < 0.05$ for $24$ of the $86$
cells in the module shown in figure \ref{fig:within-between-c}, and $129$ of
the $373$ cells in total.

The fact that only a little more than a third of the cells have a statistically
significant between-field to within-field variability ratio reflects
a continuum in the cell population, from cells where this ratio is of order
unity to cells to cells with very large field amplitude variations. This
property is also clearly visible in figure \ref{fig:within-between-c}. However,
the number of statistically significant cases greatly exceeds the chance level
(i.e.\ $5 \ \%$ of the population). By interpreting rejection or not of the
null hypothesis as a Bernoulli trial with success probability $0.05$, the
number of rejections in a population of cells should follow a binomial
distribution under the null-hypothesis, and thus we can define an aggregate
$p$-value $p_{\text{tot}}$ for the number of rejections in the population. For
the module shown in figure \ref{fig:synt-real-ratemap-amplitudes},
this comes out to $p_{\text{tot}} = 3.5 \cdot 10^{-12}$, and for all cells in
the data set we find $p_{\text{tot}} = 7.5 \cdot 10^{-71}$. This leads us to
conclude that the discrete translational symmetry of the grid cell network with
respect to physical space is broken not only by geometrical distortions of the
lattice due to the environment~\cite{Krupic2015}, but also in terms of the
firing field strength at each site.

Supplementary figure \ref{fig:supp-within-between-c} contains figures similar
to figure \ref{fig:within-between-c} for all modules in the data set,
containing between $2$ and $93$ cells after discarding cells for which fewer
than three firing fields were detected. On average, about $35 \ \%$ of the
cells in a module show statistically significant amplitude variations. We find
$p_{\text{tot}} < 0.05$ for $9$ of the $13$ modules, and $p_{\text{tot}}
< 0.005$ for $8$ of these.

\subsection{Basic aspects of the recurrent excitatory-inhibitory models}

Networks models of grid cells based on continuous attractor dynamics and
integration of velocity involve two key steps. The
first one is to generate a translationally periodic pattern of activity in
a network of neurons, and the second is to
translate this pattern along with the animal movement using speed and head
direction input \cite{mcnaughton2006path}. The fact that the fields of a single
grid cell appear due to the translation of a regular pattern, guarantees that
in these models each firing field of any given grid cell is the same as another
field of the same cell.  The analysis in the previous section thus demonstrates
that this simple mechanism cannot be the whole story and other mechanisms are
in play to generate the field-to-field variations in the firing rates of a grid
cell. 

As described in the Introduction, recent experimental evidence also shows that
inhibitory neurons in the MEC do not exhibit spatially
periodic firing \cite{buetfering2014parvalbumin}, as predicted by existing
continuous attractor models.  In this section we consider a two population
network of excitatory and inhibitory neurons and show that, with appropriate
connectivity, field-to-field variability can lead to irregular firing, and that
this comes at the cost of the ability to integrate velocity properly. 

We consider a simple two-population models of the
form~\cite{wilson1972excitatory}:
\begin{align}
  &\tau \frac{ds_i}{dt} + s_i = g\big( \sum\nolimits_b J_{ib} u_b + I_{it}
\big)_+ \ ,\\
  &\tau \frac{du_a}{dt} + u_a = g\big( \sum\nolimits_j K_{aj} s_j \big)_+ \ ,
\label{eq:motion}
\end{align}
where $s_i$ and $u_a$ represent the firing rates of the excitatory neuron (grid
cell) $i$ and inhibitory neuron (interneuron) $a$, respectively. $(\cdots)_+$
is the threshold-linear function, which outputs zero if its argument if equal
to or less than zero and is a linear function for positive arguments.  $g$ is
the single neuron gain, $\tau$ the neuronal time constant, $J_{ib}$ and
$K_{aj}$ the connectivity from interneurons to grid cells and grid cells to
interneurons, and $I_{it}$ the external input. 

We consider different forms of external input $I_{it}$ during the simulations
reported below. Classically in continuous attractor networks of grid cells, one considers a velocity dependent of the typical form
\begin{equation} 
I_{it} = I^{v}_{it} \equiv \textbf{constant} + \alpha \times \textbf{speed}(t)
\cos(\theta_t - \theta_i) \ ,
\label{vinput}
\end{equation}
where the superscript $v$ emphasizes the velocity dependence of the input,
$\textbf{speed}(t)$ and $\theta_t$ are the time-varying speed and direction of
the animal as it moves around, $\alpha$ the velocity modulation and  $\theta_i$
the preferred direction of neuron $i$, patterned as in
Burak and Fiete~\cite{burak2009accurate} to ($0$, $\pi/4$,
$\pi/2$, $3 \pi/4$).

However, as noted in the Introduction, with pure velocity input we cannot see
any variability in the firing field amplitudes of grid cells: even if the connectivity is heterogeneous
such that in the stationary pattern the peak activity at some positions in the network ends up being different from another one,
the translation of the pattern will lead the fields of each grid cell to have
exactly the same amplitudes, with the heterogeneity only causing 
the rates being different from one cell to another, and not different within one cell.
We therefore consider the case in which the 
hexagonal firing pattern and the variability in the fields of one cell
arise from a different type of input to the excitatory neurons in the model,
the place-like input, that could originate from a combination of sensory input
as well as the hippocampus or other areas with spatially selective neurons
projecting to the MEC. Assuming a population of place-like neurons whose rates
are denoted $P_j(t)$, we take the input to be of the form
\begin{equation}
I_{it} = I^{p}_{it} \equiv \textbf{constant} + \sum\nolimits_{j} H_{ij}P_j(t)
\ ,
\label{pinput}
\end{equation}
where the superscript $p$ indicates the place-like selectivity of the cells
contributing to this type of input, and $H_{ij}$ are weights that connect the
place-like neurons to the excitatory grid cell population. 
The weights are chosen from place-like inputs that are active 
at hexagonally arranged fields in the environment. The amplitude
of these $H_{ij}$ are drawn from a Gaussian distribution of mean $\mu=0.5$ and
standard deviation $\sigma$ with negative values set to zero, as suggested in
\cite{brunel2004optimal}.  We also dilute the
input by setting a fraction, $f_d$, of these connections to zero, allowing us
to study not only the effect of variability in the amplitude of the place-like
input but also the sparsity of the input.  In what follows, we will study the
effect of $\sigma$ and $f_d$ on the quality of the grid cells generated by the
model as well as the field-to-field variability. But before doing that, we
discuss in the next section two ways of wiring up the neurons in the model,
namely the choices for $J$ and $K$ in Eq. \ref{eq:motion}.

Examples of networks with inputs of the form in equation \ref{vinput} and
\ref{pinput} can be seen in accompanying videos 1 and 2.

\subsection{Two ways of generating the connectivity matrices}

Having described the types of input to the network that we consider,
in this section we detail two ways of designing the connectivity pattern
between neurons in the model. To do so, we first note that as $du_a/dt
\rightarrow 0$, we have $u_a = g \sum\nolimits_j K_{aj} s_j $, since $K_{aj}$
is non-negative and $g(\cdot)_+$ is a zero-threshold linear activation
function. We can therefore approximate the term $\sum\nolimits_b J_{ib} u_b
\approx g \sum\nolimits_{bj} J_{ib} K_{bj} s_j$, to end up with the following
effective connectivity between the excitatory neurons:
\begin{equation} \nonumber
W_{ij} = g \sum\nolimits_b J_{ib} K_{bj} \ .
\end{equation}

As a simple effective connectivity, similar to that used in the literature~\cite{burak2009accurate,couey2013recurrent}, we let 
\begin{equation}
  W_{ij} = W_0 \Theta(R_{max}-d_{ij}) \Theta(d_{ij} - R_{min}) \ ,
\end{equation}
where $\Theta(\cdot)$ is the Heaviside step function, $R_{max}$ the outer ring
of the radial extent of the connectivity, $R_{min}$ the inner ring, $W_0$ the
strength of the inhibitory connections and $d_{ij}$ the Euclidean distance
between cells $i$ and $j$.

For the simulations reported below, where the external input includes the velocity terms
the Euclidean distance, $d_{ij}$ is biased in the preferred direction
$\theta_i$ of the grid cell~\cite{burak2009accurate},  $d^2_{ij} = (x_i-x_j-l
\cos(\theta_i))^2 + (y_i-y_j-l \sin(\theta_i))^2$, with $x_i \in [1,N_x]$, $y_i
\in [1,Ny]$ representing the position of neuron $i$ in a two dimensional $N_x
\times N_y$ neural sheet with periodic boundary conditions and spatial offset
$l$. Each grid cell in the model is therefore connected with other neurons
within a radius between $R_{\min}$ and $R_{\max}$, forming an inhibitory ring
connectivity. For simulations where place-like input is used, no bias exists in the connectivity.

Our two network architectures now amount to the different ways of choosing the connectivity
matrices $\mathbf{J}$ and $\mathbf{K}$ to obtain the prescribed effective
connectivity $\mathbf{W}$. In principle, this problem is under-constrained and
has an infinite number of solutions.  Conveniently, it has been shown that grid
cells in superficial layers of MEC are primarily connected through
inhibition~\cite{couey2013recurrent, pastoll2013feedback}, making the matrix
$W_{ij}$ strictly non-positive while also justifying the lack of direct
connectivity between the excitatory neurons $s_i$.  Furthermore, by Dale's law,
$K_{ib}$ is strictly non-negative and $J_{aj}$ is non-positive, hence we
rewrite the factorization problem as $(-\mathbf{W})
= g(-\mathbf{J})\mathbf{K}$, where all three matrices are non-negative.

We considered two possible ways of generating the connectivity.  First, we
initialized both $-\mathbf{J}$ and $\mathbf{K}$ to random numbers drawn from
a uniform distribution on $[0,0.1]$ and solved for these iteratively using
methods for factorizing non-negative matrices
(NMF)~\cite{lee1999learning,paatero1994positive}. NMF is an iterative method
that converges to a local minimum of the cost function $\sqrt{\sum_{ij}
\big(W_{ij} - \sum_k J_{ik} K_{kj} \big)^2}$ (see section 
\ref{sec:twopopcanmodel_supp}-\ref{supp:NMF} for
additional details).  This minimum is not necessarily the global minimum, so
different initializations may result in different connectivity patterns. 
We refer to this approach as the \emph{fully factorized} model. 
For the second approach, which we label the \emph{partially factorized}
model, we set the connections to the interneurons $\mathbf{K}$ to sparse ($10
\ \%$) random excitatory connectivity, and solved for $\mathbf{J}$ only using NMF.  

\begin{figure}[h!]
    \centering
    \includegraphics[width=0.84\linewidth]{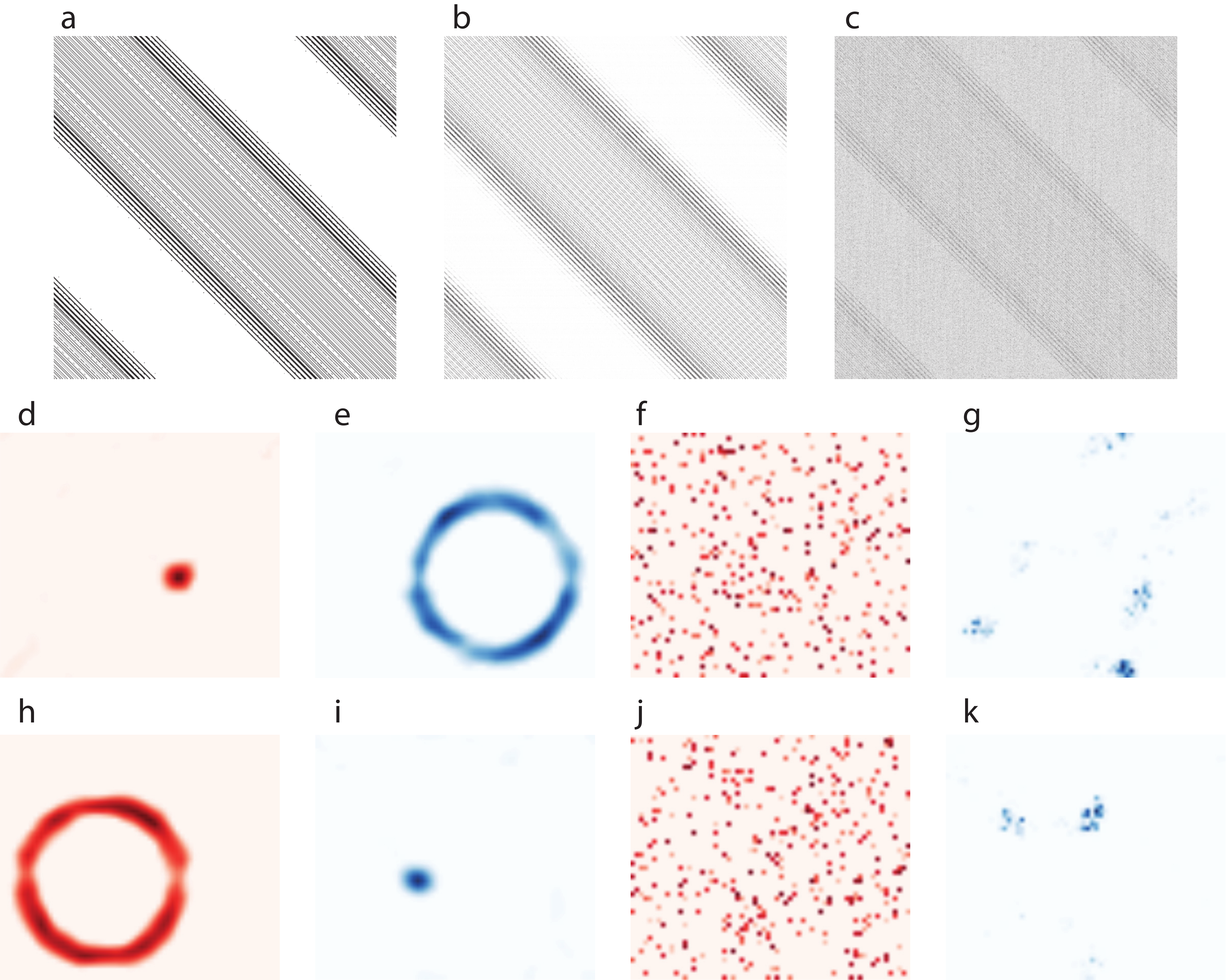}
     \caption{\textbf{Approximate connectivity using non-negative matrix factorization}
     (\subfiglabel{a}) The desired effective connectivity matrix between grid cells, $W_{ij}$ in Eq. 5, with $\sigma$, $f_d$ 
      and offset, $l$, equal to zero.
      (\subfiglabel{b}) the resulting effective connectivity for the fully factorized model, 
      with 200 interneurons and 896 excitatory cells and (\subfiglabel{c}) for the partially factorized model, with 896 excitatory and 4480 inhibitory cells.
     (\subfiglabel{d,h}) Examples of the connections from the $32 \times 28$ 
      excitatory cells arranged on a two dimensional lattice to two example interneurons in the fully factorized model and
      (\subfiglabel{e,i}) the corresponding inhibitory-to-excitatory connectivity from the same interneurons 
      to the excitatory neurons. In each of the examples the aptitude of the connections are normalized to the peak value of the coupling matrix
      with excitatory couplings colored red and inhibitory blue.
      In \subfiglabel{d,e,h,i} the excitatory cells are arranged according to spatial phase. The interneuron in \subfiglabel{d} receives input from excitatory cells with similar phase but and 
      inhibits a number of excitatory cells \subfiglabel{e}. Since this inhibitory neuron is activated by grid cells with similar phase, 
      it is going to exhibit gird-like firing. The opposite is true for the example inhibitory neuron \subfiglabel{h,i} which will show anti-grid-like tuning.
      \subfiglabel{f,g,j,k} Same as \subfiglabel{d,e,h,i} but for the partially factorized model.
    \label{fig:factorizedconnectivities}
    }
\end{figure}

Figure \ref{fig:factorizedconnectivities} shows examples
of the connectivities that we find with these two approaches. 
As can be seen in figure \ref{fig:factorizedconnectivities} \subfiglabel{a,b,c}, 
both approaches manage to reasonably approximate the desired effective connectivity.
In the case of the fully factorized model, this resulted in interneurons that receive inputs 
from grid cells of similar spatial phase (figure \ref{fig:factorizedconnectivities} \subfiglabel{d})
or receive input from grid cells of surrounding spatial phases (figure \ref{fig:factorizedconnectivities} \subfiglabel{e}). These interneurons will
exhibit grid-like or anti-grid like firing correspondingly, similar to those in \cite{pastoll2013feedback} and \cite{widloski2014model}.
These grid-like and anti-grid-like interneurons then project back to the surrounding 
(figure \ref{fig:factorizedconnectivities} \subfiglabel{h})
and center (figure \ref{fig:factorizedconnectivities} \subfiglabel{i}), respectively.
The resulting connectivity for the partially factorized model (figure \ref{fig:factorizedconnectivities} \subfiglabel{c}) 
also manages to approximate the desired connectivity, but typically requires many more interneurons 
than the fully factorized model, e.g. in the example shown it required 5 times more interneurons than excitatory neurons. This can be attributed to the fact that
in the fully factorized model both inhibitory-to-excitatory connections and excitatory-to-inhibitory connections can be adjusted to achieve the desired connectivity 
while in the partially factorized one only the former are allowed to be adjusted. 
In the partially factorized model, the inhibitory neurons obviously receive
random input from the grid cells (figure \ref{fig:factorizedconnectivities} \subfiglabel{f,j})
while projecting back in a non-random but not obviously structured manner
(figure \ref{fig:factorizedconnectivities} \subfiglabel{g,k}).
Consistent with this connectivity scheme,  Buetfering et al.,~2014, found that grid 
cells of all phases functionally project to local interneurons \cite{buetfering2014parvalbumin}.

Using these two approaches we were able to construct different models
with variable numbers of interneurons while maintaining the same effective
weight matrix $W_{ij}$ between the grid cells.
Additional details on the construction of the grid cell model can be found in \ref{sec:twopopcanmodel_supp}.

\subsection{Response of models to velocity modulated and place-like inputs}

We subjected both models to input of the form shown in Eq.~\ref{pinput},
following an experimentally recorded trajectory of an animal exploring a square
open-field environment. In both models, we could change the degree of
firing field amplitude variance in the excitatory grid cells by changing the
dilution $f_d$ of the place-like input as well as the
standard deviation $\sigma$ of the truncated Gaussian distribution from which
the connections $H_{ij}$ in Eq.~\ref{pinput} are drawn. The
fields of a grid cell that were not directly
excited by place-like input added further
variations in the field-to-field difference of the cell: a field not
driven by place-like input would be activated only
through the interaction between the external excitatory drive that all neurons
receive (i.e. the constant term in Eq.~\ref{pinput}) and the pattern forming
effective inhibitory connectivity, leading to a lower firing rate compared to
fields driven by place-like input.

\begin{figure}[ht]
    \centering
    \makebox[\linewidth][c]{
      \begin{minipage}{1\linewidth}
        \centering
        \includegraphics[width=0.9\linewidth]{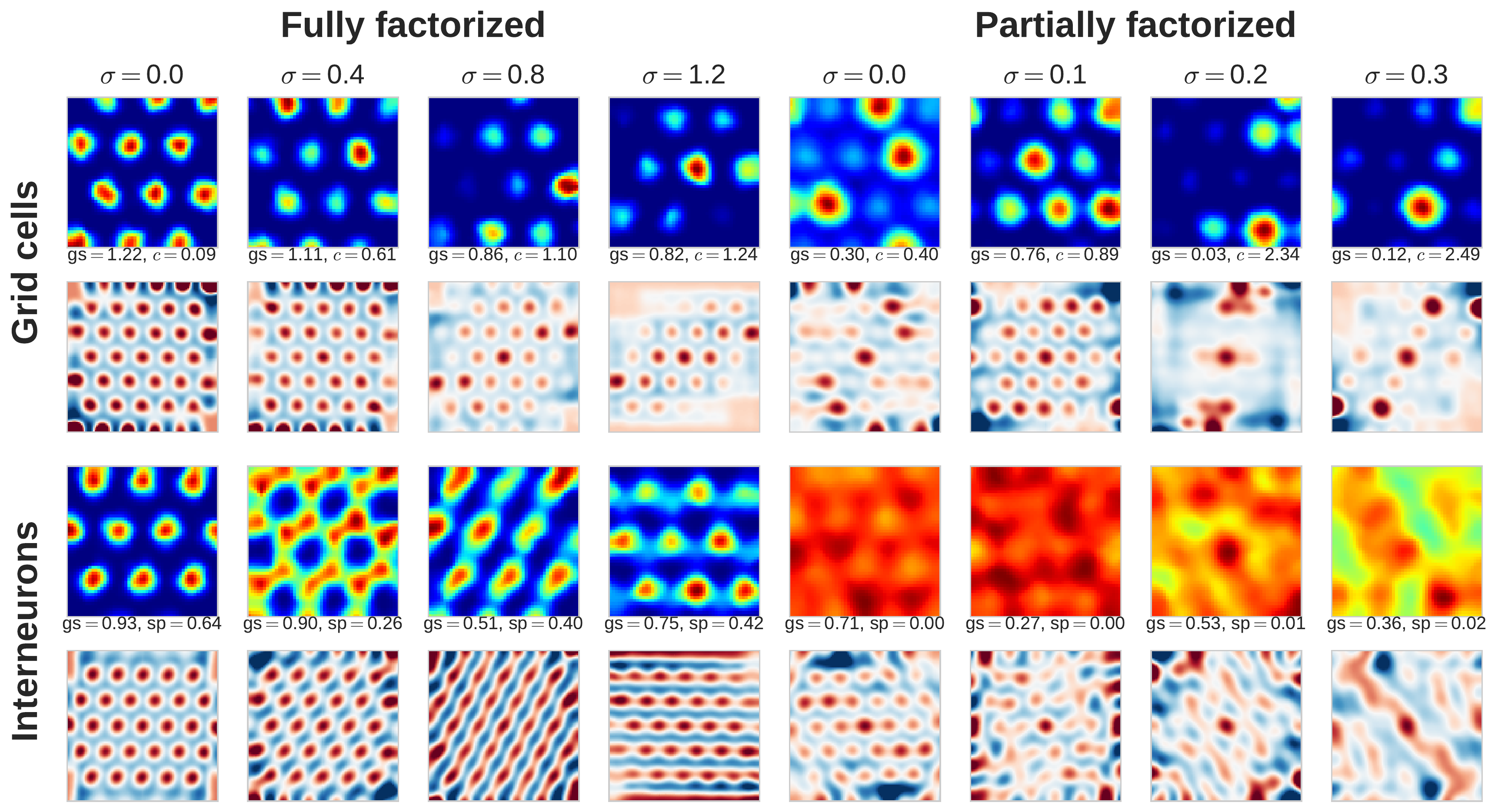}
      \caption{\textbf{Rate maps and autocorrelograms of simulated neurons.}
        Rate maps and autocorrelograms of one excitatory and one inhibitory
        neuron from each of several model simulations with different
        parameters.
        The top two rows show excitatory neurons (grid cells), with grid score
        ($\text{gs}$) and field amplitude coefficient of variation ($c$) quoted
        between the rate map (above) and autocorrelogram (below). The bottom
        two rows show inhibitory neurons (interneurons), listing grid score
        ($\text{gs}$) and sparsity ($\text{sp}$). The four left columns are
        taken from runs of the fully factorized model with different values of
        place-like input connection strengths $\sigma$, and the four right
        columns from similar runs of the partially factorized model with
        dilution $f_d = 0.2$.  Complementing the observations in figure
        \ref{fig:scores}, we see that inhibitory neurons in fully factorized
        models have a high degree of periodicity regardless of the connectivity
        variance, while interneurons in the partially factorized model lose
        their periodicity as the grid cell amplitude variability increases.
        Note that inhibitory neurons in the fully factorized model need not
        have isolated firing fields even though they are periodic; see
        Supplementary figure \ref{fig:supp-more_ratemaps} for several
        counterexamples.  \label{fig:examplesimgrids}
      }
    \end{minipage}
  }
\end{figure}


\begin{figure}[h!]
    \centering
    \includegraphics[width=0.95\linewidth]{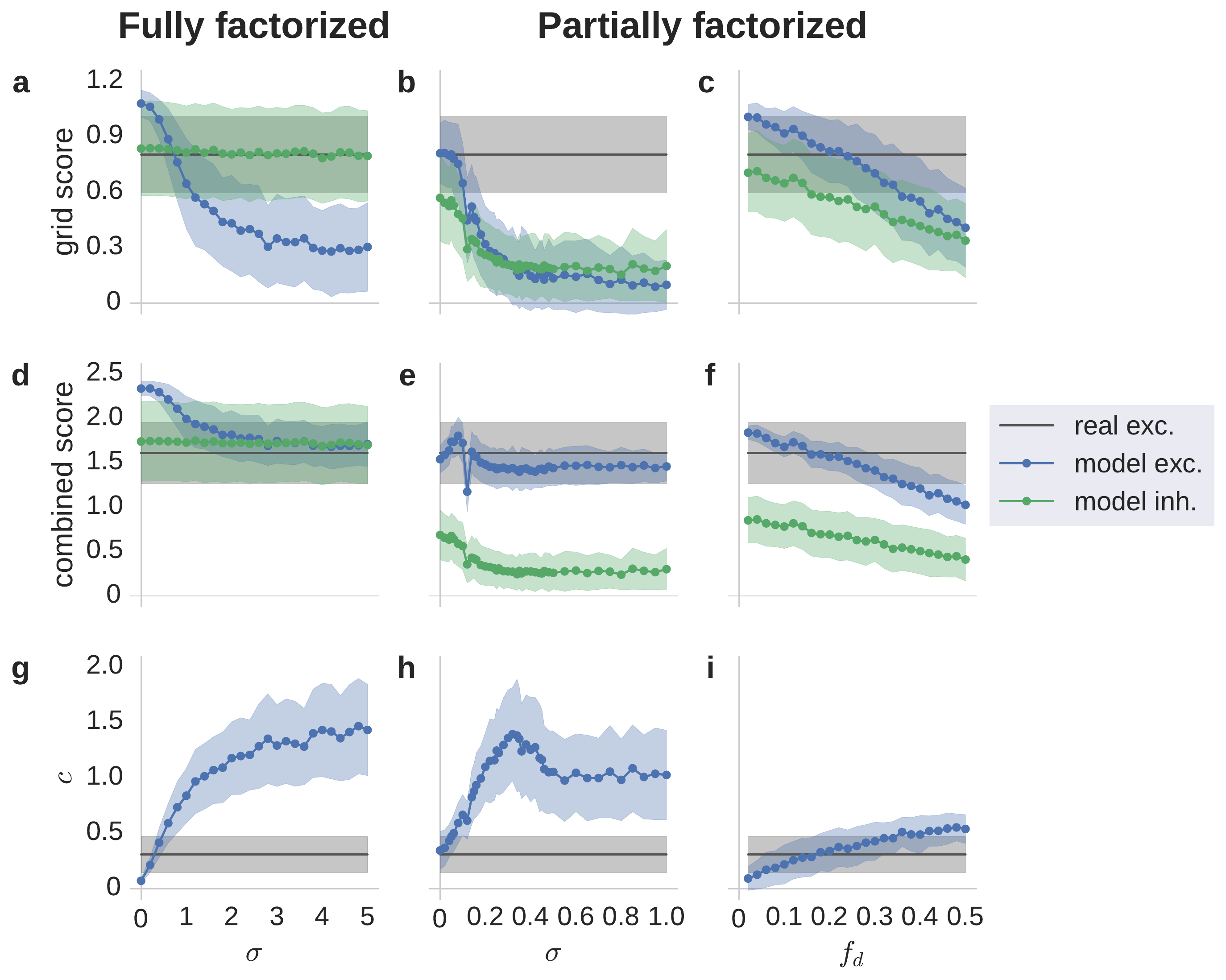}
    \caption{\textbf{Connectivity variance and the resulting selectivity of
      excitatory and inhibitory neurons.}
      \subfiglabel{a-b}: Mean and standard deviation of the grid score of
      excitatory grid cells (blue) and inhibitory interneurons (green) from
      several model simulations, plotted against the standard deviation in the
      place-like input connection strengths, $\sigma$. For comparison, the same
      quantities for real grid cells are shown in gray.
      \subfiglabel{c}: Same as \subfiglabel{b} but keeping $\sigma$ fixed and
      varying the dilution fraction $f_d$ of non-zero couplings randomly
      removed.
      \subfiglabel{d-f}: Mean and standard deviation of the combined
      score of grid cells and interneurons from
      model simulations, and real grid cells.
      \subfiglabel{g-i}: Mean and standard deviation of the firing field
      amplitude coefficient of variation $c$ of excitatory grid cells from
      model simulations, and real grid cells.
      \subfiglabel{a,d,g}: Fully factorized model. Grid scores and combined
      scores of interneurons remain high and similar to real grid cells for all
      connectivity variances, reflecting their periodic and relatively sparse
      firing in this model. Grid scores of grid cells decrease with increasing
      input variance, but only due to increasing field amplitude variation, not
      geometric distortions of the firing pattern
      (see figure \ref{fig:examplesimgrids}).
      \subfiglabel{b,e,h}: Partially factorized model with $f_d = 0.2$. 
      Grid scores of model grid cells are similar to real data for zero input
      variance, and decrease with increasing input variance. In this model,
      grid scores of interneurons also decrease as $\sigma$ is increased,
      reflecting their waning periodicity (see figure
      \ref{fig:examplesimgrids}). Due to their non-sparse firing,
        interneurons are clearly separated from grid cells by the combined
      score.
      \subfiglabel{c,f,i}: Partially factorized model with $\sigma = 0$.
      Without dilution ($f_d = 0$), both the interneurons and grid cells are
      predictably periodic with little field-to-field variations. By diluting
      the projections from the place-like input, the field
      amplitudes of the grid cells become increasingly
      variable, while the grid and combined scores of the
      interneurons decrease as in \subfiglabel{b,e,h}.}
    \label{fig:scores}
\end{figure}

In figure \ref{fig:examplesimgrids}, we show how the firing properties of the
cells change as we vary $\sigma$ with fixed $f_d = 0.2$ for both
fully factorized and partially factorized models,
by showing rate maps and cross-correlograms of
selected cells. As can be seen, in the fully
factorized model, the excitatory neurons but not the inhibitory neurons become
more irregular as $\sigma$ is increased.  This is, however, not the case for
the partially factorized model. In this model, both inhibitory and excitatory
neurons become more irregular as $\sigma$ is increased.

In figure \ref{fig:scores}, we show the trends exemplified in figure \ref{fig:examplesimgrids} at the population level.  As can be seen
in figure \ref{fig:scores} \subfiglabel{a-b}, the grid score for the excitatory neurons but not the inhibitory neurons drop as $\sigma$ is increased
in the fully factorized model. In fact, there is no value of $\sigma$ where, in this model, we can see grid cells
with grid score similar to the real data and irregular inhibitory neurons (see also figure \ref{fig:examplesimgrids}).  
This is, however, not the case for the partially factorized model. In this model, the grid score for both inhibitory and excitatory neurons 
drop by increasing $\sigma$. In fact, for the parameters we chose already at $\sigma$ close to zero and dilution level $f_d = 0.2$, 
we can have interneurons with irregular firing (figure \ref{fig:examplesimgrids}) while having excitatory neurons with realistic grid score.
As we show in figure \ref{fig:scores} \subfiglabel{c}, the irregularity in the fields can also be increased
by keeping $\sigma$ at zero and simply varying the 
dilution $f_d$ of the place-like input.

The grid score on its own suffers from the fact that it can be low due to field
strength variations even if the pattern is regular (see also figure \ref{fig:examplesimgrids}). 
We therefore also analyzed our results using a modified score which
is a linear sum of the grid score as well as the spatial sparsity of the
firing, which we will denote the \emph{combined score} (see section
\ref{supp:Scores} for further details). For grid cells most of the firing is confined to
isolated fields, so taking into account the spatial sparsity helps separate
such cells from e.g.\ interneurons, which may exhibit some degree of
periodicity but often fire all over space. As can be seen in figure
\ref{fig:scores} \subfiglabel{d-f}, in the fully factorized model both grid
cells and interneurons maintain high scores, while the interneurons of the
partially factorized model have substantially lower scores. Inspection of the
firing rate maps of individual cells (see Supplementary figure
\ref{fig:supp-more_ratemaps} for more examples)
clearly substantiates the conclusion from this figure that in the fully
factorized model, increasing the variance of the input does yield variable
field amplitudes in the grid cell but the inhibitory
neurons maintain a high degree of regularity and sparsity, while for the
partially factorized model, the inhibitory neurons have non-sparse firing, and
also lose their regularity when the variance of the input to the excitatory
neurons increases. Note that, as shown in figure \ref{fig:scores}
\subfiglabel{g-i}, in both networks the coefficient of variation
in the field-to-field firing in the excitatory neurons can be reached at small
values of $\sigma$ and $f_d$, but as described above, it is the behavior of the
inhibitory neurons that sets the two architectures apart from each other. 

\subsection{Integrating velocity}

\begin{figure}[h!]
  \centering
  \includegraphics[width=0.2\linewidth]{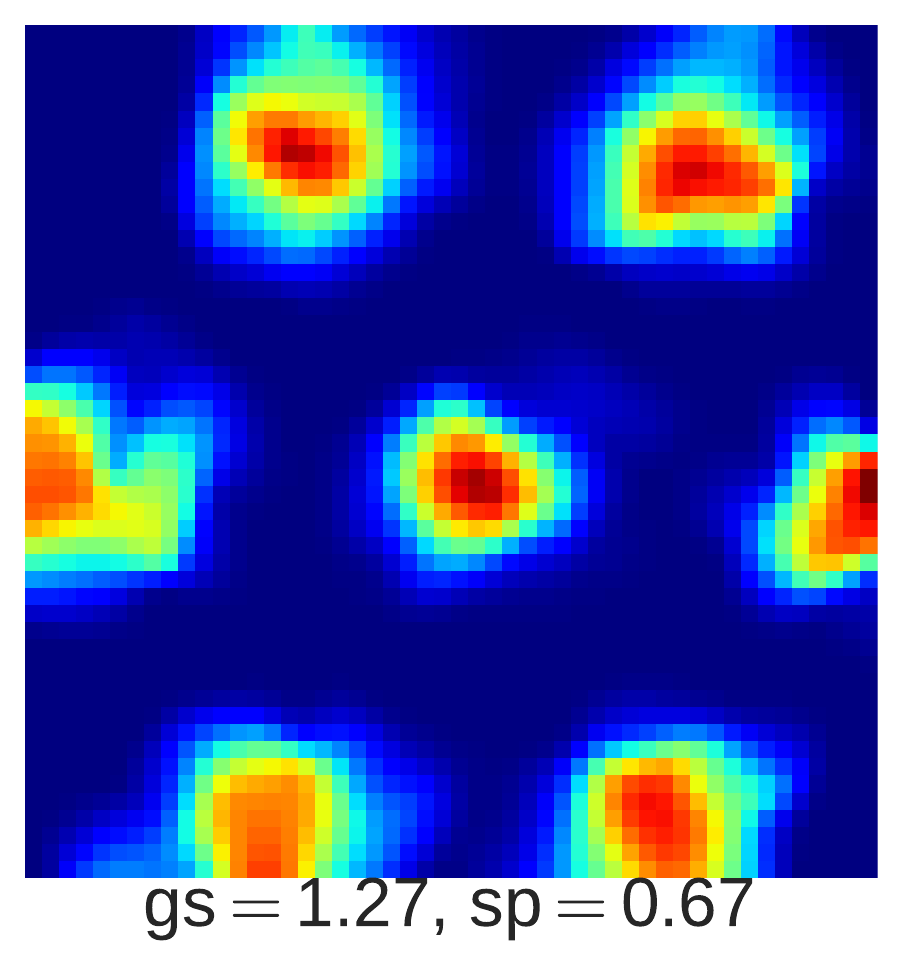}
  \includegraphics[width=0.2\linewidth]{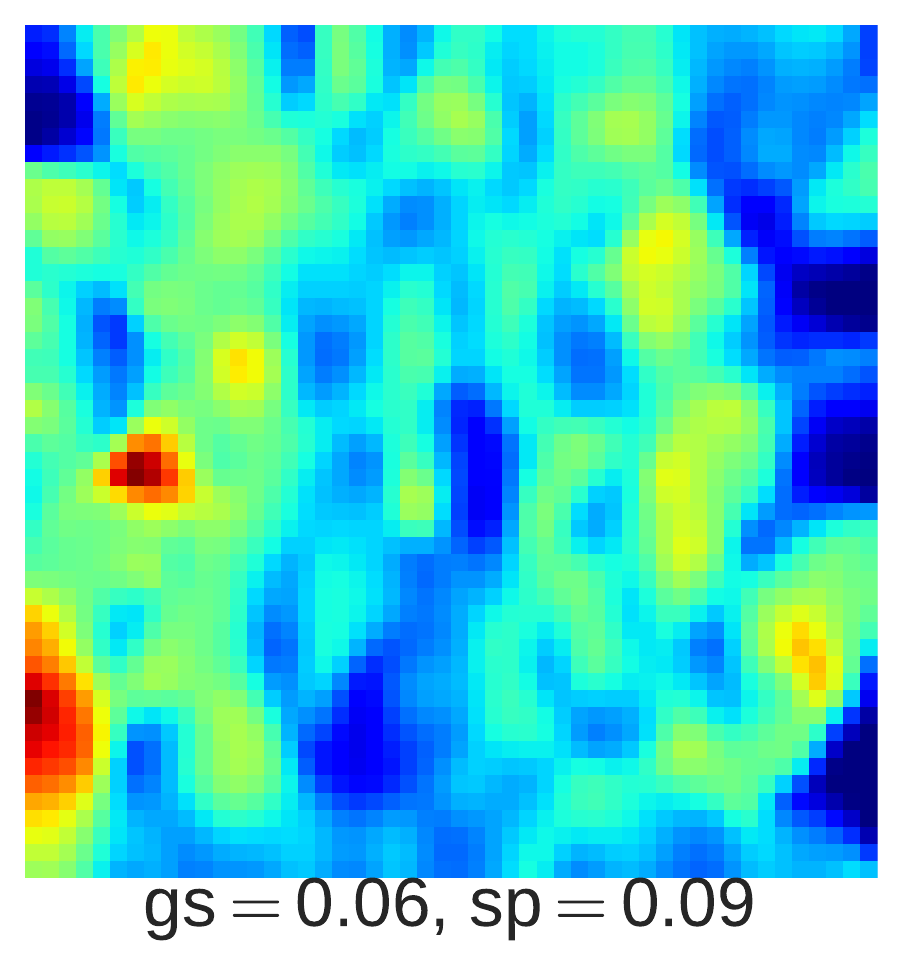}\\
  \includegraphics[width=0.2\linewidth]{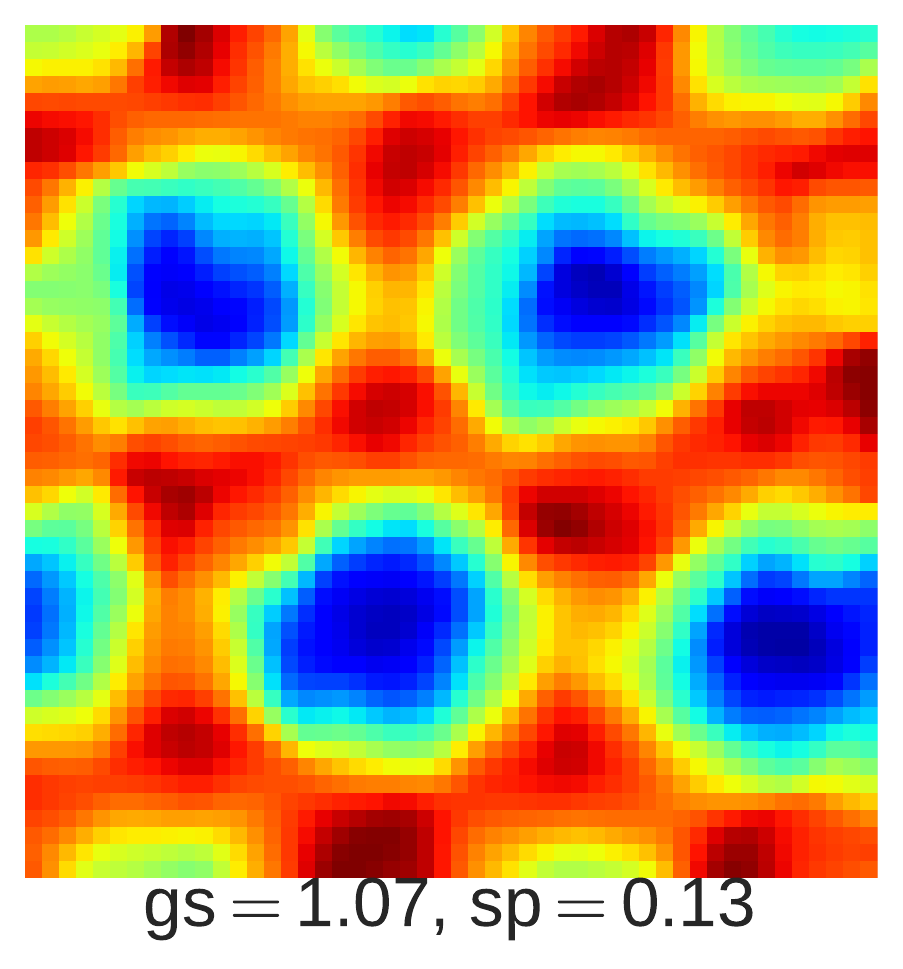}
  \includegraphics[width=0.2\linewidth]{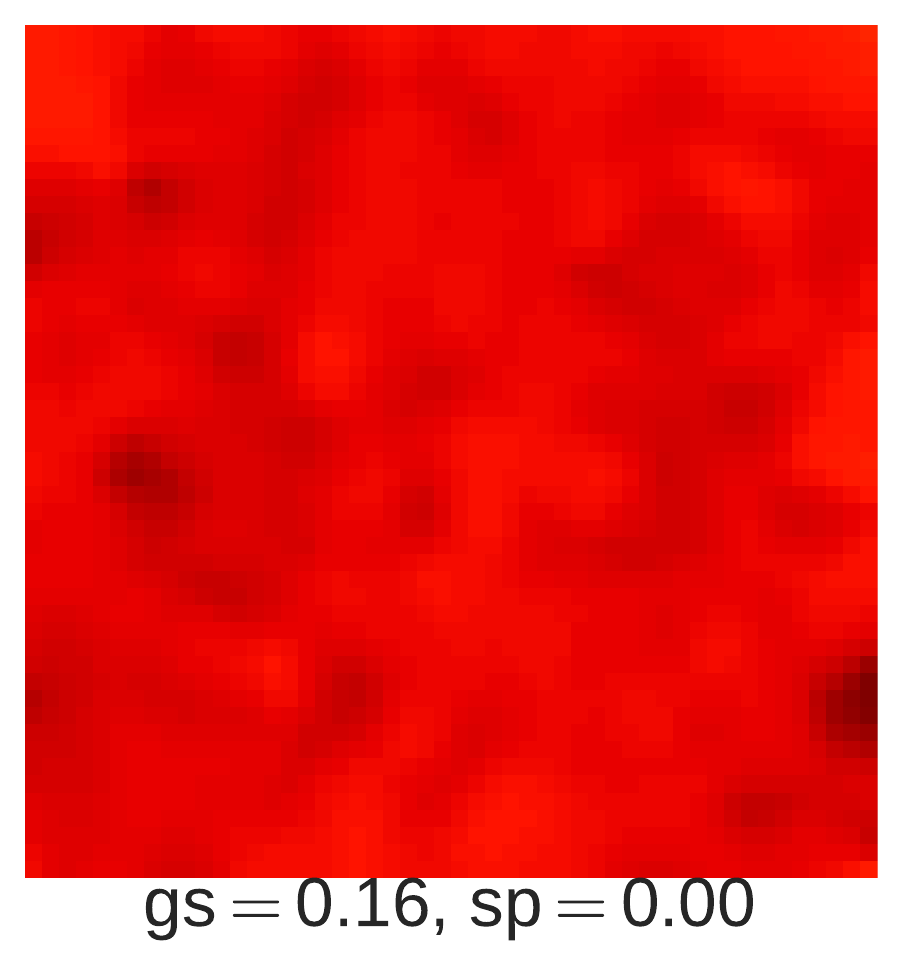}
  \caption{\label{fig:failedpathintegration}
    \emph{Top left:} Rate map of a simulated grid cell arising from stable path
    integration of velocity input in the fully factorized model.
    \emph{Bottom left:} Rate map of a simulated interneuron during the same
    simulation run.
    \emph{Top right:} Rate map of a simulated would-be grid cell that fails to
    develop a grid pattern because path integration is too unstable.
    \emph{Bottom Right:} Rate map of an interneuron during the same simulation
    run. No discernible grid-like structure was observed after a ten minute
    session.
}
\end{figure}

After establishing that with the place-like input both the fully factorized and
the partially factorized models generate grid cells, we wondered whether these
two architectures can also generate grid cells using only velocity-head
directional input (Eq. \ref{vinput}). While the fully factorized model managed
to generate hexagonal grid cells only using input of the form of Eq.
\ref{vinput}, the partially factorized model failed to do so, e.g. figure
\ref{fig:failedpathintegration}.  Testing various configurations with up to
$20$ times more interneurons than grid cells we were unable to find
a configuration that could accurately integrate the velocity and generate grid
cells in the partially factorized model.  As can be seen in figure
\ref{fig:failedpathintegration} (right), with a ratio of interneurons to grid
cells that is completely biologically implausible, the model was only able to
produce random, unreliable firing fields when attempting to integrate over
velocity for ten minutes. At the same time, the fully-factorized model
performed beyond expectations: with as little as 200 interneurons (comprising
approximately $5 \ \%$ of the network) and $64 \times 56$ grid cells, the model
could accurately path integrate over 10 minutes of tracked position data
(figure \ref{fig:failedpathintegration}, left).

\begin{figure}[ht]
    \centering
    \includegraphics[width=0.69\linewidth]{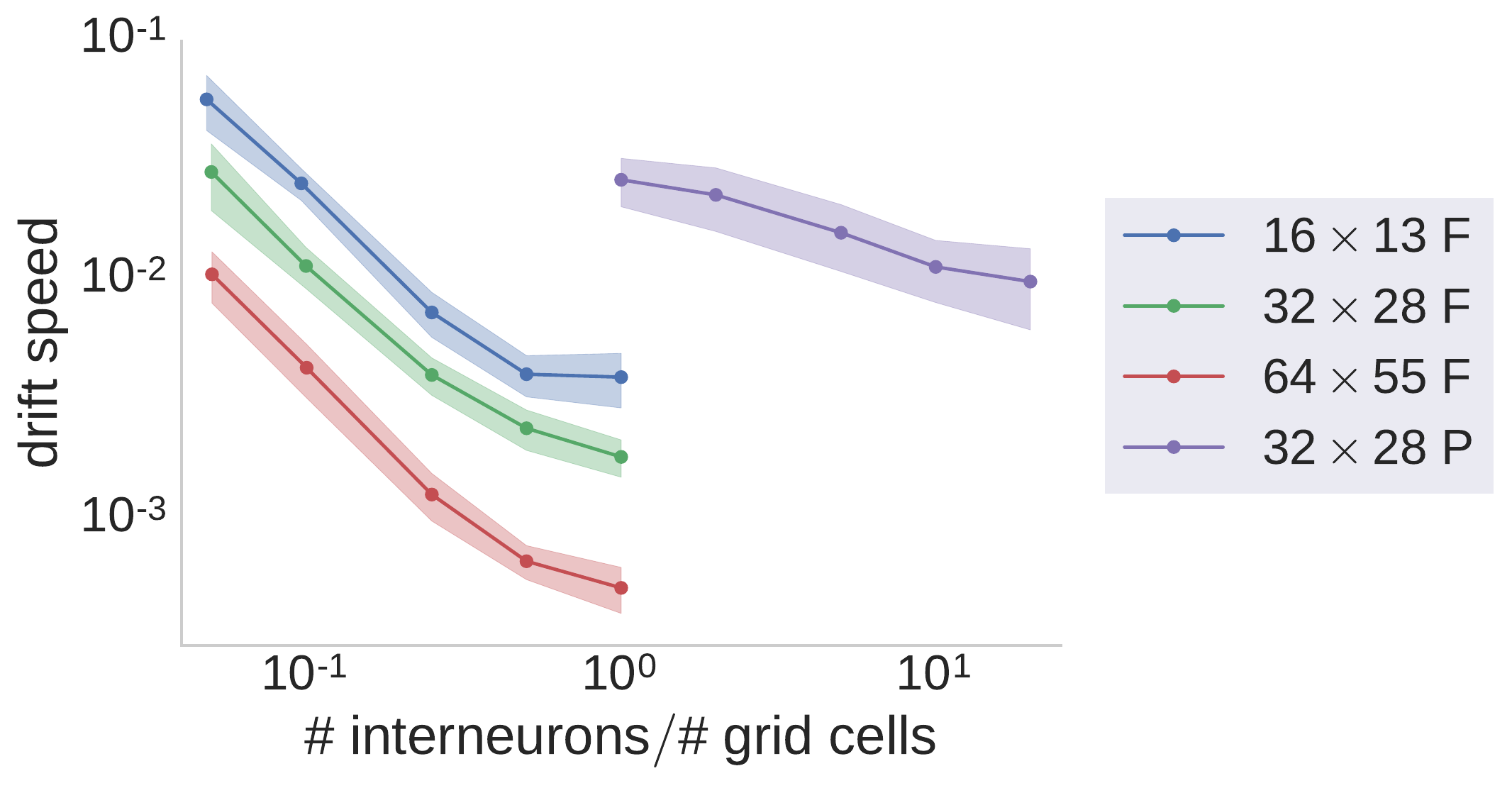}
    \caption{\textbf{Stable continuous attractor dynamics.}
      Mean and standard deviation of the measured drift speed ($y$-axis),
      defined as $(\text{displacement on neural sheet}/(\text{time step} \cdot
        \text{length of neural sheet})$. The drift speed is plotted against the
        ratio of the number of interneurons to the number of grid cells
        ($x$-axis) for various configurations of both the fully (F) and
        partially (P) factorized model.  The legend shows the size $N_x \times
        N_y$ of the neural sheet.
    \label{fig:driftfig}
    }
\end{figure}

To understand why the two models performed so differently under path
integration, we considered the issue of drift~\cite{zhang1996representation}.
Since even minute levels of randomness in connectivity is known to cause drift,
drift is a likely candidate for the failure of the 
partially factorized model for generating grid cells by integrating velocity
because of the presence of random sparse connectivity in this network.
For each drift simulation the pattern was initialized randomly, with the
external drive $I_t$ set to a constant value, allowing the network to
converge to a static pattern. Once formed, the pattern was shifted randomly
around the neural sheet. In an idealized continuous attractor, the
shifted pattern should stay in place indefinitely due to the translational
invariance property~\cite{zhang1996representation}, but any inhomogeneity in
the networks, makes the
pattern prone to drifting.  We measured the drift as the average
distance traveled from 200 shifted locations during 100 time steps, allowing us
to estimate the average drift velocity during the first second after shifting
the pattern (see figure \ref{fig:driftfig}).

For the fully factorized models, the resulting drift decreased exponentially
with the number of interneurons included in the model (see figure \ref{fig:driftfig})
while the partially-factorized model decayed significantly more slowly,
requiring several orders of magnitude more interneurons to achieve 
a similar degree of stability.

\section{Field variability and remapping}

The fact that the firing rate of individual fields of a cell may shows strong
and robust variations may have multiple implications.  One is that it will
increase the information that the firing of each individual cell carries about
the position of the animal: if all fields are identical, it is impossible to
distinguish between different sites by observing the firing of the cell, and
this ambiguity could be reduced by introducing varying field amplitudes. This
could have implications for theories on how grid cell parameters are chosen for
optimally encoding space \cite{mathis2012optimal, wei2015principle}.

\begin{figure}
  \centering
    \includegraphics{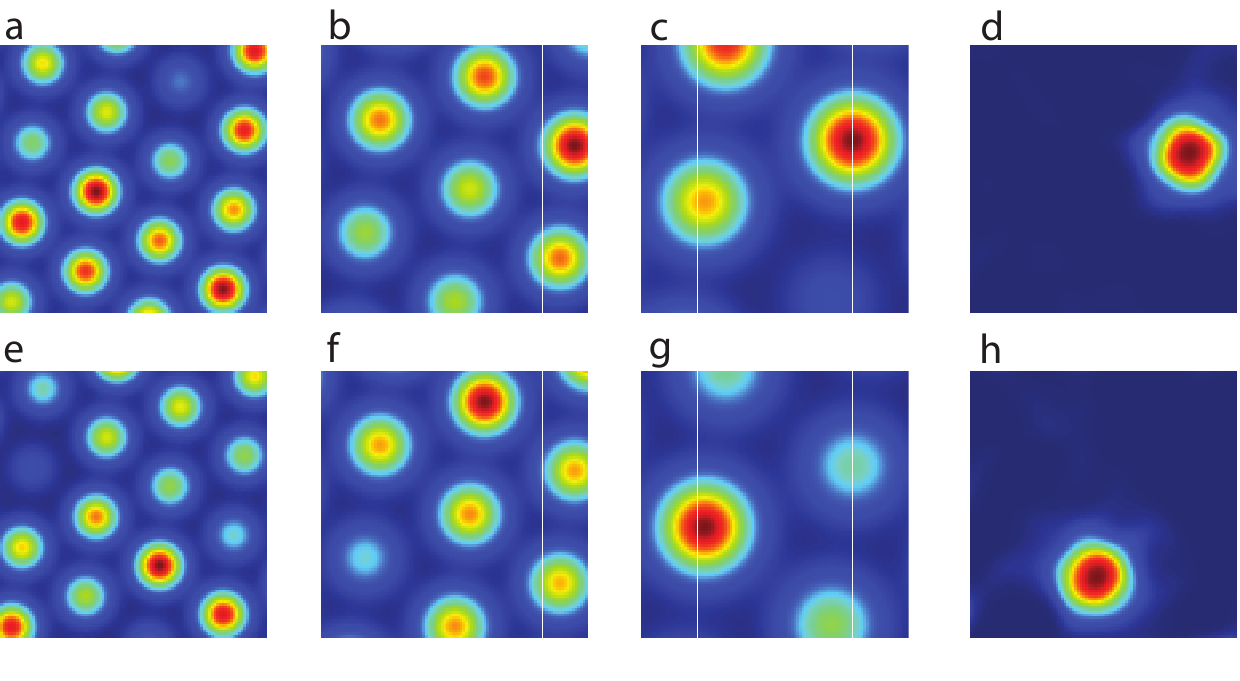}
    \caption{\label{fig:feedforward}
      \textbf{Rate-remapping of grid cells sufficient to globally remap a place cell.}
      \subfiglabel{a-c}: Example simulated grid cells from 3 modules with
      imposed field-to-field variability.
      \subfiglabel{d}: A place cell with only input from grid cells of the type in \subfiglabel{a-c} 
      (25 from each of the 3 modules) and a constant external input.
      \subfiglabel{e-g}: Same cells as in \subfiglabel{b} but with different
      field amplitudes. 
      This manipulation has been accomplished chemogenetically in rats
      \cite{kanter2016}. Here the peak rates of the fields were
      sampled from a Gaussian distribution with a mean of 1 and standard deviation of 0.3.
      \subfiglabel{h}: The same place cell as in \subfiglabel{d} but with the input grid cells 
      rate-remapped (as shown in \subfiglabel{e-g}).
      Thus with the same connections and constant external input a place cell can globally 
      remap with just a rate-remapping of the grid cells.
      Here the grid cell selectivity was simulated with Gaussian bumps centered on the vertices of a hexagonal lattice
      while the firing rate of the place cell is determined by the input with a logistic activation function.
    }
\end{figure}

In addition, variability among firing fields suggest the possibility of an
additional role for grid cells in distinguishing between different
environments, i.e.\ encoding contextual information. Previous work assuming
equal firing amplitudes~\cite{stella2012self} confirm the high spatial
information content of grid cells while suggesting that they
provide almost no contextual information (that is, they cannot be used to
discriminate between environments). Interestingly, a recent study by Kanter et
al.~\cite{kanter2016} shows that with a chemogenetic manipulation of the mouse
MEC, random shifts in the firing amplitudes of the individual grid fields can
be induced without changing the positions of fields. This observation can be
exploited in a simple feedforward model in which the output of grid cells with
different spacings are summed and thresholded by cells in the hippocampus, such
that the hippocampal cells obtain spatial tuning~\cite{solstad2006grid}. With
identical fields, the only way to obtain remapping in the place cells is to
change the relative spatial phase of different grid cells. With non-identical
fields, however, changing the relative amplitude of different fields can cause
remapping to occur, as shown in figure~\ref{fig:feedforward}. The statistical
analysis and theoretical model presented here, together with this recent
experimental finding, thus suggests an expanded role for grid cells, not only
integrating over velocity output, but over spatial and contextual cues more
generally.

\section{Summary and Discussion}

In this paper we established that the firing fields of real grid cells are not
simple translations of one another, but that their amplitudes can in fact vary
significantly. These variations are stable and reproducible, and are not merely
a consequence of experiments with finite duration and imperfect, non-uniform
sampling of the environment, or temporal fluctuations in the spiking of the
neuron.

To consider the consequences of this variability in 
firing field amplitudes within the context of
the continuous attractor model, we devised two recurrent network models of grid
cells, explicitly including both excitatory and inhibitory populations, and
using patterned (fully factorized) or random (partially factorized)
connectivity from grid cells to interneurons, respectively. We considered these
networks operating under three different conditions:  only receiving input from
place-like input, only receiving a velocity dependent signal, or receiving only
a constant input which we used for studying the drift.  While the components
necessary for the velocity input are found locally in the MEC
\cite{sargolini2006conjunctive,kropff2015speed}, the MEC has also been shown to
receive extensive input from the hippocampus as well as other areas with
discrete spatial selectivity such as the postrhinal
cortex~\cite{fyhn2004spatial,burwell2003positional}.  Indeed, even during
hippocampal inactivation, in which grid cells loose their periodic selectivity,
stable discrete spatial selectivity in the form of single fields as well as
increased directional selectivity has been demonstrated in grid
cells~\cite{bonnevie2013grid}. 

We found that the fully factorized model could generate stable grid cells by
only integrating velocity input and employing only a very small population of
interneurons, e.g.\ approximately $5 \ \%$ of the size of the excitatory
population. In this condition, the grid cells did not show noticeable firing
field amplitude variability. By adding a place-like signal via random
connectivity to the excitatory cells, we obtained grid cells that exhibited
amplitude variability in their fields, similar to what we established to occur
in the real data. The degree of this variability could be controlled by
diluting the spatial input to the grid cells or adjusting
the variance of the strength of the connections from the place-like input.

Regardless of the type of input, the interneurons in the fully factorized
models demonstrated periodic spatial coding with high grid scores, albeit with
different kinds of firing patterns such as the ``anti-grid'' pattern.
Interestingly, the interneurons maintained high grid scores and relatively low
field-to-field variability while grid cells became more variable with
increasingly variable input connections. This may indicate that the input
connection strengths to an interneuron are distributed quite uniformly among
grid cells with similar spatial phases in this model, such that the
field-to-field variations of single grid cells often cancel approximately when
summing over several cells of the same phase.

With only velocity input, the partially factorized model failed to generate the
periodic firing pattern associated with grid cells. When replacing velocity
input with a place-like input, however, the excitatory neurons acquired grid
cell firing with realistically varying firing field amplitudes, but with
amplitude variance increasing significantly faster than in the fully factorized
model as a function of the variance of the place-like input connection
strengths.  The interneurons in this model, lacking the patterned input
connectivity required to maintain their firing patterns while grid field
amplitudes fluctuate, were strongly impacted by the variance of the connections
from the place selective cells, and lost their periodicity as the variance
increased.

In order to understand why this network failed to generate grid cells with only
velocity input, we evaluated the attractor drift in the models, and observed
that while there was a rapid exponential decrease in drift with the number of
interneurons in the fully factorized models, even with a moderate number of
interneurons (figure \ref{fig:driftfig}), the decrease was much slower and did
not reach the same levels even at an order of magnitude more interneurons in
the partially factorized model.

We interpret these results by proposing two possible scenarios for the
underlying network that generated grid cells. Since both models are able to
generate grid cells firing patterns in excitatory neurons with the field
amplitude variability found in experimental data, both can be used as models
for the generation of grid cells. However, they are distinguished by their
predictions for the firing patterns of interneurons and their response to
different types of input, and may thus differ in their compatibility with
experimental data as we discuss further below.

The selectivity of interneurons in the partially factorized model (figure
\ref{fig:examplesimgrids}, right) resembled the selectivity of the majority of
recorded local interneurons in Buetfering et
al.~\cite{buetfering2014parvalbumin}.  However, the model required place
selective input in order to generate grid cells, a prediction consistent with
developmental data showing that place cell selectivity appears before grid
cells~\cite{langston2010development}.

The fully factorized model (figure \ref{fig:examplesimgrids}, left), managed to
generate grid cells with only velocity input, and firing field amplitude
variations could be observed in the network when place-like input was added.
However, this model also generated periodic spatial selectivity in the
interneurons. Although the majority of the interneurons recorded in
Buetfering et al.~\cite{buetfering2014parvalbumin} did
not show periodic firing, a few of them still did, which is consistent
with the fact that the fully factorized model can generate the hexagonal 
pattern of the grid cells by integrating velocity input using a very small
number of interneurons compared to the total size of the network.  An
interesting experimental proposal is therefore to investigate how many of the
interneurons in the MEC are actually involved in generating grid cells, e.g.\
by looking at the fraction of interneurons that project back to grid cells, and
seeing if these are among the few interneurons that show periodic firing, or by
optogenetically manipulating a fraction of interneurons and seeing how it
affects the grid cells' periodic firing.

It should be mentioned that the use of place-like input in generating grid
cells is not an entirely new idea. The so-called adaptation model originally
developed by Kropff and Treves~\cite{kropff2008emergence}, and related
models~\cite{castro2014feedforward,stepanyuk2015self}, consider the entorhinal
network not as a fixed code with discrete translation symmetry built for path
integration, but one that learns and adapts to different environments.  In
these models, external spatially modulated input provides an important role in
the formation and maintenance of the grid pattern. Here, we have not discussed
mechanisms through which the place-like input connectivity to the grid network
may be formed through development or synaptic plasticity while the animal
navigates a new environment. However, mechanisms such as those proposed by
Kropff and Treves~\cite{kropff2008emergence} might admit the establishment of
proper place-like connectivity to the grid cells. Furthermore, though it has
yet to be studied, it seems plausible that incorporating the observed field
amplitude variability within such a framework may also be achieved by assuming
that the spatial input can vary on a time scale that is faster than the
learning of the connectivity required for the pattern to form.

The model presented here shares the much of the benefits of the adaptation model, 
such as stability to the types quenched noise \cite{zhang1996representation,
tsodyks1995associative} that plague models based on integrating velocity while
still uses attractor dynamics enabling, for example, firing fields to
appear for cells where external spatial input was removed or not provided.
It is still an open question, however, what 
mechanisms for path integration could exist that are consistent
with emerging properties, such as variability among grid cell firing fields.

\section{Materials and Methods}

\subsection{Data} 
\label{supp:Data}
The $373$ identified cells used in this analysis were selected from recordings
of rats $13855$, $13960$, $14147$, $15314$, and $15708$
in Stensola et al.~\cite{stensola2012entorhinal}, and details about the data acquisition can be
found in that reference. These rats had multisite (hyperdrive) tetrode
implants, and the recordings were carried out in square open-field environments
of either $100 \times 100 \ \mathrm{cm}$ (rat $13855$) or $150 \times 150
\ \mathrm{cm}$ (others).


\subsection{Position filtering}
The position time series, which were recorded at sampling rate $25
\ \mathrm{Hz}$, were smoothed to reduce jitter using a low-pass Hann window FIR
filter with design cutoff frequency $2.0 \ \mathrm{Hz}$ and kernel support of
$13$ taps, or approximately $0.5 \ \mathrm{s}$.

For each sample, the filter kernel was renormalized to compensate for missing
samples within the kernel duration. In practice, this is accomplished by
applying the filter to the original sequence but with missing samples
replaced by zeros, and then to an indicator sequence with a value of $1$ at
every valid sample and $0$ at every missing sample. The elementwise ratio of
these two sequences is the final filtered sequence. This approach automatically
interpolates across gaps shorter than the kernel duration, and is the simplest
example of normalized convolution~\cite{knutsson1993normalized}.

All experiments provided two sequences of position samples, corresponding to
LEDs located in slightly different positions on the animal. The animal position
was taken to be the mean of the two (filtered) sequences.

The speed at each sample was computed as the average speed over the surrounding
$13$ samples, by dividing the trajectory length with the elapsed time within
this window. This was used to apply a speed filter: only samples where the
speed exceeded $5 \ \mathrm{cm}/\mathrm{s}$ were kept.

\subsection{\label{sec:ratemaps}Rate maps}
Spike locations $\ve{\xi}_j = (\xi_j, \chi_j)$ at spike times $\tau_j$ were
found by linearly interpolating the position time series $(t_i, x_i, y_i)$ at
the spike times.  The spatial firing rate at location $\ve{x} = (x, y)$ was
defined as the ratio of kernel density estimates of the spatial spike frequency
and the occupancy,
\begin{equation}
  \label{eq:spatialrate}
  f(\ve{x}) = \frac{\sum_j K_1(\ve{x} \ | \ \ve{\xi}_j)}
  {\sum_i \Delta t_i K_1(\ve{x} \ | \ \ve{x}_i)}
  \ ,
\end{equation}
where the sums are taken over the spikes and position samples, respectively.
Here, $\Delta t_i = (t_{i + 1} - t_{i - 1}) / 2 \approx 1 / F_s$, where $F_s
= 25 \ \mathrm{Hz}$ is the target sampling frequency. A triweight kernel was
used,
\begin{equation}
  K_1(\ve{x} \ | \ \ve{\xi}) =
    \frac{4}{9 \pi \sigma^2} \left[1 - \frac{\|\ve{x} - \ve{\xi}\|^2}{9
    \sigma^2} \right]^3 \ , \ \ \|\ve{x} - \ve{\xi}\| < 3\sigma \ ,
\end{equation}
with bandwidth $\sigma = 4.2 \ \mathrm{cm}$ for $150 \ \mathrm{cm}$
environments and $\sigma = 2.8 \ \mathrm{cm}$ for $100 \ \mathrm{cm}$
environments.

The triweight kernel has compact support, so we avoid the infinite
extrapolation entailed by a Gaussian. It is also a smooth, bell-shaped bump,
that eliminates some problems that in our experience tend to show up when using
finite kernels that go non-smoothly to zero (such as the uniform or
Epanechnikov kernels), in particular that the ratios of such KDEs tend to blow
up along the edges of the support.

Model simulations returned a continuous firing rate rather than discrete
spikes, providing a combined position-rate time series $(t_i, x_i, y_i, r_i)$.
Rate maps were created by considering each rate sample as a fractional spike
and otherwise proceeding as with spike trains,
\begin{equation}
  \label{eq:spatialrate-model}
  f(\ve{x}) = \frac{\sum_i r_i \Delta t_i K_1(\ve{x} \ | \ \ve{x}_i)}
  {\sum_i \Delta t_i K_1(\ve{x} \ | \ \ve{x}_i)}
  \ .
\end{equation}

All rate maps were discretized for further analysis by sampling $f(\ve{x})$ on
a uniform square grid with bin centers $\ve{x}_{ij}$. We write $f_{ij}
= f(\ve{x}_{ij})$.  Unless otherwise stated, a $50 \times 50$-bin grid was
used.

\subsection{Correlograms}
Spatial rate correlograms were computed from discretized rate maps using a
Pearson product-moment correlation coefficient of overlapping regions of the
rate maps at all integer-bin spatial lags. The cross correlogram $\rho$ of
rate maps $f$ and $g$ can be expressed as:
\begin{equation}
  \label{eq:correlogram}
  \rho_{ij} = \frac{\sum_{kl} \left(f_{kl} - \overline{f} \right) \left(g_{k + i,
  l + j} - \overline{g} \right)}
  {\sqrt{\sum_{kl} \left( f_{kl} - \overline{f} \right)^2 \sum_{kl} \left(g_{k
  + i, l + j} - \overline{g} \right)^2}} \ .
\end{equation}
All sums were taken over the overlapping region of the rate maps at the given
lag $i, j$.  A single global mean $\overline{f} = \sum_{ij} f_{ij} / N$ was
used at all lags (here $N$ is the number of bins in $f_{ij}$, usually $50
\times 50 = 2500$).

\subsection{Scores and selection}
\label{supp:Scores}
The grid score of a cell was computed from its spatial rate autocorrelogram by
considering the angular Pearson autocorrelation of the autocorrelogram at lags
of $30\dg,60\dg,90\dg,120\dg,150\dg$, as well as $\pm 3\dg$ and $\pm 6\dg$
offsets from these angles.  Only bins closer to the center than a radius
$u$ were included in the angular autocorrelations, and the grid score at
a particular $u$ was defined as the difference between the average of the
maximum correlations around $60\dg$ and $120\dg$ lags and the average of the
minimum correlations around the $30\dg, 90\dg$ and $150\dg$ lags:
\begin{equation}
  \begin{aligned}
    \text{Grid score}(u) {}={}
    &\frac{1}{2} \big[\hphantom{+}
    \max\left\{\text{acorr}(u) \text{ at } 60\dg \pm
      (0\dg,3\dg,6\dg)\right\} +\\
    &\hphantom{\frac{1}{2} \big[+}
      \max\left\{\text{acorr}(u) \text{ at } 120\dg \pm
    (0\dg,3\dg,6\dg)\right\} \big]\\
    {}-{}&\frac{1}{3} \big[\hphantom{+}
    \min\left\{\text{acorr}(u) \text{ at } 30\dg \pm
      (0\dg,3\dg,6\dg)\right\} +\\
    &\hphantom{\frac{1}{3} \big[+}
      \min\left\{\text{acorr}(u) \text{ at } 90\dg \pm
    (0\dg,3\dg,6\dg)\right\} +\\
    &\hphantom{\frac{1}{3} \big[+}
      \min\left\{\text{acorr}(u) \text{ at } 150\dg \pm
    (0\dg,3\dg,6\dg)\right\} \big] \ .
  \end{aligned}
\end{equation}
The final grid score of the cell was defined as the maximum of $\text{Grid
score}(u)$ over values of $u$ ranging from four bin widths (i.e.\
$u = 12 \ \mathrm{cm}$ for $50 \times 50$ bins in a $150 \ \mathrm{cm}$
box) to the full width of the experimental environment (half the width of the
autocorrelogram), computed at intervals of one bin width.

We also calculated the spatial sparsity score defined
by Buetfering et al.~\cite{buetfering2014parvalbumin}
(adapted from Skaggs et al.~\cite{skaggs1996theta}) to quantify
the degree of spatial tuning in the rate map:
\begin{equation}
  \label{eq:sparsity}
  \text{Sparsity score}
  = 1 - \frac{\left(\overline{f}\right)^2}{\overline{f^2}} \ ,
\end{equation}
where $f$ denotes the firing rate of the cell and the bars indicate an
occupancy weighted spatial mean, i.e.\ a mean where each bin is weighted by the
value of the occupancy map in that bin. The occupancy map is the denominator in equation
\ref{eq:spatialrate}.

The selection of cells for further analysis was done using a combination of
these scores:
\begin{equation}
  \text{Combined score} = 1.2 \cdot \text{Grid score} + 1.55 \cdot
  \text{Sparsity score} \ .
\end{equation}
Only cells for which the combined score exceeded $1.0$ were included in the
further analysis. The purpose of this selection was to only include cells for
which it was possible to make out clear and distinct firing fields in the
spatial rate maps (quantified by the sparsity score) and for which these fields
were laid out in a grid-like pattern (quantified by the grid score). The form
of the combined score allows for an additive tradeoff between the two
properties, while rejecting cells for which both scores are low. The exact
weights were determined by looking for an appropriate cutoff line through grid
score-sparsity space (see Supplementary figure \ref{fig:supp-filter}) that
separated cells that were subjectively deemed too noisy or disordered from the
others.

In order to exclude suspected multiple recordings of the same cell, an
additional selection filter was applied%
. For each pair of cells recorded from the same animal, one cell was excluded
if the following criteria were satisfied:
\begin{enumerate}
    \item The cells were recorded on the same tetrode.
    \item The cells were either
      \begin{itemize}
        \item recorded closer in depth than $75 \ \mathrm{\mu m}$ (for datasets
          where tetrode depth estimates for each session were available), or
        \item recorded closer in time than nine days (for other datasets).
      \end{itemize}
    \item The Pearson correlation of the cell rate maps at zero spatial lag
      exceeded $0.8$.
\end{enumerate}
Note that the issue of possible repeats in these datasets was also addressed in
the original reference~\cite[Supplementary Materials and
Methods]{stensola2012entorhinal}, and we have only used data that was kept at
that time.

 
\subsection{\label{sec:lattice-vectors}Lattice vectors}
The spatial rate autocorrelograms were separated into segments associated with
different peaks using a watershed algorithm. The algorithm is seeded with all
the local maxima in the correlograms, and works by interpreting the negative of
the correlogram as a topographical map and virtually ``flooding'' the image
from imaginary water sources at the seeds. Thus each source is assigned
a basin, separated from neighboring basins along the watershed lines where
water from different sources meet as the water level is increased. Segmentation
was only applied to regions of the correlogram exceeding a correlation
threshold $\rho_{\min}$, where we used $\rho_{\min} = 0.1$ as a default value
and made adjustments for certain cells in order to properly separate the
central peak and the inner ring of six peaks around it. We always discarded all
segments associated with a seed in an edge bin of the correlogram, but only
after performing the segmentation with such seeds included.

The exact location of a peak was defined as the center of mass of the
associated segment of the correlogram.

The inner six peaks provide three vectors that would ideally correspond to
primitive lattice vectors of the grid pattern of the cell. (Due to the
inversion symmetry of the autocorrelogram, there are only three unique vectors
up to a trivial sign difference.)  However, while a lattice can be defined by
any two of the vectors extracted from the autocorrelogram, the third vector
will in general not conform exactly to the same lattice.  Therefore, we derived
a projection from the peak vectors to a corresponding set of consistent lattice
vectors.

Let $\ve{p}_i, \ i = 1, 2, 3$ be three unique peak vectors from the
autocorrelogram, numbered in successive order around the origin. In
addition, let $\ve{a}_i$ be a set of vectors that define a consistent lattice
and are closely approximated by the $\ve{p}_i$. In order to define a lattice,
the $\ve{a}_i$ must satisfy
\begin{equation}
  \label{eq:lattice-criterion}
  \ve{a}_2 = \ve{a}_1 + \ve{a}_3 \ .
\end{equation}
The vector $\ve{a}_2$ must then be close to both $\ve{p}_2$ and $\ve{p}_1
+ \ve{p}_3$, so we postulate that it should be expressible as a weighted
average of the two: $\ve{a}_2 = (\mu \ve{p}_2 + \ve{p}_1 + \ve{p}_3)
/ (\mu + 1)$. Extending this reasoning to $\ve{a}_1$ and $\ve{a}_3$, we
arrive at the transformation
\begin{equation}
  \begin{pmatrix}
    \ve{a}_1 \\
    \ve{a}_2 \\
    \ve{a}_3
  \end{pmatrix}
  =
  P
  \begin{pmatrix}
    \ve{p}_1 \\
    \ve{p}_2 \\
    \ve{p}_3
  \end{pmatrix} \ ,
\end{equation}
with
\begin{equation}
  P =
  \frac{1}{\mu + 1}
  \begin{pmatrix}
    \mu & 1 & -1 \\
    1 & \mu & 1 \\
    -1 & 1 & \mu \\
  \end{pmatrix} \ .
\end{equation}
This transformation is by construction equivalent to the identity whenever the
$\ve{p}_i$ already satisfy \ref{eq:lattice-criterion}. To guarantee that the
$\ve{a}_i$ always define a consistent lattice, it is therefore sufficient to
require that $P$ is a projection matrix, i.e.\ $P^2 = P$. This gives the unique
solution $\mu = 2$. The effect of this projection is shown in Supplementary
figure \ref{fig:supp-vector-projection}.

In practice, we found it more convenient to work with all six peaks in the
inner ring in the autocorrelogram. The projection derived above is
straightforward to extend to that case, and the result is
\begin{equation}
  \begin{pmatrix}
    \ve{a}_1 \\
    \ve{a}_2 \\
    \ve{a}_3 \\
    \ve{a}_4 \\
    \ve{a}_5 \\
    \ve{a}_6
  \end{pmatrix}
  =
  \frac{1}{6}
  \begin{pmatrix}
    2 & 1 & -1 & -2 & -1 & 1 \\
    1 & 2 & 1 & -1 & -2 & -1 \\
    -1 & 1 & 2 & 1 & -1 & -2 \\
    -2 & -1 & 1 & 2 & 1 & -1 \\
    -1 & -2 & -1 & 1 & 2 & 1 \\
    1 & -1 & -2 & -1 & 1 & 2
  \end{pmatrix}
  \begin{pmatrix}
    \ve{p}_1 \\
    \ve{p}_2 \\
    \ve{p}_3 \\
    \ve{p}_4 \\
    \ve{p}_5 \\
    \ve{p}_6
  \end{pmatrix} \ ,
\end{equation}
where the vectors are numbered in successive order around the origin.

In the following, we write the Cartesian components of $\ve{a}_i$ as $(a_{x,
i}, a_{y, i})$.

\subsection{Deformation ellipse}
To quantify the deformation of the grid cell firing pattern from a perfect
triangular lattice%
, and also to define the spacing of the grid pattern, we fitted an ellipse
through the endpoints of the lattice vectors $\ve{a}_i$. This fitting problem
has an exact solution, which we derive here.

A general conic section can be written as an implicit quadratic equation in two
variables:
\begin{equation}
  \label{eq:conic}
  A x^2 + 2 B x y + C y^2 + 2 D x + 2 E y + F = 0 \ .
\end{equation}
Scaling the six parameters $A, B, C, D, E, F$ by a common factor will not
change the conic, leaving five degrees of freedom for size, shape and
orientation. The conic will be an ellipse if $B^2 - A C < 0$.

For $N$ points $(x_i, y_i), \ i = 1, \ldots, N$, we define the $N \times 6$
design matrix $M$ as
\begin{equation}
  \label{eq:ellipse-matrix}
  M =
  \begin{pmatrix}
    x_1^2 & 2 x_1 y_1 & y_1^2 & 2 x_1 & 2 y_1 & 1 \\
     & & & & &  \\
    \vdots & \vdots & \vdots & \vdots & \vdots & \vdots \\
     & & & & &  \\
    x_N^2 & 2 x_N y_N & y_N^2 & 2 x_N & 2 y_N & 1
  \end{pmatrix}
\end{equation}
The conic section with parameter vector $\ve{\theta} = (A, B, C, D, E,
F)^T$ passes through all the points if $M \ve{\theta} = 0$.  When $M$ has
rank $5$, and hence a one-dimensional null space, this equation uniquely
determines $\ve{\theta}$ up to a scaling factor, and hence there is a unique
conic section passing through all $N$ points.  In this case the system can be
solved via singular value decomposition of $M$ -- the right-singular vector
corresponding to the vanishing singular value is the parameter vector
$\ve{\theta}$.

If we construct $M$ from the lattice vectors $\ve{a}_i = (a_{x, i}, a_{y, i})$,
and use the fact that $\ve{a}_{i + 3} = -\ve{a}_i, \ i = 1, 2, 3$ (assuming
cyclically wrapping indices), we can write
\begin{equation}
  \label{eq:ellipse-block-form}
  M =
  \begin{pmatrix*}[r]
    A & B + C \\
    A & -B + C
  \end{pmatrix*} \ ,
\end{equation}
which is row equivalent to
\begin{equation}
  \label{eq:ellipse-block-reduced}
  M' =
  \begin{pmatrix*}[r]
    A & C \\
    0 & -B
  \end{pmatrix*} \ .
\end{equation}
Here,
\begin{equation}
  \label{eq:ellipse-blocks}
  \begin{split}
    A = {} &
    \begin{pmatrix}
      a_{x, 1}^2 & 2 a_{x, 1} a_{y, 1} & a_{y, 1}^2 \\
      a_{x, 2}^2 & 2 a_{x, 2} a_{y, 2} & a_{y, 2}^2 \\
      a_{x, 3}^2 & 2 a_{x, 3} a_{y, 3} & a_{y, 3}^2
    \end{pmatrix} \ , \\
    B = {} &
    \begin{pmatrix}
      2 a_{x, 1} & 2 a_{y, 1} & 0 \\
      2 a_{x, 2} & 2 a_{y, 2} & 0 \\
      2 a_{x, 3} & 2 a_{y, 3} & 0
    \end{pmatrix} \ , \\
    C = {} &
    \begin{pmatrix}
      0 & 0 & 1 \\
      0 & 0 & 1 \\
      0 & 0 & 1
    \end{pmatrix} \ .
  \end{split}
\end{equation}
Provided that $\ve{a}_1$, $\ve{a}_2$ and $\ve{a}_3$ are not all parallel, we
see that $\mathrm{rank} \ B = 2$. As long as none of them are parallel, we also
find that $A$ is full-rank (the easiest way to see this is to divide row $i$ in
  $A$ by $a_{x, i}^2$, and consider the linear independence of the resulting
  rows). Hence, $\mathrm{rank} \ [A \ \ C] = 3$, and thus $\mathrm{rank}
  \ M = 5$. This is exactly the condition required to solve exactly for the
  conic section passing through the lattice vector endpoints. Note that this
  derivation only relies on inversion symmetry among the $\ve{a}_i$, so the
  conic can just as easily be fitted through the unprojected peak vectors
  $\ve{p}_i$%
\footnote{What we have shown here is really that six points with inversion
symmetry through a common center, or equivalently, a center and three
peripheral points, uniquely define a conic section, as long as the points lie
on three different rays through the center.}.

We have not shown that the fitted conic will be an ellipse, but the inversion
symmetry among the $\ve{a}_i$ rules out the possibility of obtaining
a parabola, and a hyperbola will not be obtained from points that are vertices
of a convex polygon. Both these properties should always be satisfied for the
inner ring of peaks in a grid cell autocorrelograms.

From the analytic geometry parametrization $\ve{\theta} = (A, B, C, D, E,
F)^T$, one can derive the equivalent canonical parameters $(x_c, y_c, a, b,
\theta)$ of the ellipse (see e.g.~\cite{weisstein}).  Here, $(x_c, y_c)$ are
the coordinates of the center of the ellipse, $a$ and $b$ are the lengths of
the semi-major and semi-minor axes, and $\theta \in [-\pi / 2, \pi / 2)$ is the
  angle between the major axis of the ellipse and the coordinate $x$-axis.
  Defining the translated and rotated coordinates
\begin{equation}
  \label{eq:trans-rot}
  \begin{split}
    x' &=  (x - x_c) \cos \theta + (y - y_c) \sin \theta \ , \\
    y' &= -(x - x_c) \sin \theta + (y - y_c) \cos \theta \ ,
  \end{split}
\end{equation}
the ellipse satisfies
\begin{equation}
  \label{eq:ellipse-canonical}
  \frac{{x'}^2}{a^2} + \frac{{y'}^2}{b^2} = 1 \ .
\end{equation}

\subsection{Grid spacing}
We defined the grid spacing $\lambda$ of a cell as the geometric mean of the
lengths of the semiaxes of the deformation ellipse, $\lambda = \sqrt{a b}$.
Thus defined, $\lambda$ is the radius of a circle with the same area as the
ellipse, and it is therefore preserved under any area-preserving deformation of
the grid pattern, such as the shearing described
in Stensola et al.~\cite{stensola2015shearing}.

An ellipse fitted through a ring of grid cell autocorrelogram peaks, and
a circle with the same area, are shown in Supplementary figure
\ref{fig:supp-ellipse}.


\subsection{Grid cell feature space}
We defined a space of grid cell features in which Euclidean distance was used
as a proxy for the similarity of the cells' firing patterns. The feature vector
of a cell can be written as
\begin{equation}
  X = (w_{\lambda} \log \lambda, \tilde{a}_{x, 1}, \tilde{a}_{y, 1},
  \tilde{a}_{x, 2}, \tilde{a}_{y, 2}, \tilde{a}_{x, 3}, \tilde{a}_{y, 3})^T
     \ .
\end{equation}
Here, we define $\tilde{a}_{x, i} = a_{x, i} / \lambda$, and similarly for
$\tilde{a}_{y, i}$. These are dimensionless versions of the grid lattice
vectors, and characterize the orientation and deformation of the grid pattern
independently of the spacing.  This way, the grid pattern shape and size are
decoupled in feature space, and their relative importance is tuned by the
weight $w_{\lambda}$. The grid spacing $\lambda$ only enters through its
logarithm, so that all differences and distances in feature space are
dimensionless and independent of the measurement units used ($\log \lambda_2
- \log \lambda_1 = \log \lambda_2 / \lambda_1$).

\subsection{Module clustering}
Modules were identified by detecting clusters in the grid cell feature space
using the mean-shift clustering algorithm~\cite{comaniciu2002mean}.
The algorithm works by randomly choosing a candidate centroid, and iteratively
updating it until convergence by replacing it with the centroid of all points
within distance $h$.  This is repeated for a number of different starting
locations, and after eliminating near-duplicates, each resulting centroid
corresponds to a cluster in the dataset. A sample which was only visited by
(i.e.\ within distance $h$ of) centroid trajectories that converged to the same
final centroid is taken to be part of the corresponding cluster, while
remaining points are declared outliers. The algorithm takes a single parameter
$h$, the bandwidth, and determines the number of clusters automatically.

We discarded outliers and clusters with less than five cells, and accepted the
remaining clusters as grid cell modules. Using the feature space weight
$w_{\lambda} = 3.3$ and the mean shift bandwidth to $h = 0.35$, we reproduced
the modules from
Stensola et al.~\cite{stensola2012entorhinal} in all
animals%
\footnote{By this we mean that we found the same number of modules, with
approximately the same average grid parameters, in all animals. Since we have
filtered out a number of cells in previous steps, and our clustering algorithm
generates outliers, the allocation of cells to modules was not identical.}.

In order to assert the stability of the clusters, the clustering was repeated
$10$ times on each dataset. The clustering was identical in all $10$ runs on
all datasets when using the mentioned parameters.


\subsection{\label{sec:firing_field}Average firing field shape}
We used a small region around the central peak in the autocorrelogram as
as a model for the shape and size of a typical firing field of the cell. We
defined this region using a separate correlation threshold
$\rho_{\text{field}}$, and modeled the firing field shape as a 2D Gaussian bump
with variance $\Sigma^2 = -A / 2 \pi \log (\rho_{\text{field}})$, where $A$ is
the area of the region around the central peak where autocorrelation exceeds
$\rho_{ff}$. This definition is designed to match the value of the Gaussian at
distance $R = \sqrt{A / \pi}$ from the origin, relative to the peak value, to
the correlation threshold $\rho_{\text{field}}$, i.e.\
\begin{equation*}
  e^{-R^2 / 2 \Sigma^2} = \rho_{\text{field}} \ .
\end{equation*}
We used the threshold $\rho_{\text{field}} = 0.55$ for all cells. We defined
the radius of the average firing field as $u_f = 1.6 \Sigma$.

\subsection{Idealized spatial tuning profile}
\label{supp:Idealized}
We derived an idealized spatial tuning profile from a grid cell by tessellating
space with the lattice defined by the primitive lattice vectors $\ve{a}_i$, and
placing a circular Gaussian bump with variance $\Sigma^2$ at each node. The
lattice was translated to the location that maximized the zero-lag cross
correlation between the rate map of the actual cell and the idealized profile,
and scaled to have the same mean firing rate as the recorded cell. The tuning
profile is meant to represent an idealization of the spatial tuning of the real
grid cell, with the same lattice parameters, spatial phase, firing field size
and mean firing rate, but with firing fields that are identical.


\subsection{\label{sec:firing_field_detection}Firing field detection}
Firing field centers were detected in grid cell rate maps using the same
watershed segmentation algorithm used to detect peaks in the autocorrelograms
(see section \ref{sec:lattice-vectors}). The detection was performed on rate maps
created with $1.5$ times the bandwidth of the ordinary rate maps, i.e.\ $\sigma
= 6.3 \ \mathrm{cm}$ for $150 \ \mathrm{cm}$ environments and $\sigma = 4.2
\ \mathrm{cm}$ for $100 \ \mathrm{cm}$ environments. Segmentation was applied
only to regions of these rate maps where the firing rate exceeded the $25$th
percentile of the distribution of rates in the rate map.

Candidate firing fields were defined as circular regions of radius $u_f$ around
the detected firing field centers. The amplitude $r$ of a field was defined as
the number of spikes recorded within this field region, divided by the time
spent there.

A final filtering step to distinguish true firing fields from possible noise
blobs was performed by comparing the candidate fields to the firing fields in
the idealized spatial tuning profile derived from the rate map, using the same
radius $u_f$ for the idealized and candidate fields. We computed the overlap
fraction $q$ of each pair of a candidate and an idealized field, defined as the
ratio of the overlapping area of the fields to the field area $\pi u_f^2$.
Pairs with $q > 0.25$ were sorted by the product $r q$, where $r$ is the
amplitude of the candidate field, and unique pairs were kept starting from the
top of this list. Thus false positives among detected fields were eliminated,
and each detected field was associated with an idealized counterpart.

For rate maps created from sampled spikes, firing fields were detected in the
exact same way, using the same idealized fields as for the corresponding real
rate map (i.e.\ no new idealized spatial tuning profile was created for the
synthetic rate map). Corresponding fields in the real and synthetic rate maps
were identified by association with the same idealized field, and only fields
detected in both the real rate map and all synthetic rate maps from the same
cell were used further analysis.

For rate maps created from model simulations, no lattice parameters were
extracted from the autocorrelogram and no idealized tuning profile was created.
Hence, candidate fields were used directly without any filtering. This is
unproblematic, since model cell rate maps are not noisy.

Supplementary figure \ref{fig:supp-synt-real-ratemap-amplitudes} shows the
detected firing fields and corresponding amplitudes for a selection of real
grid cells and an accompanying synthetic rate map per real cell.

\subsection{Sampled spike trains}
\label{supp:SampledSpikes}
We used the idealized spatial tuning to sample spike trains corresponding to
this spatial tuning and the real animal trajectory, assuming Poisson spiking
statistics and no covariates except spatial location.  To this end, the real
trajectory of the rat was divided into $\Delta t = 5 \ \mathrm{ms}$ bins, and
the target firing rate $f_i$ at in bin $i$, centered on time $t_i$, was found
by evaluating the tuning profile at the corresponding location $(x_i, y_i)$. In
each bin, a Bernoulli trial was performed with success probability $f_i \Delta
t$, and if successful a spike was placed at $t_i$.

Rate maps were computed from these spike trains in the same way as for real
cells.


\subsection{Statistical analysis of firing field amplitudes}
\label{sec:stanalysis_supp}
Only cells for which $3$ or more fields were detected (counting only fields
detected in the real rate map and all synthetic rate maps derived from it) were
used in the statistical analysis of firing field amplitudes.  We computed the
sample coefficient of variation of the amplitudes of a cell as
\begin{equation}
  c = \frac{s}{\overline{r}} = \frac{1}{\overline{r}}
  \sqrt{\frac{1}{k - 1} \sum_{i = 1}^k \left(\tilde{r}_i
  - \overline{\tilde{r}}\right)^2} \ .
  \label{eq:CVdef}
\end{equation}
Here, $k$ is the number of fields, $r_i, i = 1, \ldots, k$ are the amplitudes
of the fields, and $\overline{r} = \sum_i r_i / k$ is the mean amplitude. The
usual sample variance is given by $s^2$.

To compare the temporal stability of the field amplitudes with the
field-to-field variations, we extracted firing field amplitudes using only data
from the first and second halves of the experiment, respectively (using the
same fields as for the full-length data). We denote these amplitudes as
$r_{ij}$, where the index $i$ runs over the $k$ firing fields and $j$ runs over
the partitions $1, 2$. First, we define the following sample means:
\begin{align}
  \overline{r}_{i \cdot} &= \frac{1}{2} \left(r_{i1}
+ r_{i2} \right) \ , \\
\overline{r}_{\cdot j} &= \frac{1}{k} \sum_{i = 1}^k r_{ij}
\ , \\
  \overline{r}_{\cdot \cdot} &= \frac{1}{2k} \sum_{i = 1}^k
\left(r_{i1} + r_{i2}\right) \ .
\end{align}
The between-field variability is defined as,
\begin{equation}
  s^2_{B} = \frac{1}{k - 1} \sum_{i = 1}^k 2 \left(\overline{r}_{i \cdot}
  - \overline{r}_{\cdot \cdot}\right)^2 \ ,
\end{equation}
and the within-field variability as
\begin{equation}
  \label{eq:within-field}
  s^2_{W} = \frac{1}{k - 1} \sum_{j = 1}^2 \left[\left( \sum_{i = 1}^k
  \left(r_{ij} - \overline{r}_{i \cdot}\right)^2 \right)
- k \left(\overline{r}_{\cdot j} - \overline{r}_{\cdot
\cdot}\right)^2
\right] \ .
\end{equation}
The second term in equation \ref{eq:within-field} is a repeated-measures correction
included to remove contributions to the within-field variability due to a
coherent shift in amplitudes between the two halves of the experiment. We do
this because we are only interested in capturing random within-field
fluctuations that may explain the measured between-field variability, not
coherent variations that would not contribute to this and may even be due to
experimental artifacts such as as the cell not being detected by the tetrode
for parts of the experiment. We compensate by subtracting $n - 1$ from the
degrees of freedom of the estimator (i.e.\ dividing by $(k - 1)(2 - 1) = k - 1$
instead of $k(2 - 1) = k$), such that $s^2_{W}$ remains unbiased.

A key point here is that if all the $r_{ij}$ are independent and identically
distributed, both $s_B^2$ and $s_W^2$ are unbiased estimators of the variance
of this distribution. Hence, if $s_B^2$ is significantly larger than $s_W^2$,
this may indicate that there is a component of variability between the field
amplitudes that cannot be explained by fluctuations that affect between- and
within-field variabilities alike. This is made quantitative via the
$F$-statistic,
\begin{equation}
  F = \frac{s_B^2}{s_W^2} \ ,
\end{equation}
which may be familiar from one-way ANOVA. For each cell, we computed a $p$-value
based on this statistic by ranking it against the same statistic computed from
the $1000$ simulated spike trains associated with the cell. (Notably, we did
not use critical values of the $F$-distribution to get the $p$-values, as this
would be entail unjustified normality assumptions)

We also define dimensionless coefficients of variability, useful for plotting
and comparison:
\begin{equation}
  \begin{aligned}
    c_B = \frac{s_B}{\overline{r}_{\cdot \cdot}} \ , \\
    c_W = \frac{s_W}{\overline{r}_{\cdot \cdot}} \ .
  \end{aligned}
\end{equation}
The $F$-statistic can be expressed in terms of these quantities:
\begin{equation}
  F = \frac{c_B^2}{c_W^2} \ .
\end{equation}

\subsection{Two population continuous attractor network model}
\label{sec:twopopcanmodel_supp}

To investigate the firing properties of interneurons in the two population model with place-like 
external input, we constructed networks
consisting of $32 \times 28$ grid cells and $200$ interneurons for the fully factorized model, and 
$4480$ interneurons for the partially factorized model.
The activity was translated during exploration via place-like input from $180  \times 180 $ units
each encoding one unique position of the environment and projecting to a single grid cell.
The values of the non-zero connections from the place-like neurons was
drawn from a Gaussian distribution with mean $0.5$, truncated at $0.0$.
The values for the standard deviation ranged from
$\sigma = 0$ to $\sigma = 5$
for the fully factorized
model, and $\sigma = 0$ to $\sigma = 1$
for the partially factorized model.
The path of the animal in the simulation was taken from two adjacent 10 minute
recordings and resampled into $10 \ \mathrm{ms}$ bins.

In the models to check the drift of the network, we used $16 \times 13$, $32 \times 28$ and $64 \times 55$
grid cells in the fully factorized case and $32 \times 28$ for the partially factorized model. 
The number of interneurons was chosen as a fraction of the total grid cell population.
In each of these, the neurons were assumed to lie on a twisted torus~\cite{guanella2007model}
with connectivity as described in the main text and
$R_{\min} = 0.5 \times N_y$ and $R_{\max} = N_y$.
The other parameters were chosen as 
$\tau = 10 \ \mathrm{ms}$, 
$\textbf{constant} = 0.1$, $W_0 = 0.1$ and $l=0$.

For the path integration model we went with a four bump model with $64 \times 56$ grid cells,
$200$ interneurons, $R_{\min} = 15$, $R_{\max} = 20$, $\tau = 10 \ \mathrm{ms}$, $l=4$, 
$\alpha=0.0075$, $\textbf{constant} = 0.1$ and $W_0 = -0.05$.
As mentioned above, no place-like input were used for error correction during path integration.
A 10 minute recording of the animal was resampled into 10ms bins and used to insure that the
statistics of the movement were consistent with those of an actual animal.

\subsection{NMF}
\label{supp:NMF}
The matrices $\mathbf{J}$ and $\mathbf{K}$ were first set to small random values and the updated 
iteratively according to the multiplicative learning rules
\begin{align}
  &\mathbf{K} = \mathbf{K} \times (\mathbf{J}^T \mathbf{W})  / ( \mathbf{J}^T\mathbf{J} \mathbf{K}) \\
  &\mathbf{J} = \mathbf{J} \times (\mathbf{W} \mathbf{K}^T)  / ( \mathbf{J} \mathbf{K} \mathbf{K}^T ) 
\end{align}
where the signs $\times$ and $/$ represent element-wise multiplication and division, respectively. 
The above equations are iterated until convergence, i.e.\ the change in the value of the cost
function $\sqrt{\sum_{ij} \big(W_{ij} - \sum_k J_{ik} K_{kj} \big)^2}$ is below $0.0001$.

\section*{Acknowledgments}
We thank Hanne and Tor Stensola as well as Edvard and May-Britt Moser for making the data available. We are also grateful to
Edvard Moser for his comments on this work, and also Trygve Solstad for discussions at the early stages of this project.
This work was supported by the  Kavli Foundation, Norwegian Research Council (Centre of Excellence Scheme), and a Starr Foundation 
fellowship at the Simons Center for Systems Biology, IAS to YR.

\bibliographystyle{unsrt}
\bibliography{refs}

\clearpage
\newgeometry{top=0.425in,left=0.9in,right=0.9in,footskip=0.75in}
\section*{Supplementary Figures}

\setcounter{figure}{0}
\renewcommand{\figurename}{Supplementary fig}

\begin{figure}[h!]
  \centering
    \includegraphics[width=0.42\linewidth]{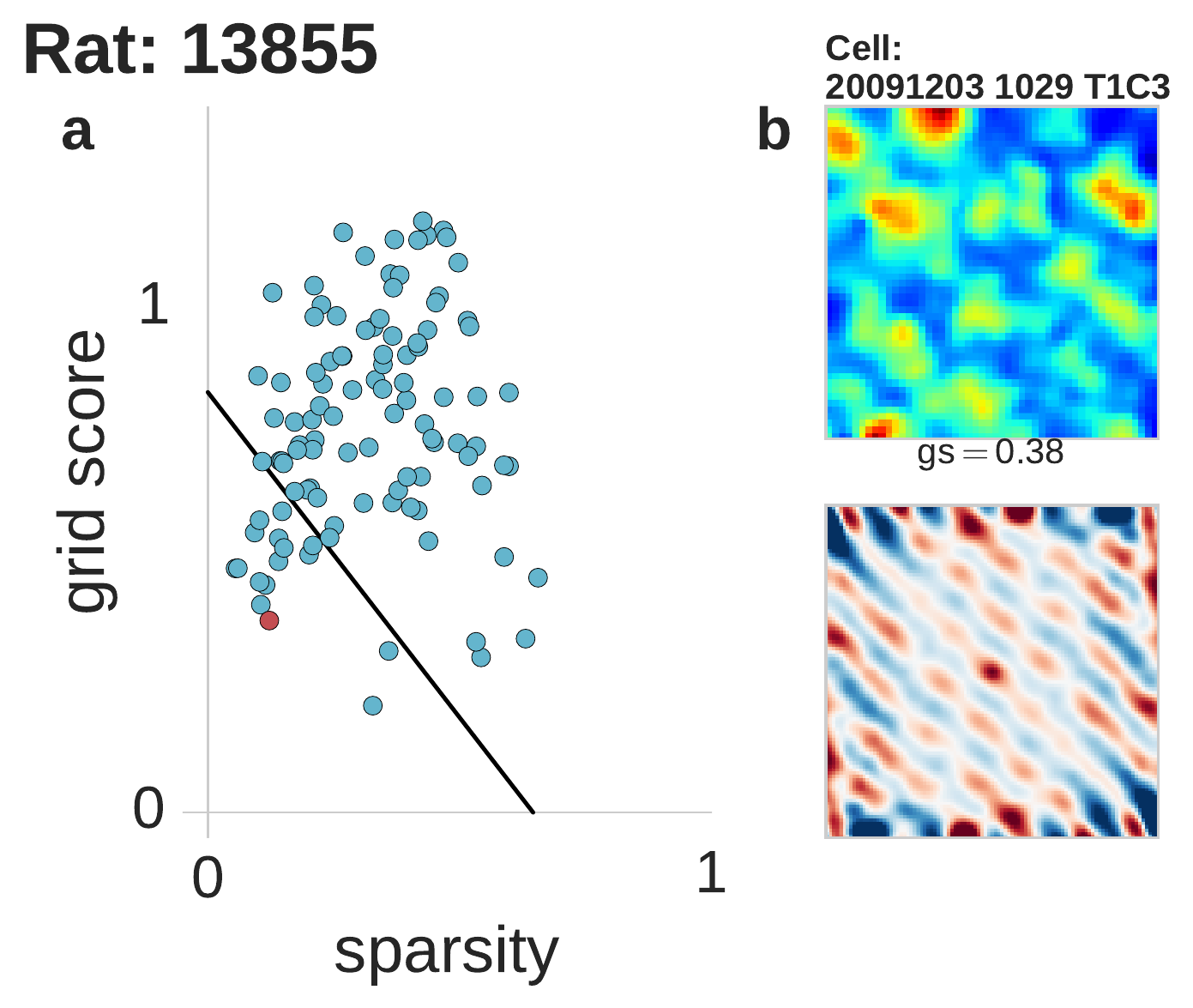}
  \hfill
    \includegraphics[width=0.42\linewidth]{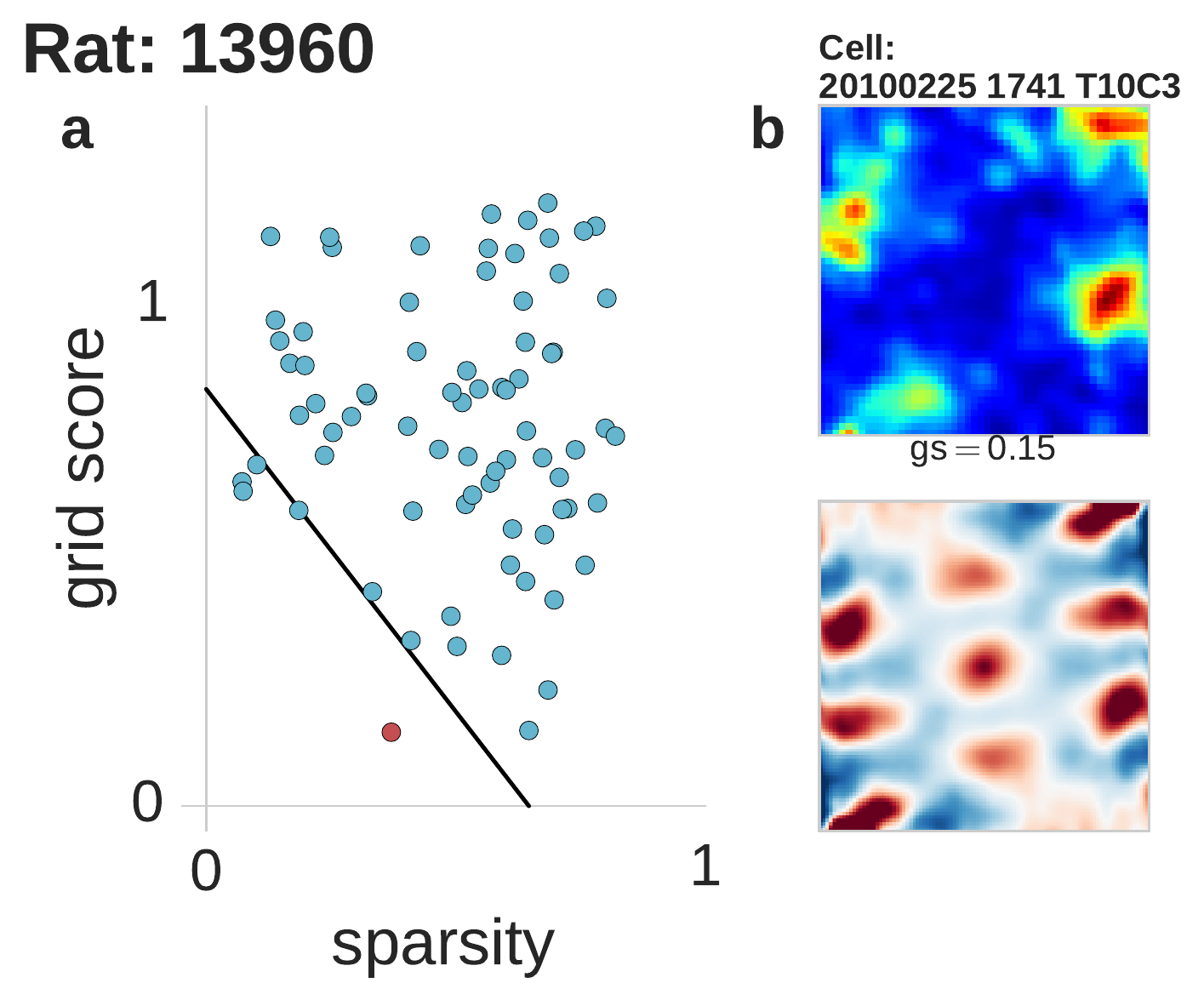}
  \\\vspace{10pt}
    \includegraphics[width=0.42\linewidth]{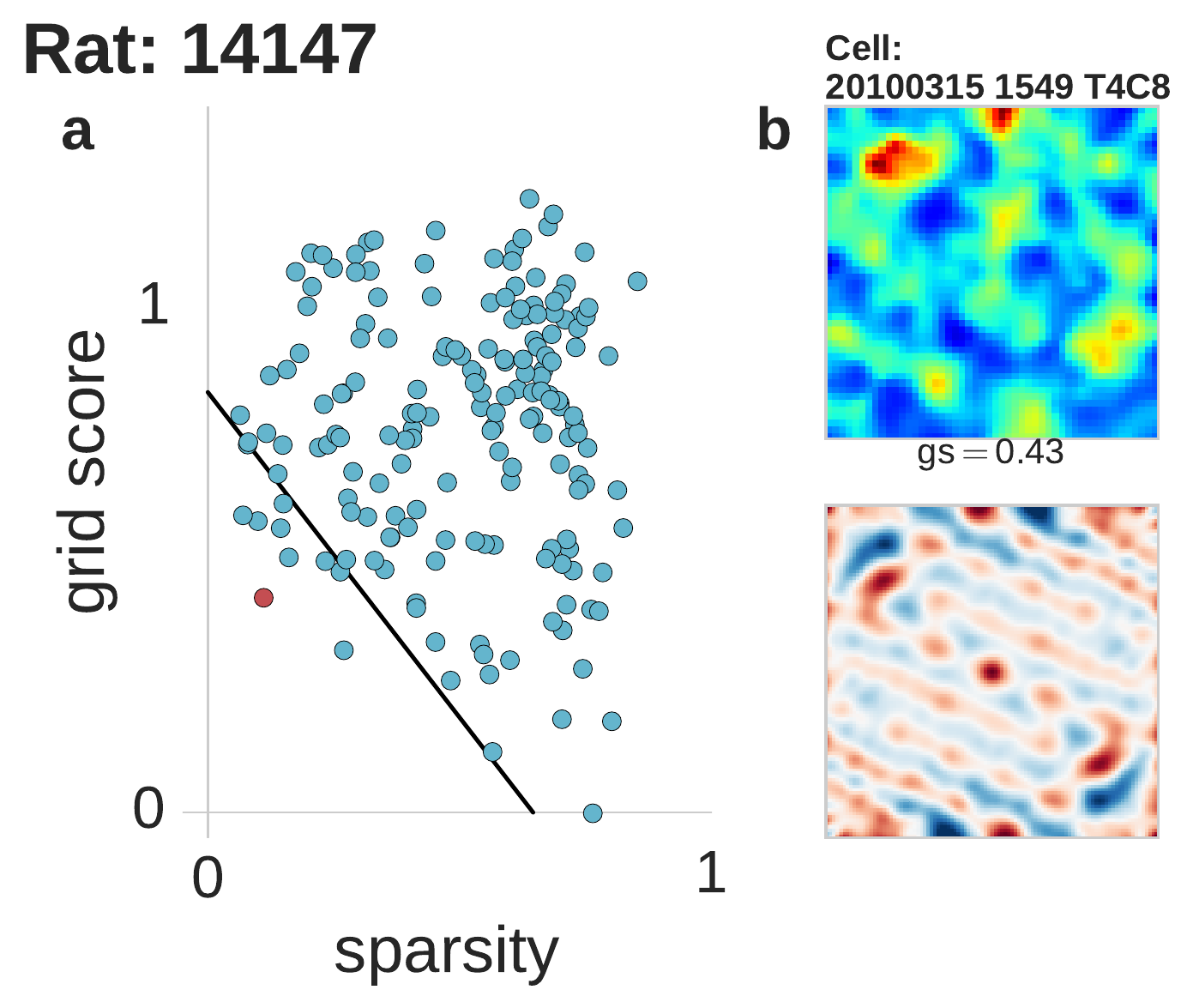}
  \hfill
    \includegraphics[width=0.42\linewidth]{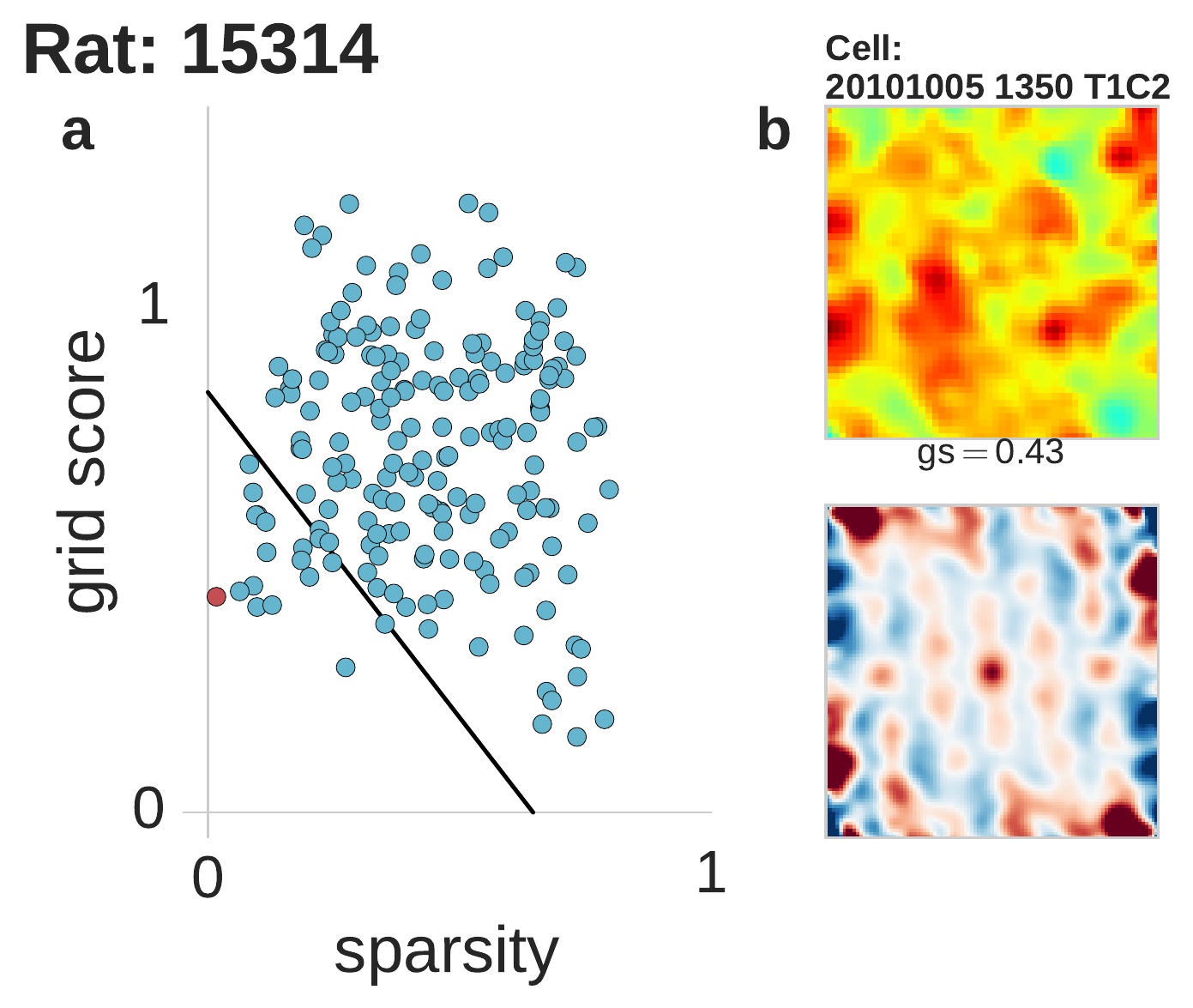}
  \\\vspace{10pt}
    \includegraphics[width=0.42\linewidth]{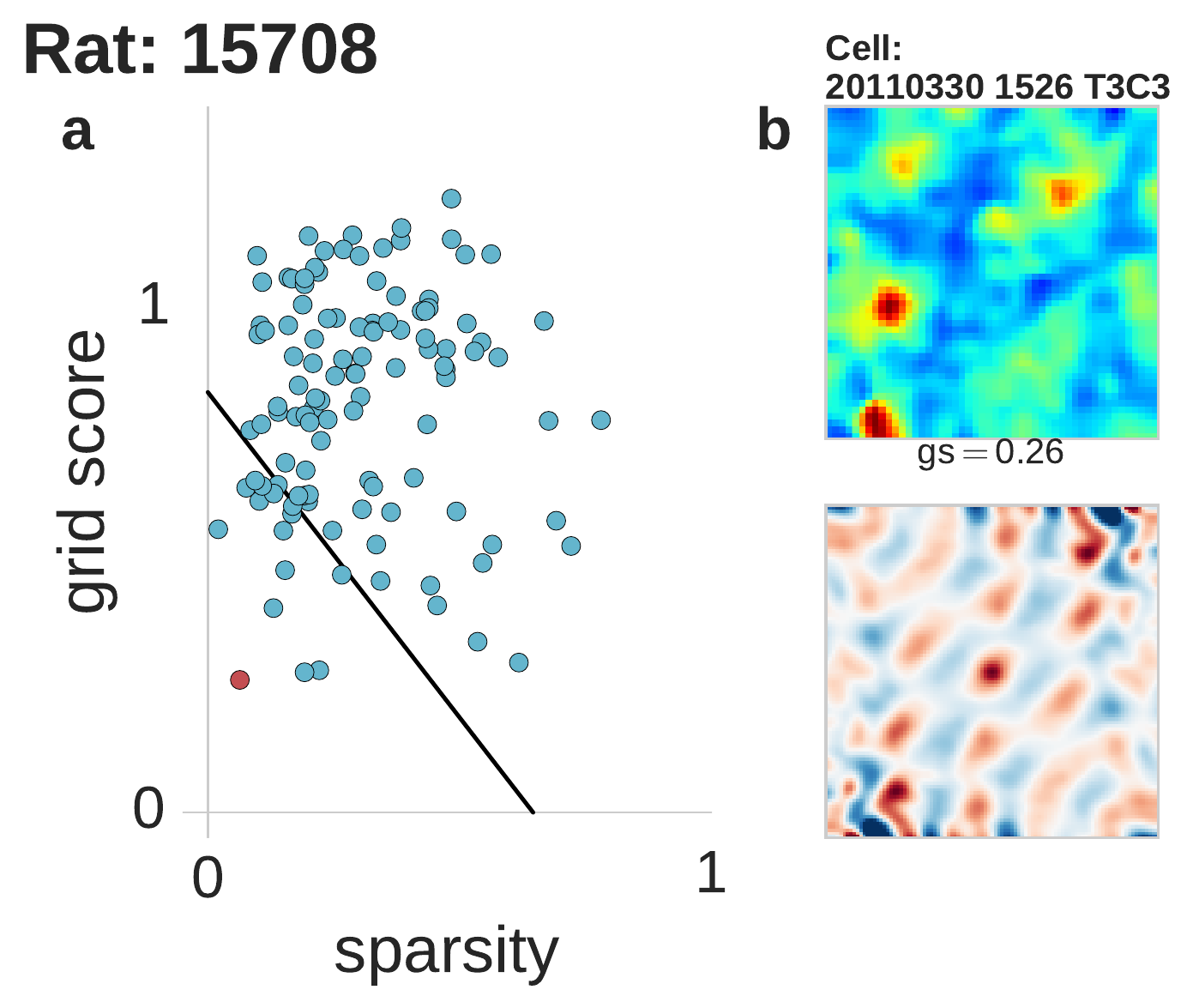}
    \caption{\label{fig:supp-filter}
      \textbf{Grid score--sparsity filtering.}
      \subfiglabel{a}: All available cells scattered in sparsity--grid
      score space. The combined score cutoff defined in the Materials and Methods section
      is drawn with a black line: only cells above/to the right of the
      line were kept. The discarded cells tend to have noisy ratemaps, where
      firing fields and a clear triangular lattice may be difficult to make
      out.
      \subfiglabel{b}: Ratemap and autocorrelogram of the cell with lowest
    combined score. This cell is also marked with a red dot in \subfiglabel{a}.
  The grid score ($\text{gs}$) and sparsity ($\text{sp}$) of the cell are
quoted below the ratemap.}
\end{figure}

\begin{figure}
  \centering
  \begin{minipage}{0.49\linewidth}
    \centering
    \includegraphics[width=\linewidth]{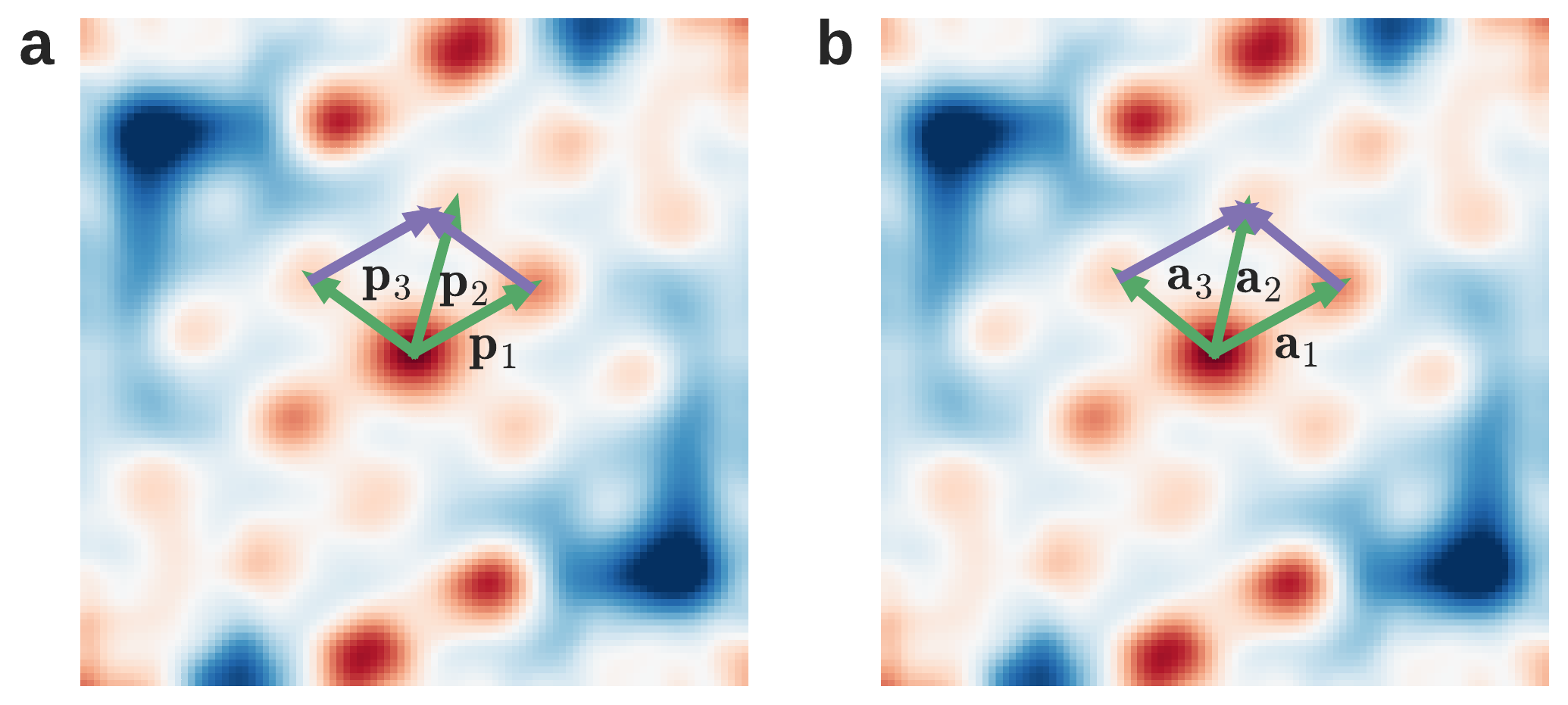}
    \caption{\label{fig:supp-vector-projection}
      \textbf{Projection from peaks to lattice vectors.}
      \subfiglabel{a}: A typical grid cell autocorrelogram with three detected
      peakcenters drawn as vectors $\ve{p}_1, \ve{p}_2, \ve{p}_3$ from the
      origin. The purple vectors demonstrate that $\ve{p}_1 + \ve{p}_3 \neq
      \ve{p}_2$.
      \subfiglabel{b}: The same autocorrelogram, but now with vectors
      $\ve{a}_1, \ve{a}_2, \ve{a}_3$ that are related to $\ve{p}_1, \ve{p}_2,
      \ve{p}_3$ by the lattice projection derived in the Materials and Methods section.
      These vectors satisfy $\ve{a}_1 + \ve{a}_3 = \ve{a}_2$, and hence they
      define a consistent lattice.
    }
  \end{minipage}
  \begin{minipage}{0.49\linewidth}
    \centering
    \includegraphics[width=0.6\linewidth]{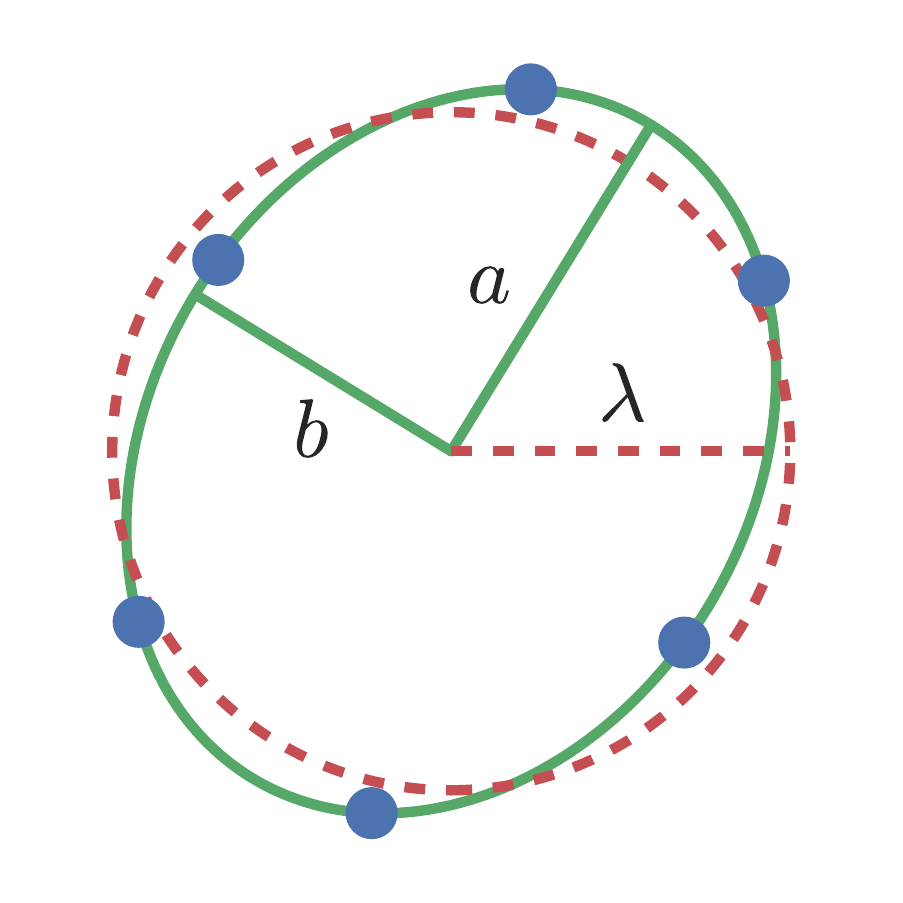}
    \caption{\label{fig:supp-ellipse}
      \textbf{Deformation ellipse and spacing of a grid pattern}
      The ellipse (green curve) has been fitted through a ring of
      autocorrelogram lattice points (blue dots) taken from the autocorrelogram.
      The semiaxes are labeled $a$ and
      $b$.  A circle with the same area as the ellipse is plotted
      concentrically (red dashed curve). The radius of this circle, $\lambda
      = \sqrt{a b}$, defines the grid spacing.}
  \end{minipage}
\end{figure}

\begin{figure}
  \centering
  \frame{%
    \includegraphics[width=0.32\linewidth]{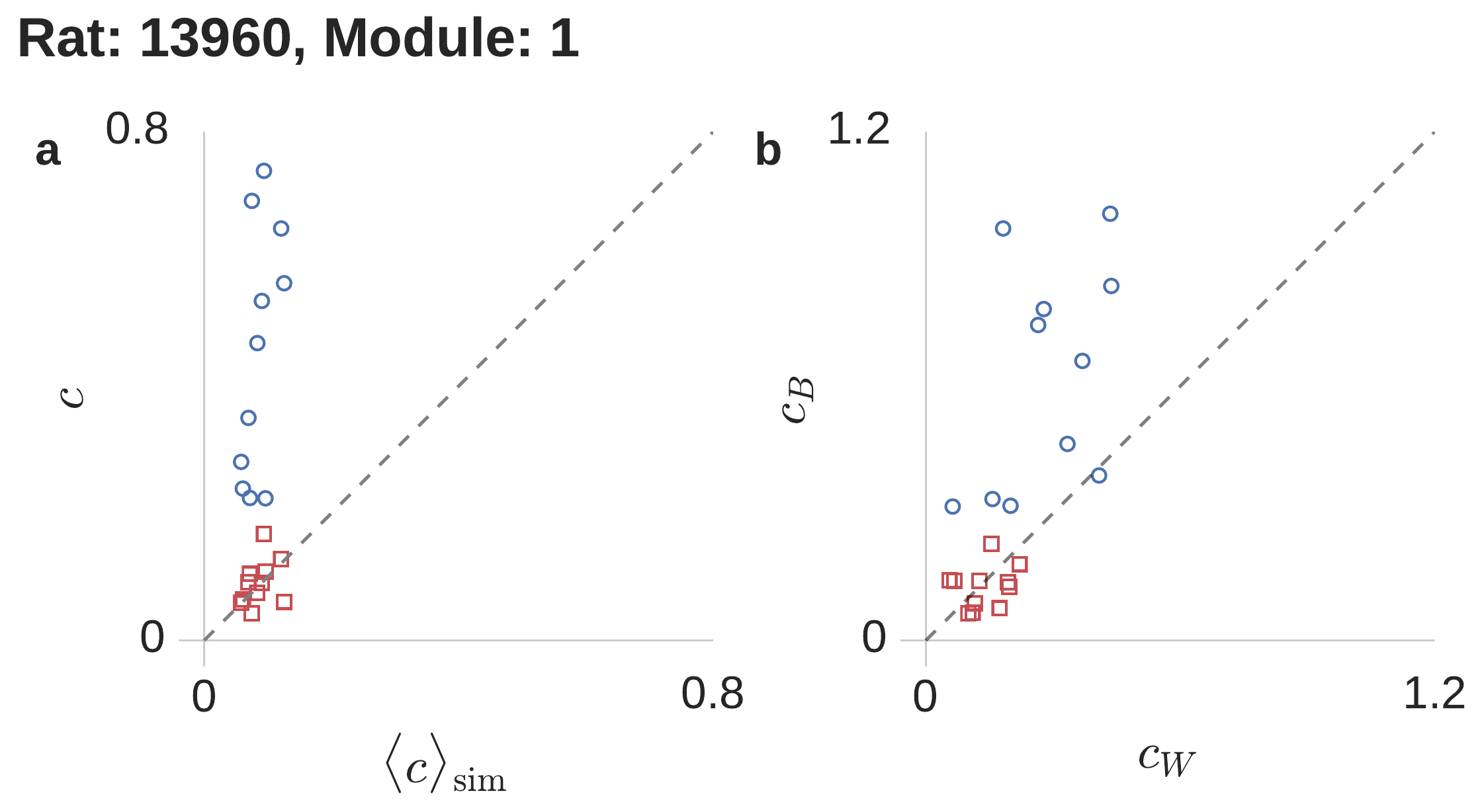}%
  }
  \hfill
  \frame{%
    \includegraphics[width=0.32\linewidth]{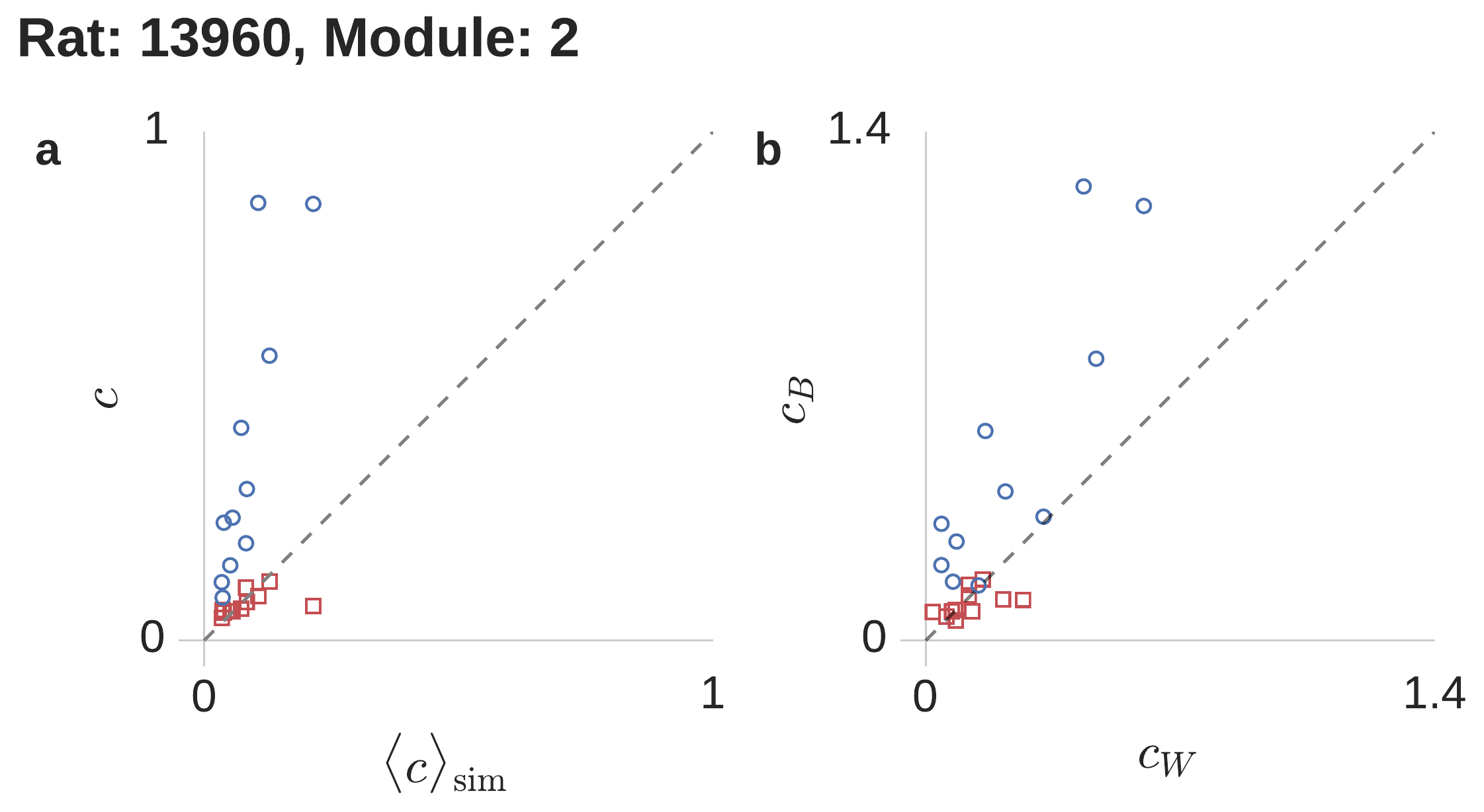}%
  }
  \hfill
  \frame{%
    \includegraphics[width=0.32\linewidth]{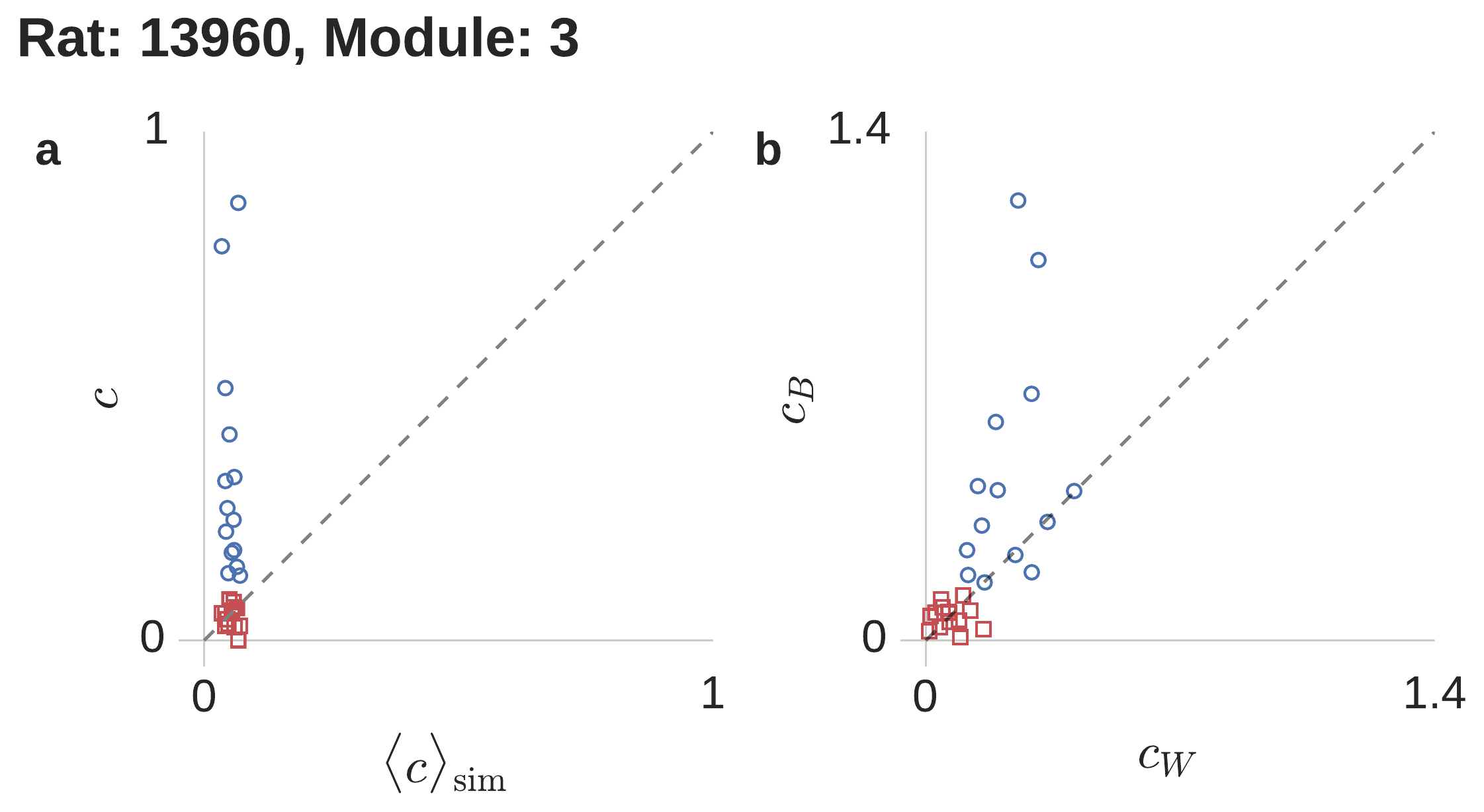}%
  }
  \frame{%
    \includegraphics[width=0.32\linewidth]{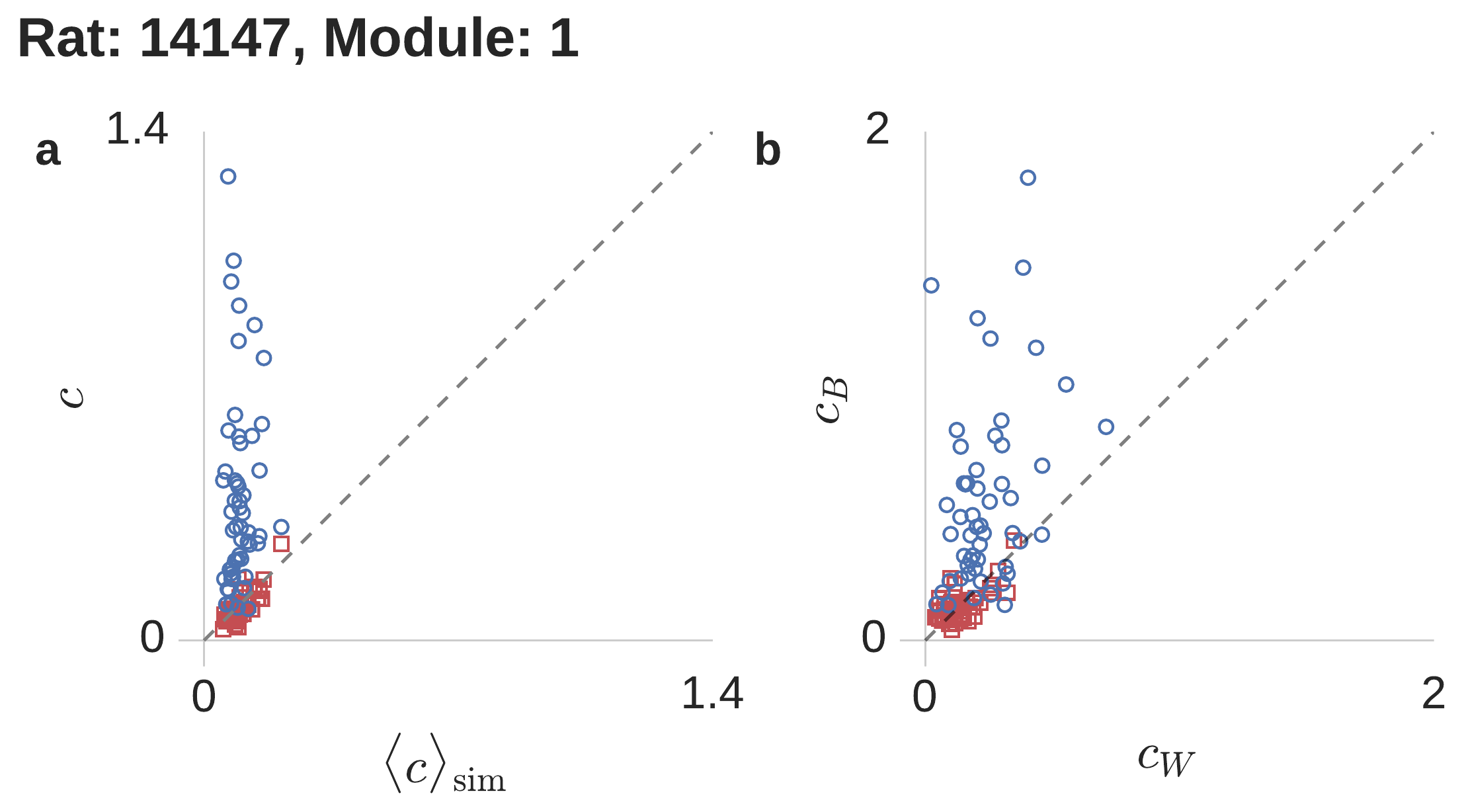}%
  }
  \hfill
  \frame{%
    \includegraphics[width=0.32\linewidth]{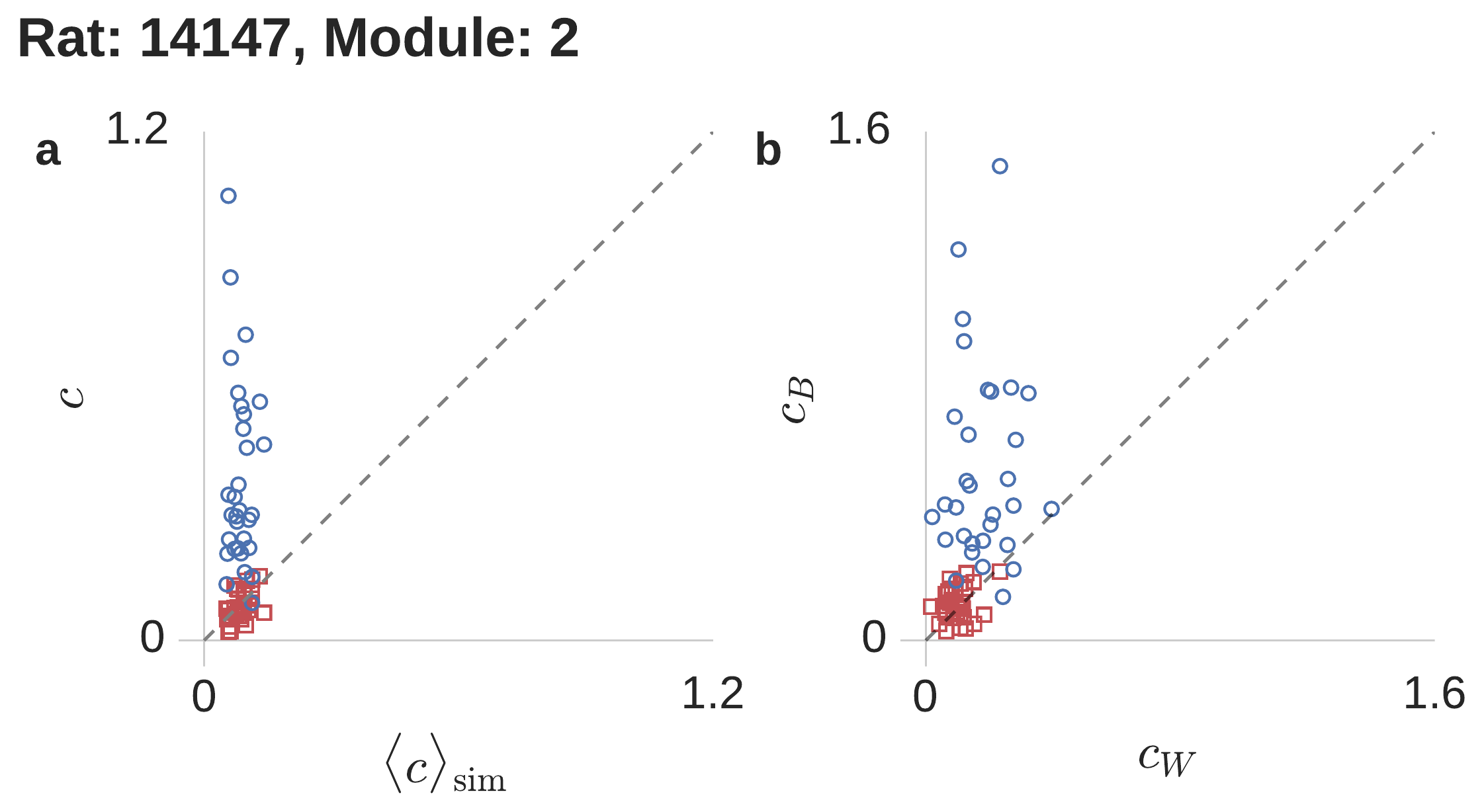}%
  }
  \hfill
  \frame{%
    \includegraphics[width=0.32\linewidth]{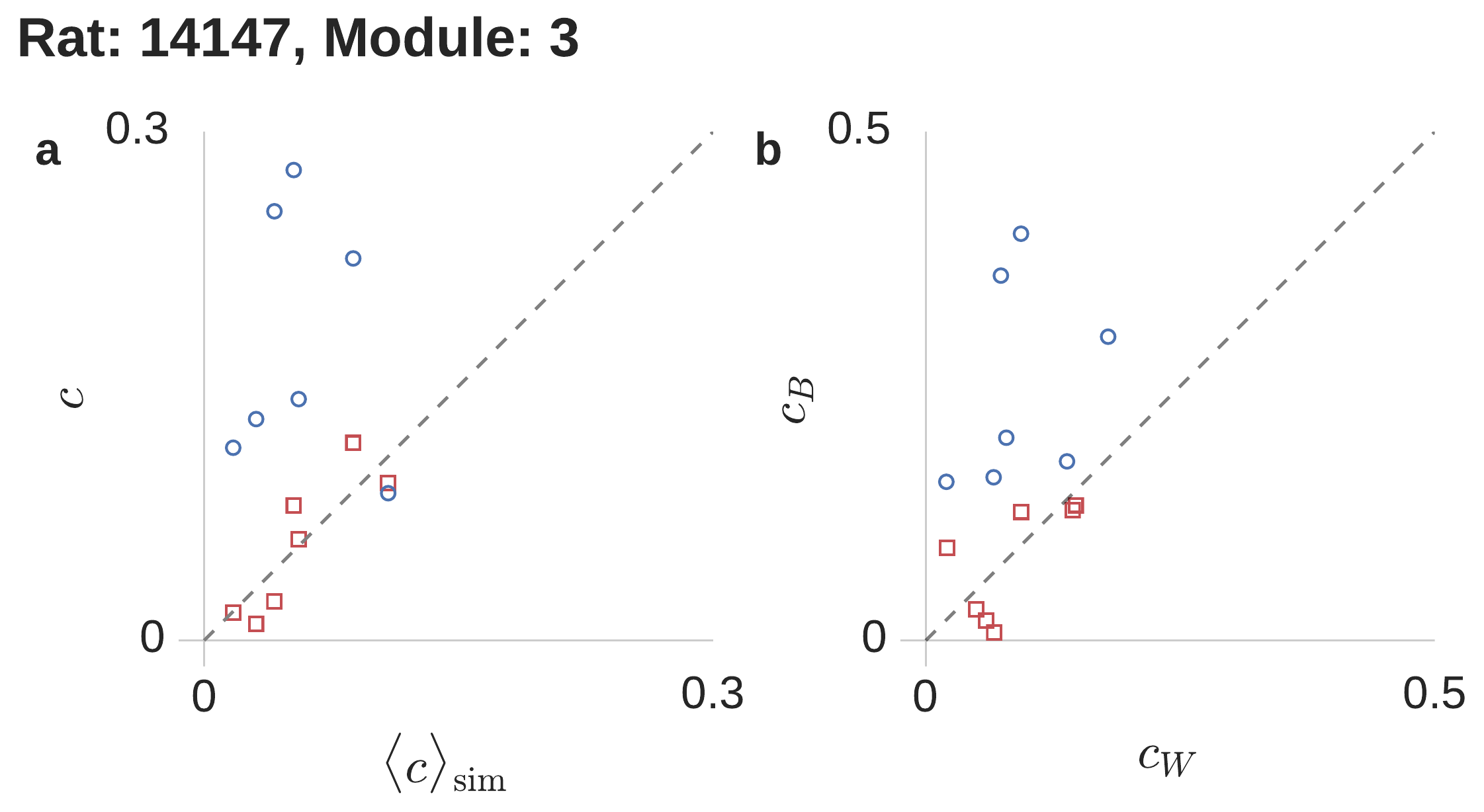}%
  }
  \frame{%
    \includegraphics[width=0.32\linewidth]{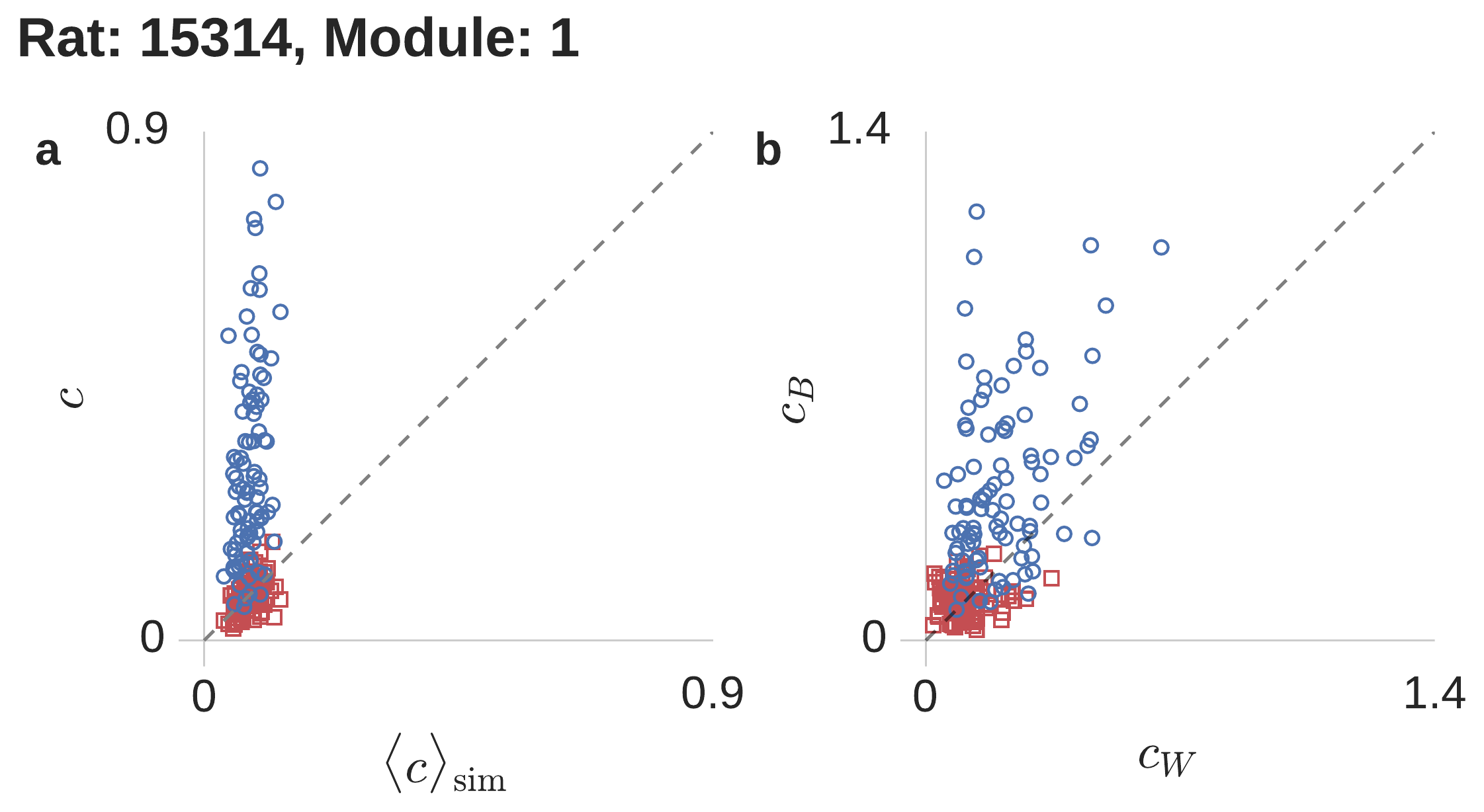}%
  }
  \hfill
  \frame{%
    \includegraphics[width=0.32\linewidth]{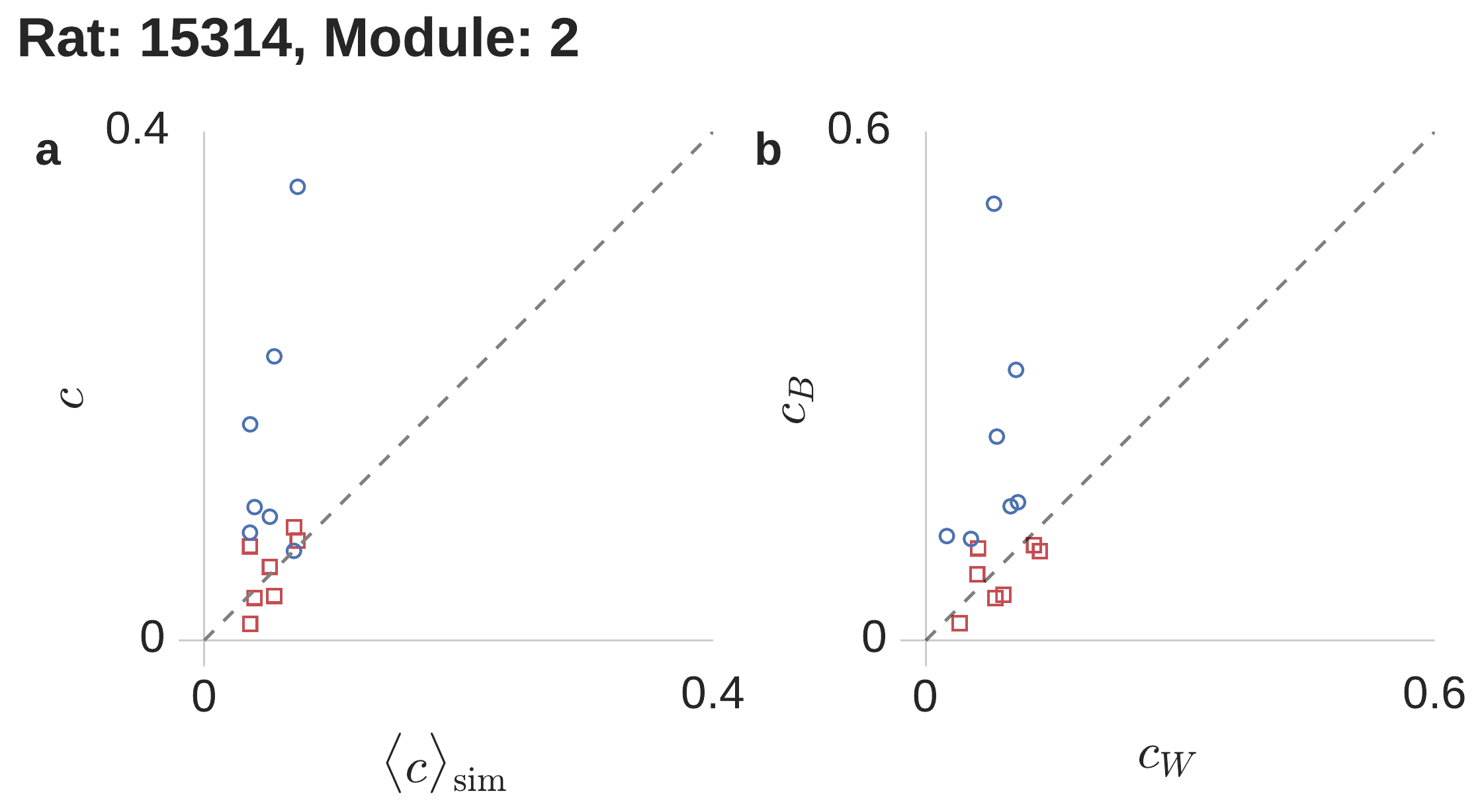}%
  }
  \hfill
  \frame{%
    \includegraphics[width=0.32\linewidth]{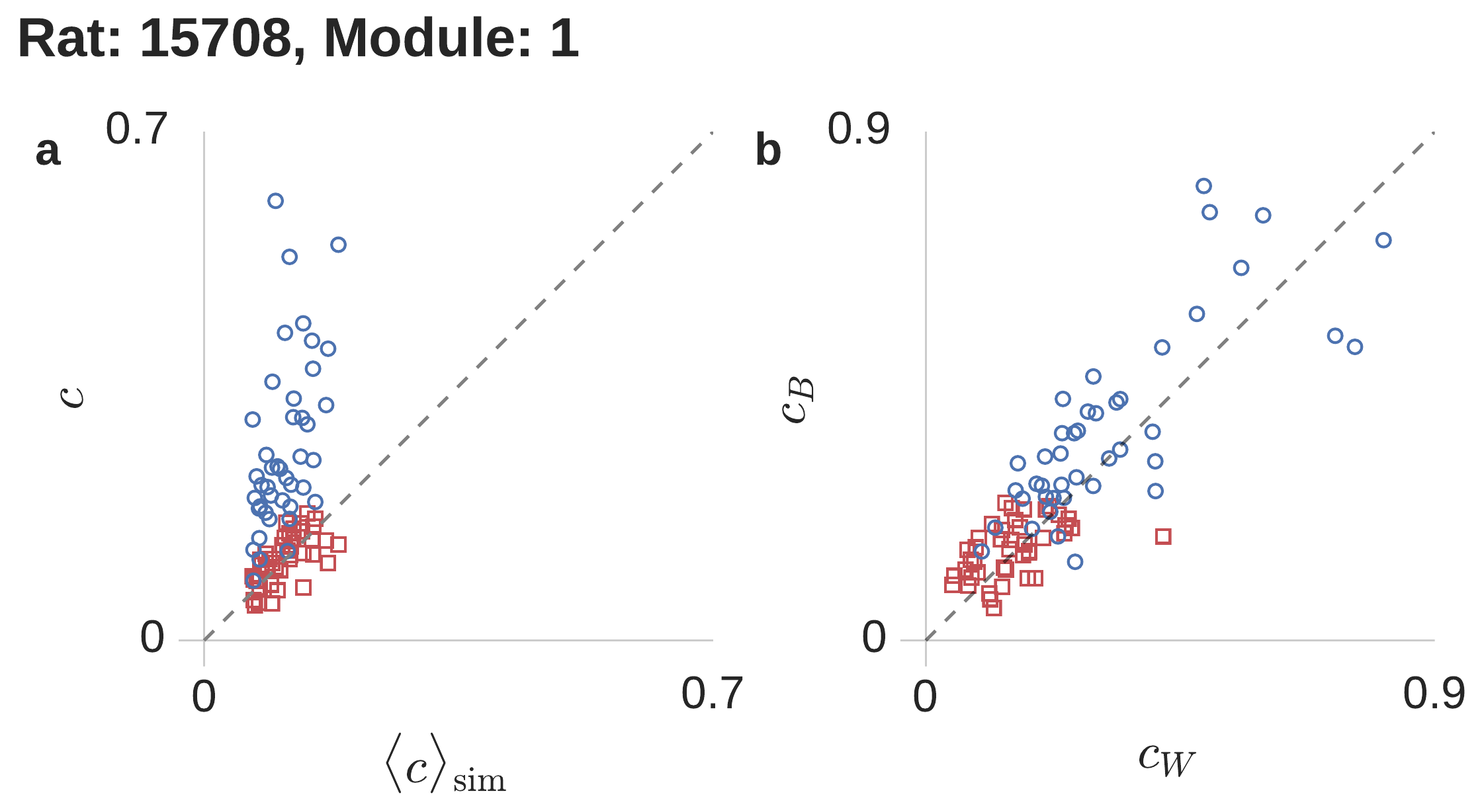}%
  }
  \frame{%
    \includegraphics[width=0.32\linewidth]{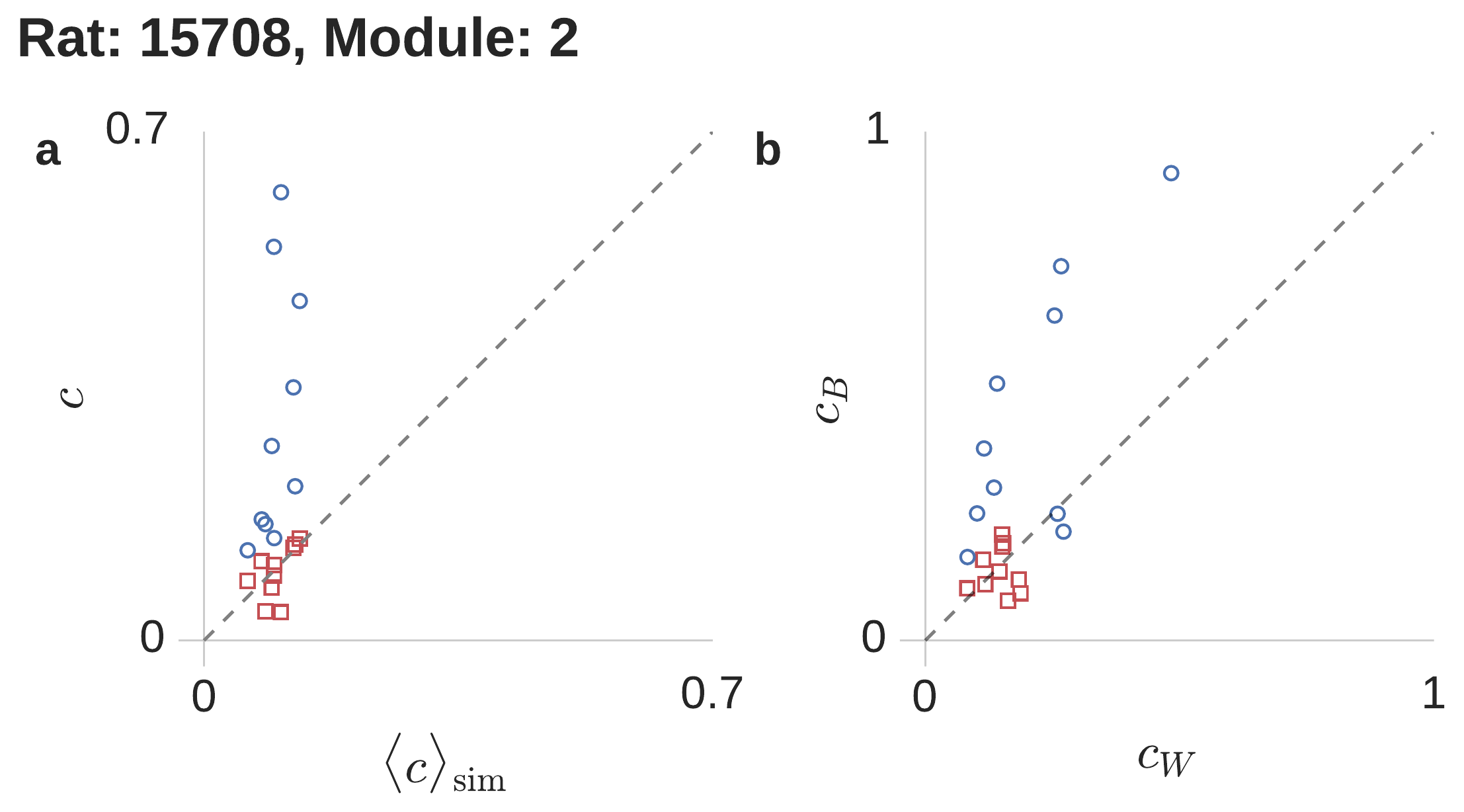}%
  }
  \hfill
  \frame{%
    \includegraphics[width=0.32\linewidth]{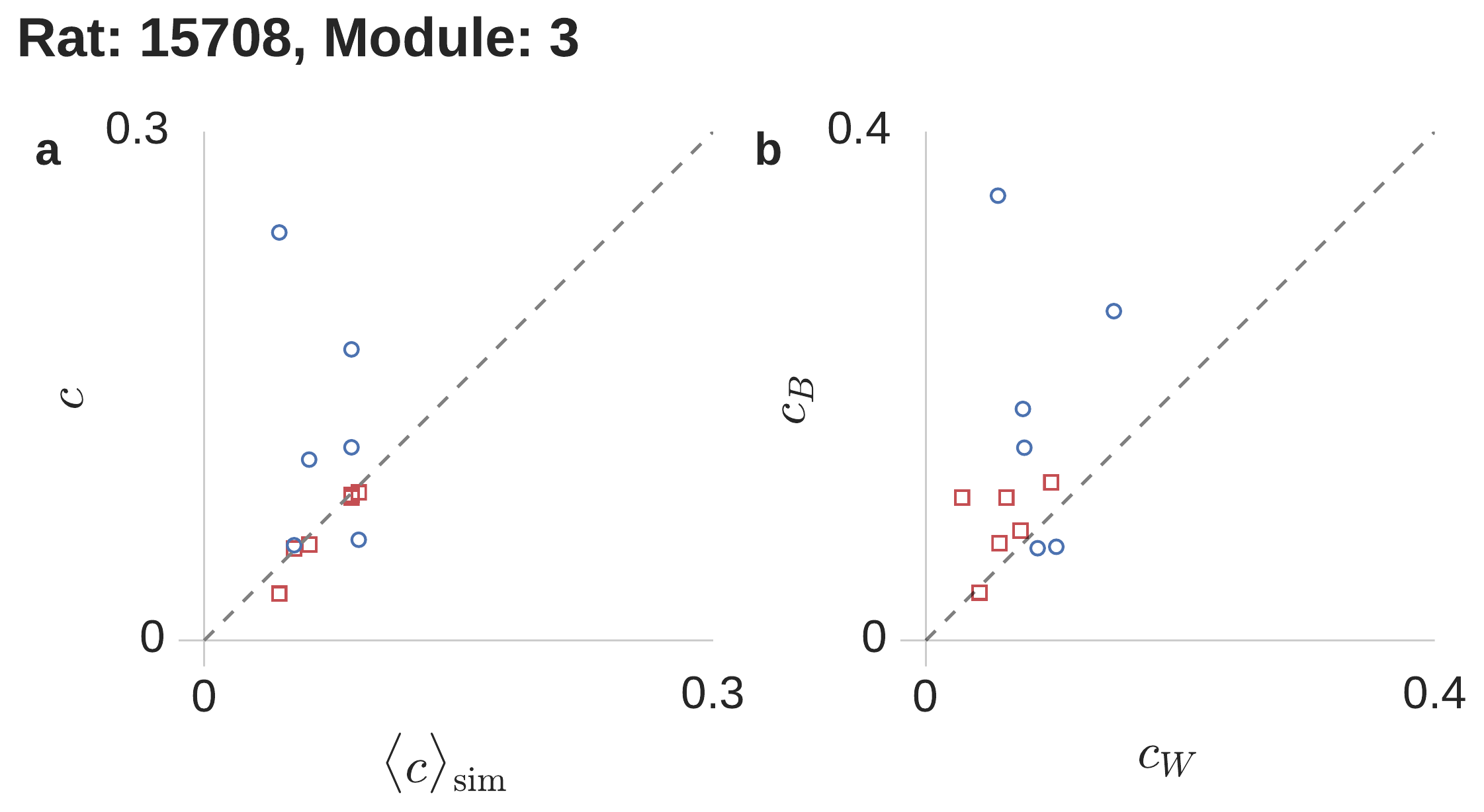}%
  }
  \hfill
  \frame{%
    \includegraphics[width=0.32\linewidth]{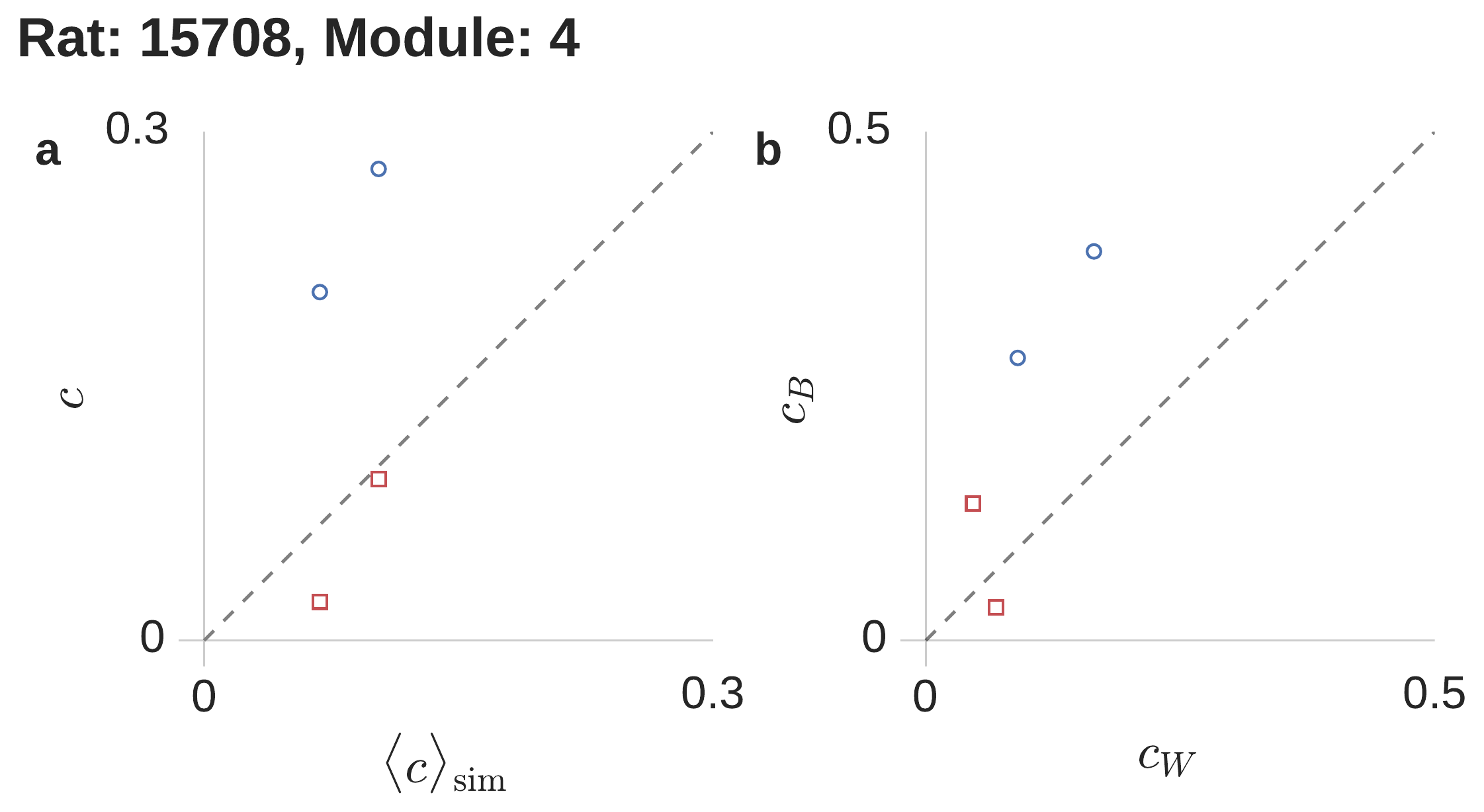}%
  }
\caption{\label{fig:supp-within-between-c}
      \textbf{Firing field amplitude variability.}
      \subfiglabel{a}: Blue circles show the field amplitude coefficient of
      variation of each cell in a module, scattered against the mean
      coefficient of variation from the $1000$ sampled spike trains
      corresponding to that cell. Red squares show the same quantity for one
      particular sampled spike train per real cell.
      \subfiglabel{b}: Between-field coefficient of variability scattered
      against within-field coefficient of variability from the same real cells
      and synthetic spikes as in \subfiglabel{a}. Variabilities are computed
      from field amplitudes measured in the first and second halves of the
      experimental sessions separately.
}
\end{figure}

\begin{figure}
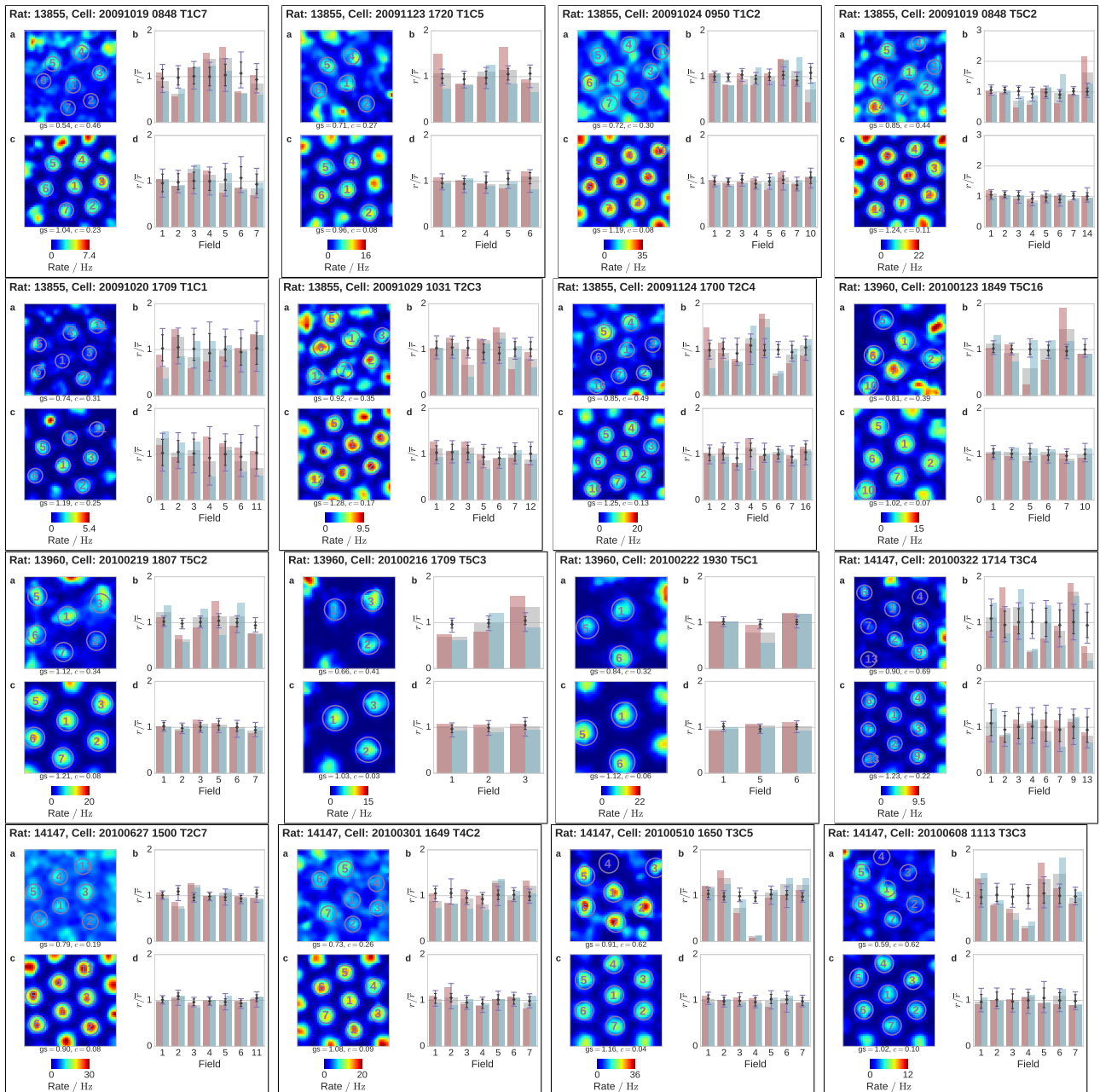

  \centering
  \input{ratemaps-Snipp-mod1.tex}
  \input{ratemaps-Labbetuss-mod1.tex}
  \input{ratemaps-Labbetuss-mod2.tex}
  \input{ratemaps-Labbetuss-mod3.tex}
  \input{ratemaps-Flekken-mod1.tex}
    \caption{\label{fig:supp-synt-real-ratemap-amplitudes}
      \textbf{Example grid cell ratemaps and firing field amplitudes, with
      simulated counterparts.}
      The cells used in this figure are a subset of the cells constituting the
      respective modules (approx.\ every 13th cell in each module).
      \subfiglabel{a}: Grid cell rate map with detected firing fields
      circled and numbered. The grid score ($\text{gs}$), sparsity
      ($\text{sp}$) and field amplitude coefficient of variation ($c$) are
      quoted below the rate map.
      \subfiglabel{b}: The normalized amplitude of each field in
      \subfiglabel{a} is shown (gray wide bars in the background) together
      with the amplitudes measured in only the first and second halves of the
      session (red and light blue bars in the foreground). On top of this,
      markers and error bars visualize the distribution of normalized
      amplitudes from $1000$ synthetic spike trains, with the black diamond
      showing the mean amplitude, the small black whiskers showing the central
      $95 \ \%$ of the full-length amplitudes (to be compared with the gray
      wide bars), and the large purple whiskers showing the central $95 \ \%$
      of the half-length amplitudes (to be compared with the red and light blue
      narrow bars). For each cell and synthetic spike train, amplitudes are
      normalized to the mean of the full-length amplitudes from that cell or
      spike train when creating this plot.
      \subfiglabel{c-d}: Equivalent to \subfiglabel{a-b}, but using sampled
      spikes generated from the real animal trajectory and a tuning profile
      consisting of identical firing fields. Another $1000$ synthetic spike
      trains like this were used to generate the error bars in \subfiglabel{b}
      and \subfiglabel{c}, and to perform the statistical analysis of the
    variability.}
\end{figure}
\begin{figure}\ContinuedFloat
  \centering
  \input{ratemaps-Flekken-mod2.tex}
  \input{ratemaps-Flekken-mod3.tex}
  \input{ratemaps-Kvek-mod1.tex}
  \input{ratemaps-Kvek-mod2.tex}
  \input{ratemaps-Joppe-mod1.tex}
  \input{ratemaps-Joppe-mod2.tex}
  \input{ratemaps-Joppe-mod3.tex}
  \input{ratemaps-Joppe-mod4.tex}
  \caption{Continued.}
\end{figure}

\begin{figure}
  \centering
  \frame{
    \begin{minipage}{0.49\linewidth}
      \centering
      \textsf{\textbf{Fully factorized},} $\sigma = {}$\textsf{0.0}\\
      \includegraphics[width=\linewidth]{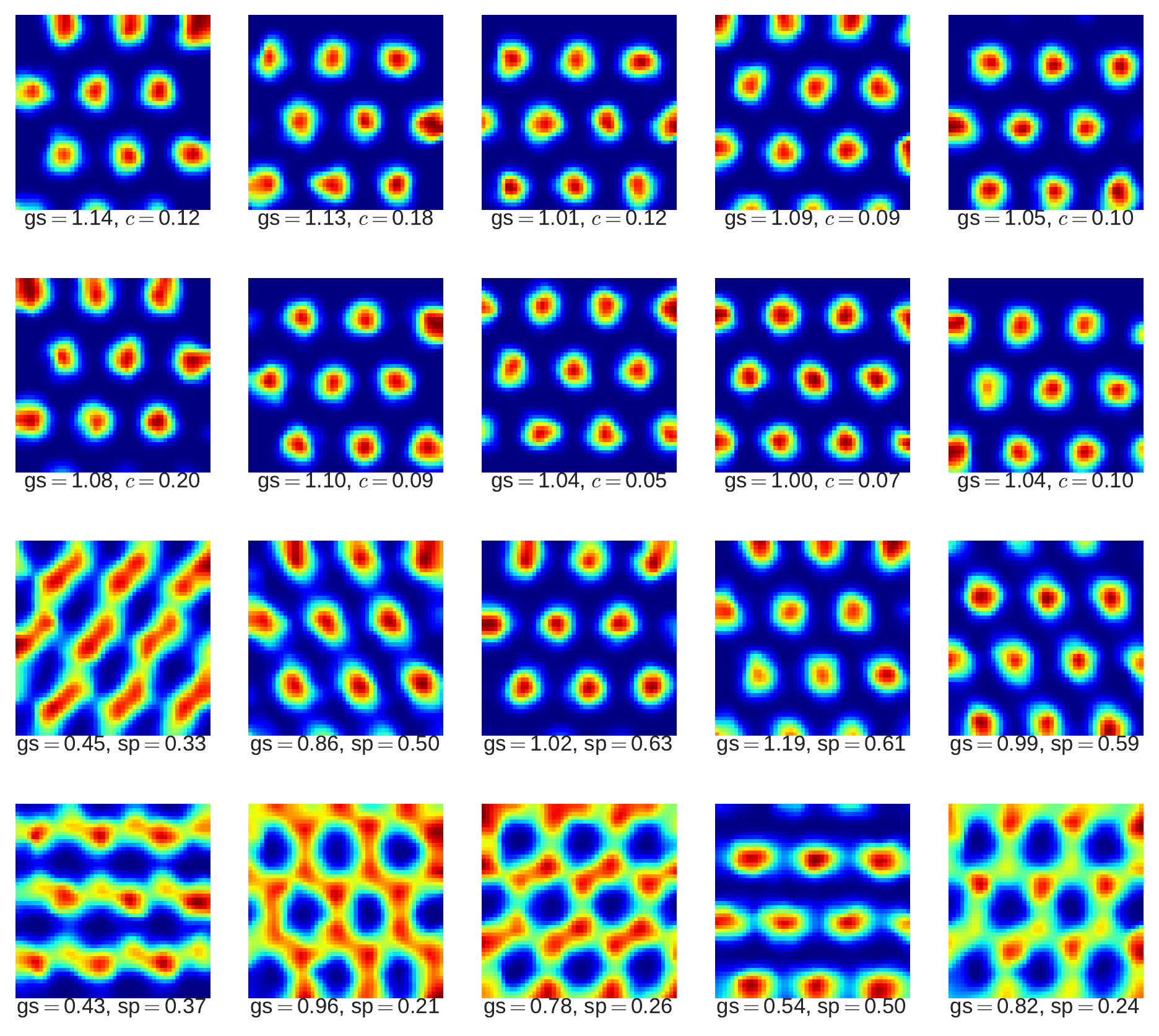}
    \end{minipage}
  }
  \frame{
    \begin{minipage}{0.49\linewidth}
      \centering
      \textsf{\textbf{Fully factorized},} $\sigma = {}$\textsf{0.4}\\
      \includegraphics[width=\linewidth]{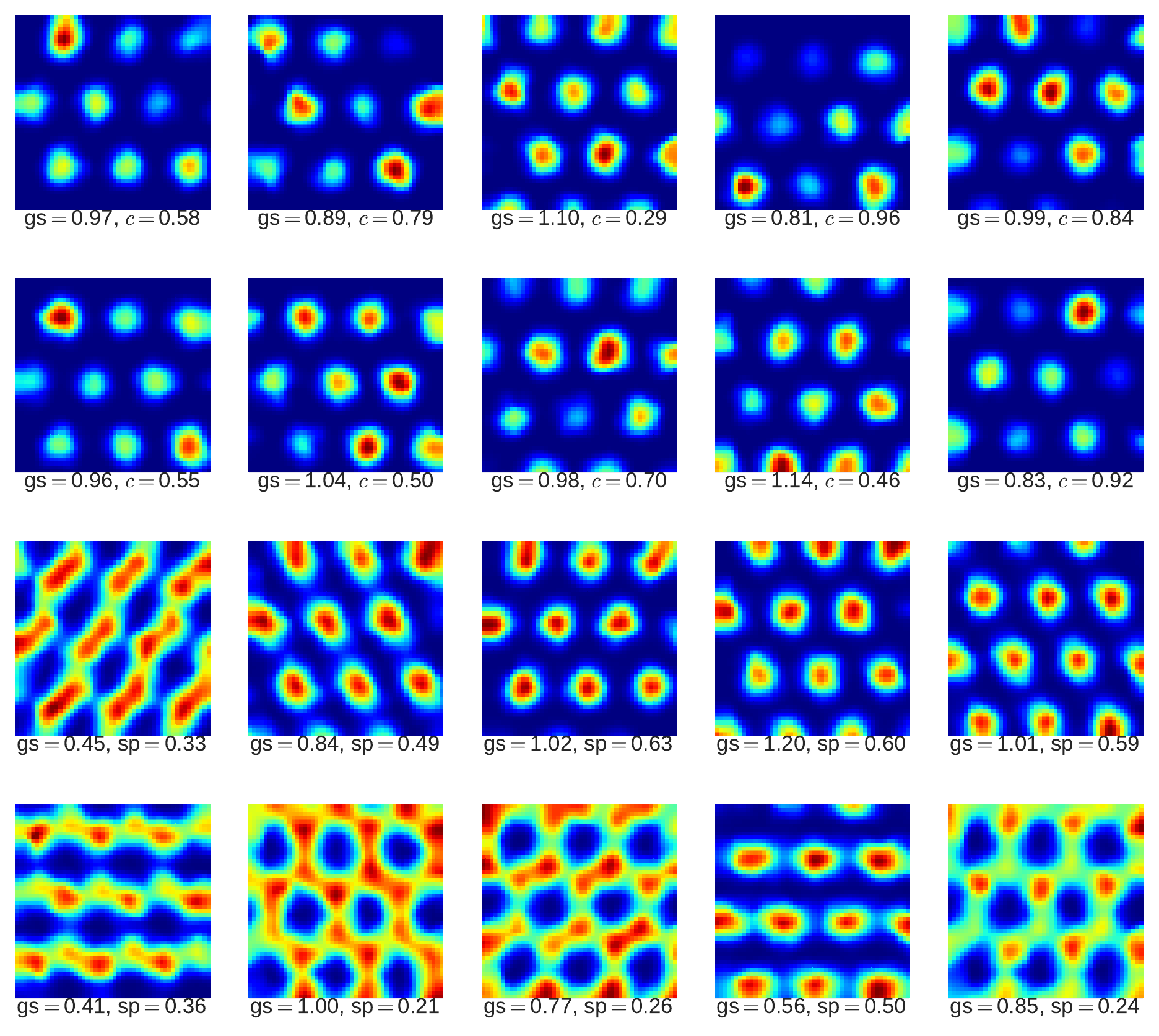}
    \end{minipage}
  }
  \frame{
    \begin{minipage}{0.49\linewidth}
      \centering
      \textsf{\textbf{Fully factorized},} $\sigma = {}$\textsf{0.8}\\
      \includegraphics[width=\linewidth]{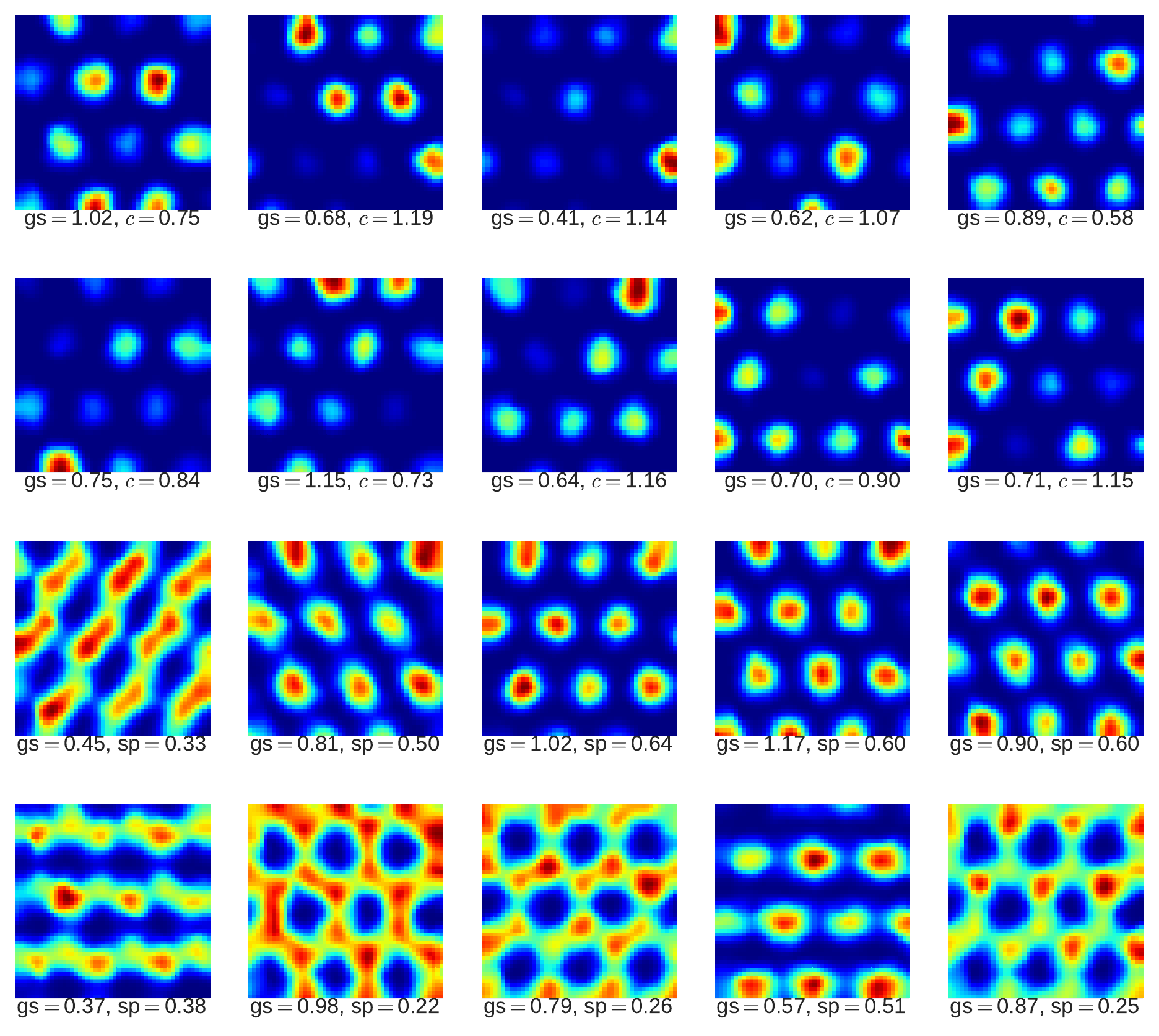}
    \end{minipage}
  }
  \frame{
    \begin{minipage}{0.49\linewidth}
      \centering
      \textsf{\textbf{Fully factorized},} $\sigma = {}$\textsf{1.2}\\
      \includegraphics[width=\linewidth]{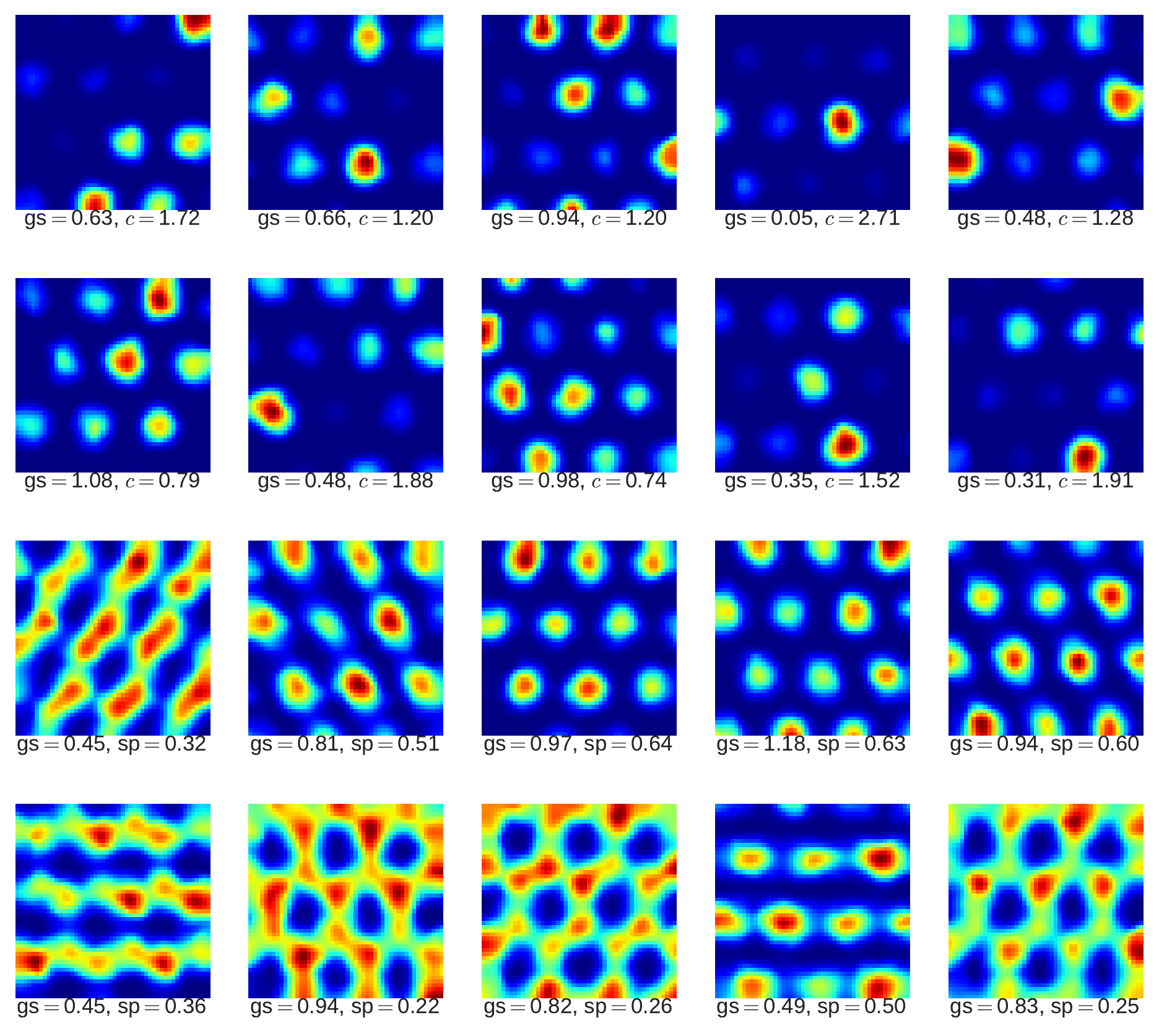}
    \end{minipage}
  }
  \caption{\label{fig:supp-more_ratemaps}
    Additional examples of rate maps from the model simulation runs used in figure 4.
    In each frame, the two top rows are excitatory
    neurons (grid cells) and the two bottom rows are inhibitory neurons
    (interneurons).
  }
\end{figure}
\begin{figure}\ContinuedFloat
  \centering
  \frame{
    \begin{minipage}{0.49\linewidth}
      \centering
      \textsf{\textbf{Partially factorized},} $\sigma = {}$\textsf{0.0}\\
      \includegraphics[width=\linewidth]{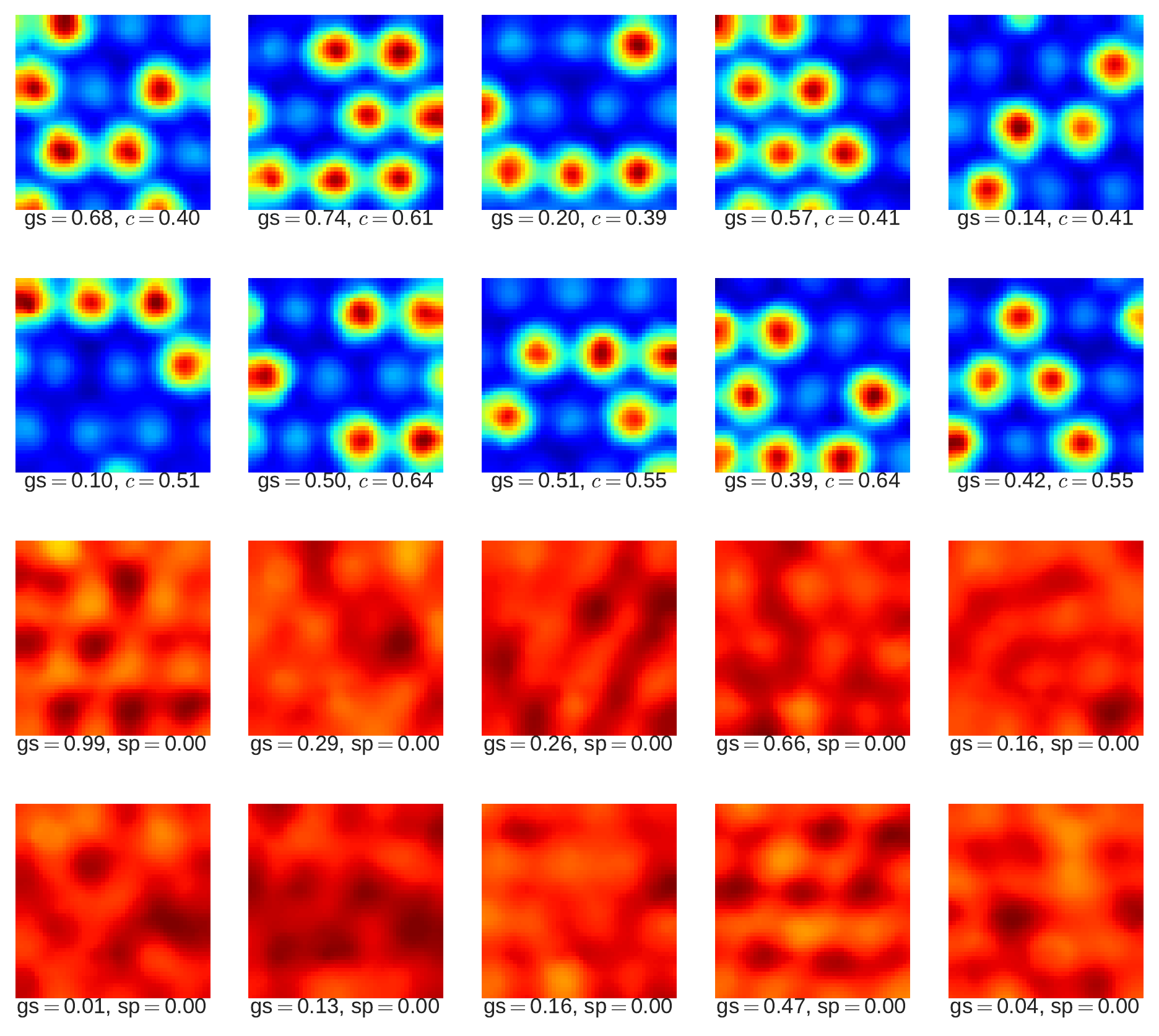}
    \end{minipage}
  }
  \frame{
    \begin{minipage}{0.49\linewidth}
      \centering
      \textsf{\textbf{Partially factorized},} $\sigma = {}$\textsf{0.1}\\
      \includegraphics[width=\linewidth]{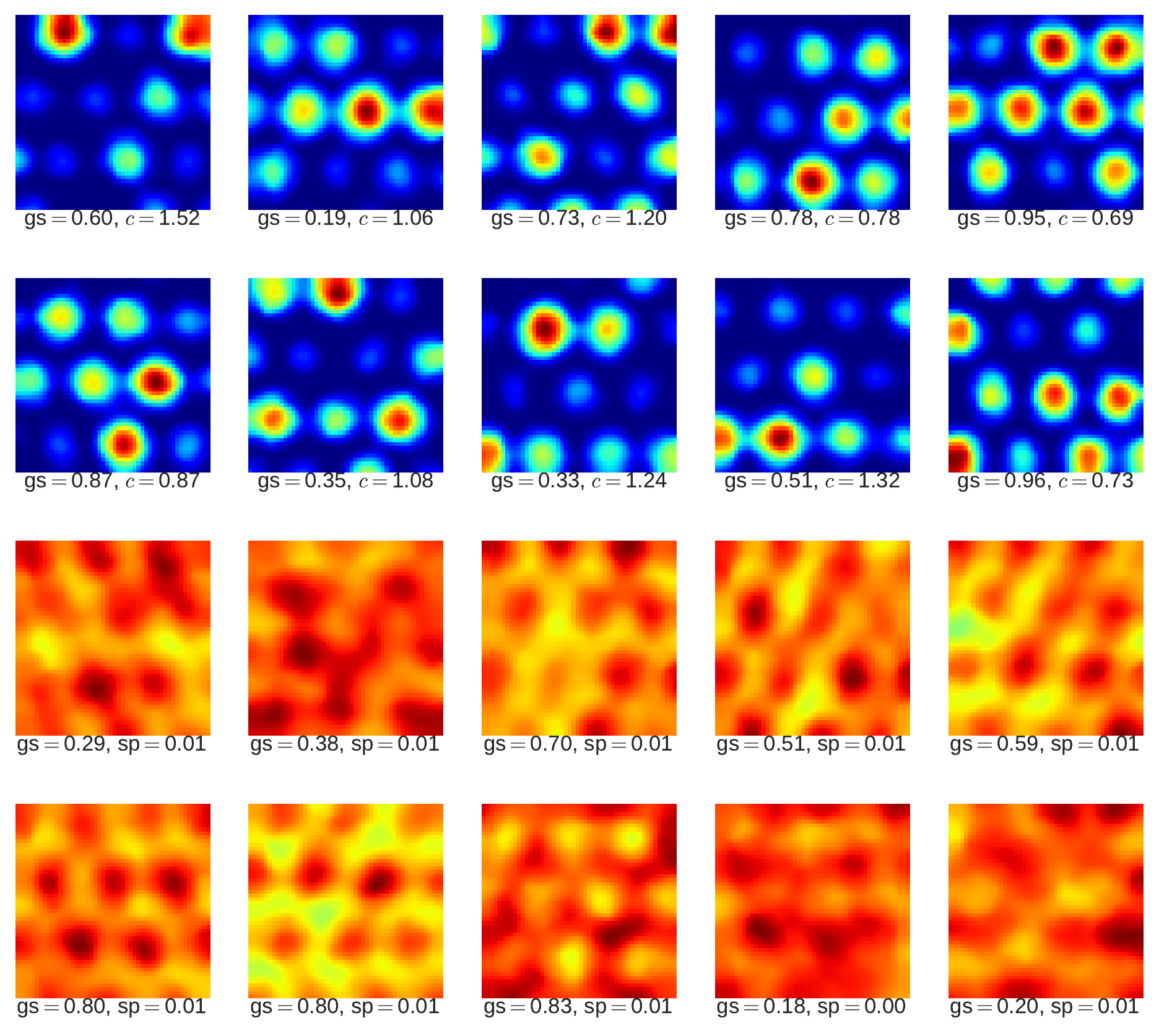}
    \end{minipage}
  }
  \frame{
    \begin{minipage}{0.49\linewidth}
      \centering
      \textsf{\textbf{Partially factorized},} $\sigma = {}$\textsf{0.2}\\
      \includegraphics[width=\linewidth]{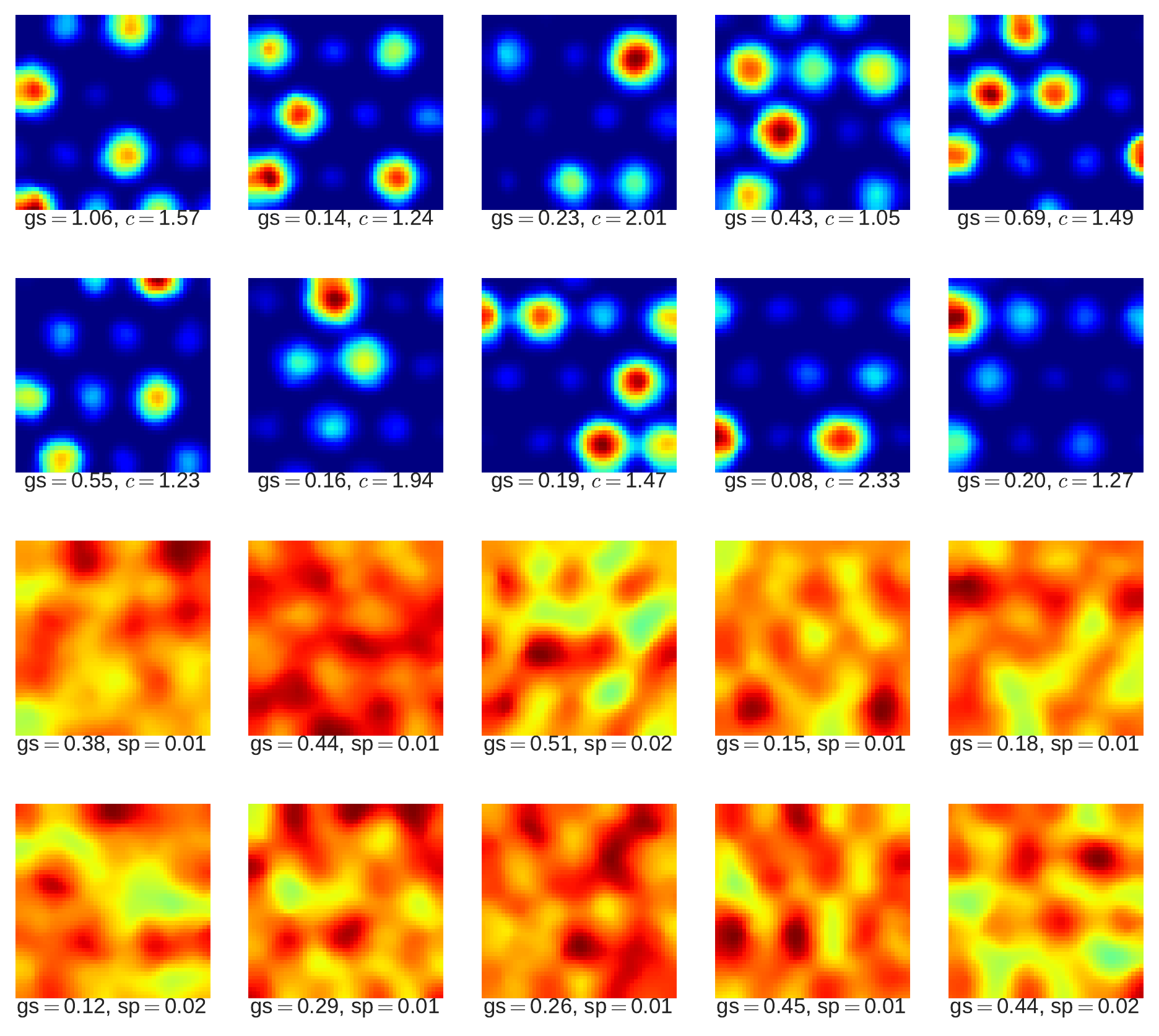}
    \end{minipage}
  }
  \frame{
    \begin{minipage}{0.49\linewidth}
      \centering
      \textsf{\textbf{Partially factorized},} $\sigma = {}$\textsf{0.3}\\
      \includegraphics[width=\linewidth]{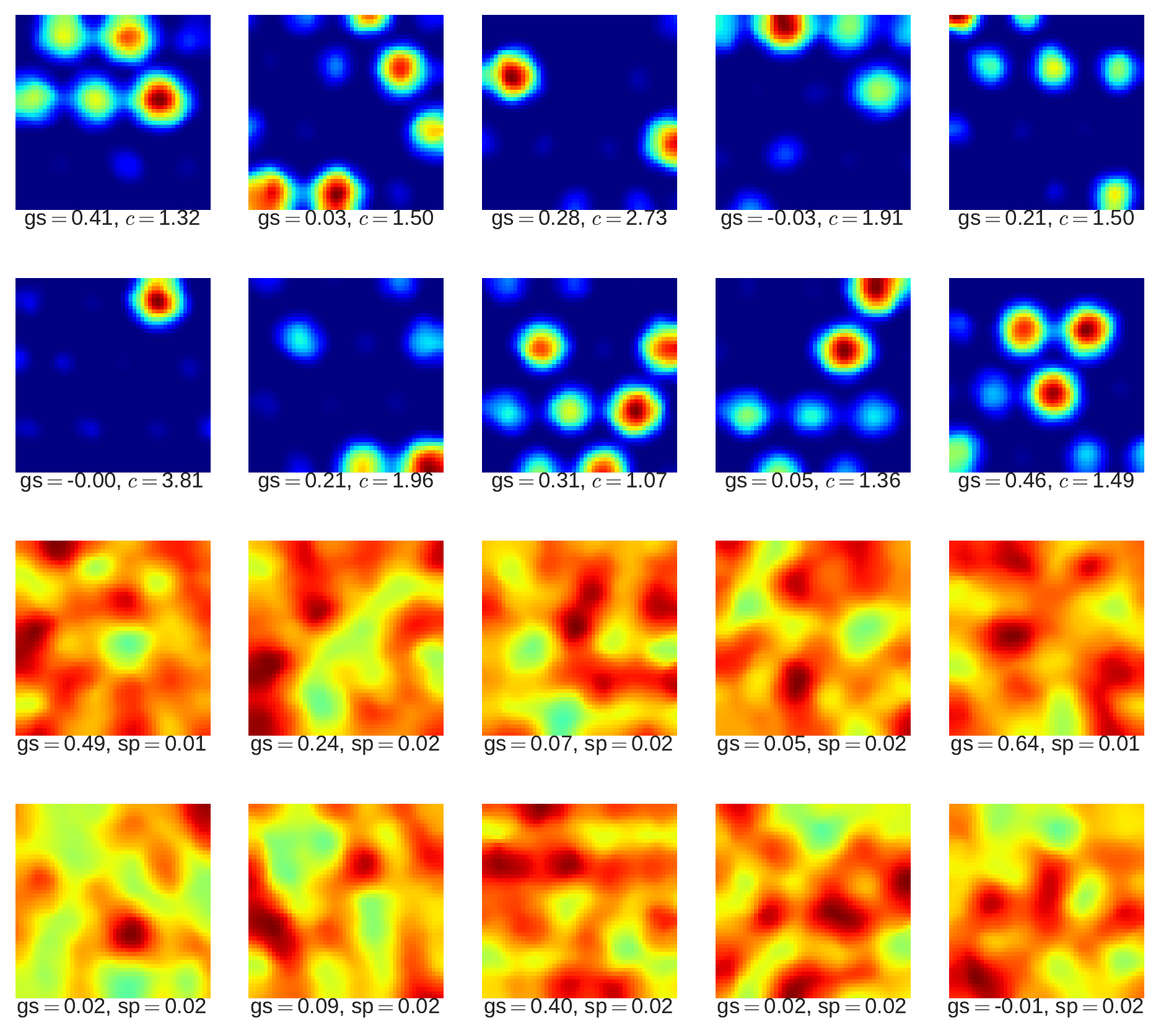}
    \end{minipage}
  }
  \caption{Continued.}
\end{figure}

\begin{figure}
  \centering
  \includegraphics[width=\linewidth]{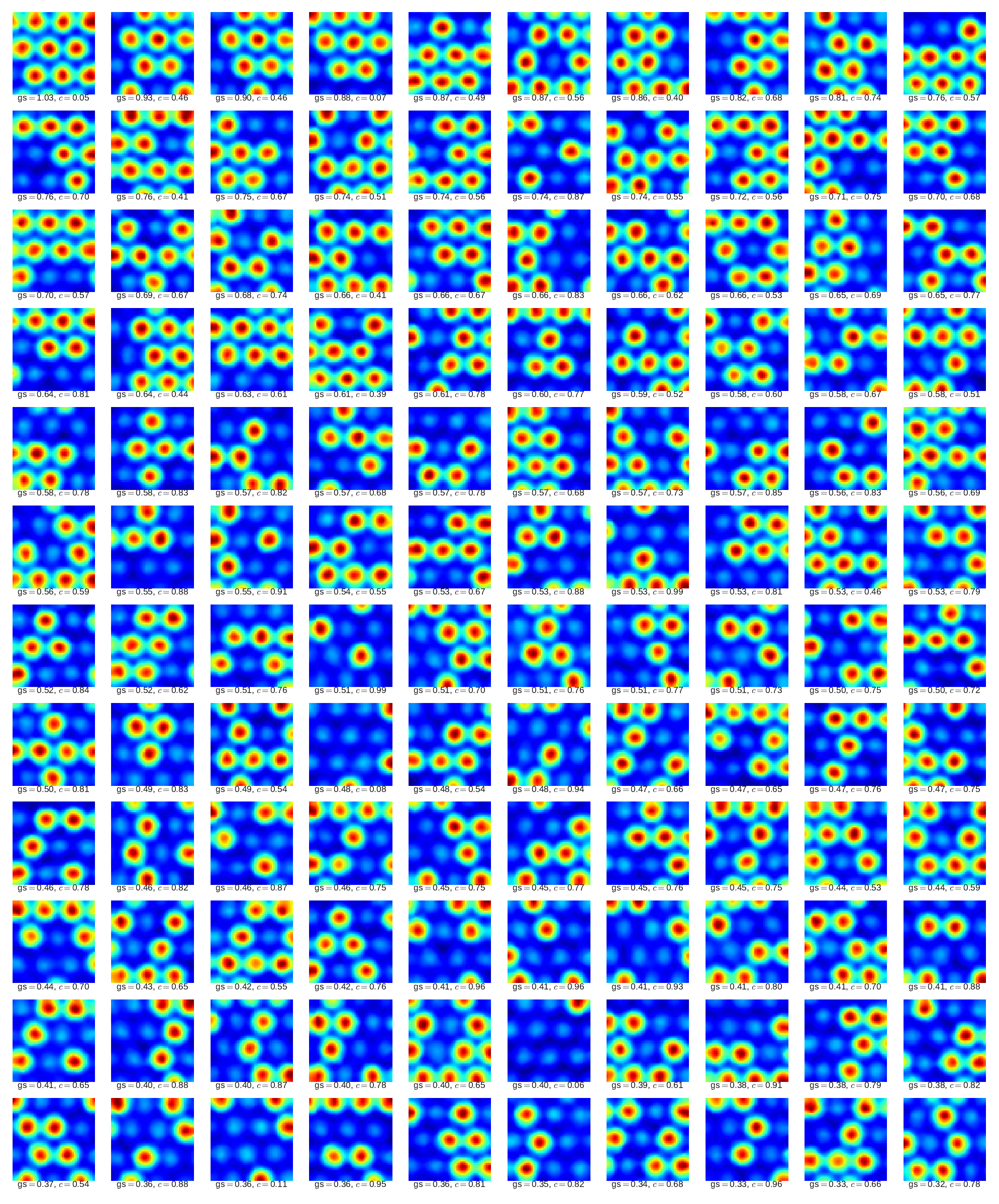}
  \caption{\label{fig:all_ratemaps_partially_factorized}
    Ratemaps of all $200$ excitatory neurons (grid cells) in a run of the
    partially factorized model with zero connectivity variance ($\sigma
    = 0$) and dilution fraction $f_d = 0.5$.  Grid score ($\text{gs}$) and
    field amplitude coefficient of variation ($c$) quoted beneath each ratemap.
    The ratemaps are sorted by grid score, in decreasing order.
  }
\end{figure}
\begin{figure}\ContinuedFloat
  \centering
  \includegraphics[width=\linewidth]{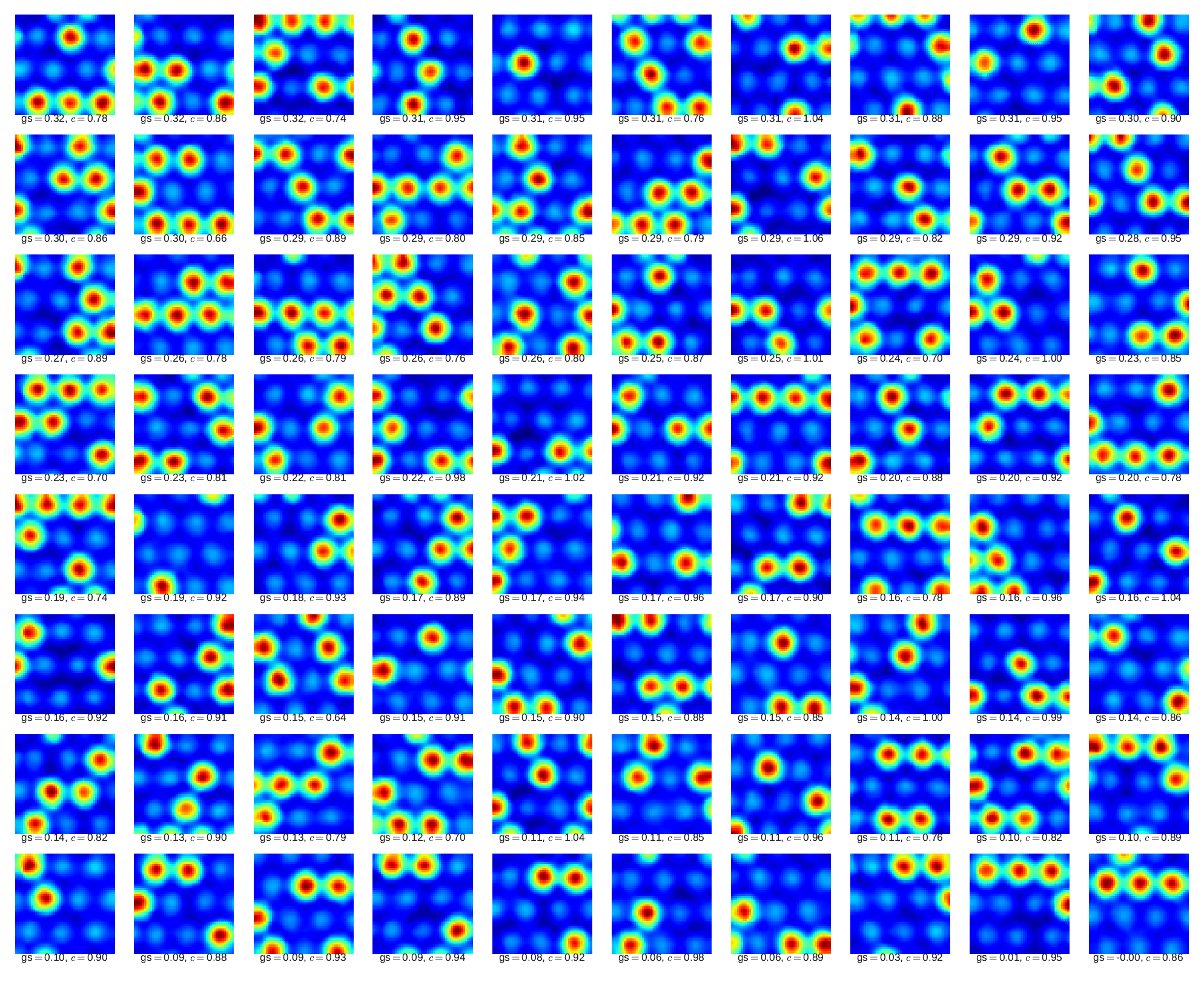}
  \caption{Continued.}
\end{figure}
\end{document}